\documentclass{jfm_arxiv}
\usepackage{graphicx,epstopdf}
\usepackage{caption,subcaption}
\usepackage{newtxtext}
\usepackage{newtxmath}
\usepackage{natbib}
\usepackage{hyperref}
\usepackage{amsmath,amssymb}
\usepackage{float}
\usepackage{booktabs}
\usepackage{enumitem}
\usepackage{tabularx}
\hypersetup{
    colorlinks = true,
    urlcolor   = blue,
    citecolor  = blue,
}
\usepackage{tikz}
\usetikzlibrary{arrows.meta}
\usepackage{pgfplots}
\pgfplotsset{compat=1.16}

\newcommand{\RomanNumeralCaps}[1]

\title{Physics-informed neural networks for two-dimensional wall-reactive solute dispersion in canonical shear flows}

\author{Nanda Poddar\aff{1,2}
\corresp{\email{nandapoddarcr7@gmail.com,
   nandap@srmist.edu.in}}
\and Subham Dhar\aff{3}}

\affiliation{\aff{1}Department of Mathematics, College of Engineering and Technology, SRM Institute of Science and Technology, Kattankulathur, Tamil Nadu 603203, India

\aff{2}School of Mathematical and Statistical Sciences, University of Galway, Galway H91TK33, Ireland

\aff{3}Department of Civil Engineering, National Taiwan University, Taipei City 10617, Taiwan}

\begin{document}
\maketitle

\begin{abstract}
The dispersion of reactive solutes in shear flows is governed by the interplay between advective stretching, transverse diffusion, and boundary exchange kinetics. While classical analytical methods and grid-based numerical solvers have extensively characterised these transport mechanisms, accurately resolving the spatiotemporal evolution of solute plumes in asymmetric reactive environments remains computationally demanding. In this study, we introduce a physics-informed neural-network (PINN) framework to simulate two-dimensional wall-reactive solute dispersion in canonical shear flows (Couette, Poiseuille, and Couette--Poiseuille) bounded by absorbing walls. By embedding the governing convection--diffusion equation and Robin boundary conditions into a unified loss function, the mesh-free PINN reconstructs the spatiotemporal concentration field. The network predictions are validated against an alternating-direction implicit (ADI) finite-difference benchmark, showing close agreement across non-reactive, symmetric, and asymmetric reactive regimes. The computations are carried out at \(\mathrm{Pe}=10\) for impermeable walls, symmetric absorption \((\beta_1,\beta_2)=(1,1)\), and tenfold asymmetric wall-reactivity contrasts \((\beta_1,\beta_2)=(0.2,2)\) and \((2,0.2)\). Leveraging the differentiable nature of the trained PINN, we extract wall-resolved transport diagnostics, including the apparent axial dispersion coefficient, cumulative wall-removal dynamics, and localised uptake fluxes. The results show that the imposed shear profile governs the streamwise organisation of reactive uptake, while unequal wall reactivities induce transverse asymmetry that modifies the macroscopic spreading rate. Overall, this framework establishes PINNs as an interpretable mesh-free tool for analysing boundary-coupled reactive transport in shear flows.
\end{abstract}


\section{Introduction}
\label{sec:intro}
The dispersion of reactive solutes in fluid flows is a ubiquitous phenomenon governing mass transport across a wide spectrum of natural and engineered environments, from groundwater aquifers and environmental flows to chemical reactors and microfluidic devices. The spatiotemporal evolution of these solute plumes is dictated by a complex interplay among advection, molecular diffusion, and chemical kinetics. In shear flows, this coupling is fundamentally amplified; velocity gradients continuously stretch and fold the concentration fields, profoundly altering both the effective macroscopic dispersion and the localised boundary reaction rates. Consequently, a rigorous understanding of these coupled hydro-chemical mechanisms is essential for the accurate prediction of reactive mixing and the optimal design of transport-reliant engineering systems.

The transport of solutes in flowing fluids has been the focus of extensive research for more than seven decades due to its significance in environmental, geophysical, biological, and industrial applications. Understanding the combined effects of molecular diffusion and velocity gradient on solute dispersion was the main goal of early research. Groundbreaking research by \cite{Taylor1953} showed that shear-induced velocity gradients greatly increase longitudinal dispersion beyond molecular diffusion alone. The theoretical basis of solute dispersion in laminar flows was subsequently established by \cite{Aris1956}, who extended Taylor's analysis using moment techniques. The initial transient development of diffusion and the approach to normality were subsequently rigorously examined \citep{Lighthill1966, Chatwin1970}, while \cite{Gill1970} formulated the exact analysis of unsteady convective diffusion. The method of moments was substantially formalised by \cite{Barton1983}, enabling systematic computation of transport coefficients. Three-time regimes are characterised by \cite{Latini2001} for Poiseuille flow. Further explorations of shear dispersion in Poiseuille flows, and straight pipes have since established a robust theoretical framework for characterising concentration distributions \citep{Stokes1990,  Wu2014a, Wu2014, Guan2024}.

However, the conservative tracer and impermeable-boundary assumptions of classical Taylor-Aris theory are rarely satisfied in real-world transport systems. Solutes constantly exchange with surrounding tissues through permeable vessel walls in physiological systems, whereas dissolved species undergo adsorption, absorption, or chemical reactions at the confining interfaces in many environmental applications. These boundary interactions significantly change the mean transport rate and the longitudinal spreading of the solute cloud \citep{Aris1959, Sankarasubramanian1973}. The impact of homogeneous, heterogeneous, and kinetic sorptive reactions has been extensively analysed for tubes, open channels, and annular flows \citep{Gupta1972, Smith1983, Purnama1988, Das1989,  Ng2001, Sarkar2002, Mondal2005, Barik2017, Debnath2020}. More recently, the complexities of reversible and irreversible boundary reactions have been addressed using sophisticated analytical techniques \citep{Ng2006, Ng2008, Jiang2018, Barik2022, Jiang2022}, revealing how phase-exchange mechanisms and boundary desorption govern the transient and asymptotic states of the solute cloud.

While steady models are foundational, contemporary research increasingly focuses on complex, time-dependent hydrodynamic environments. Oscillatory and pulsatile flows introduce intricate advective-diffusive coupling, heavily influencing mass transport in estuaries, blood vessels, and engineered channels \citep{Allen1982, Mazumder1992, Bandyopadhyay1999, Bandyopadhyay1999a, Paul2008, Barik2019}. Analytical and semi-analytical moment techniques have been further extended to capture higher-order statistics (such as skewness and kurtosis) in these unsteady and porous regimes \citep{Das2024b, Jiang2026}. The coupled transport physics is compounded by non-Newtonian rheology \cite{Rana2016} and active suspensions \citep{Caldag2025}. Furthermore, environmental and physiological applications demand the incorporation of vegetation drag, bulk degradation, and magnetohydrodynamic (MHD) effects, leading to highly complex governing equations across steady and oscillatory Couette–Poiseuille flows \citep{Debnath2020, Dhar2021, Poddar2021, Poddar2021a, Poddar2023, Das2024a, Poddar2024a, Poddar2024b, Saha2024}.

Despite the extensive theoretical and experimental characterisation of shear-driven dispersion, numerical simulation of the underlying reactive transport often relies on classical mesh-based discretisation \citep{Mondal2006, Mondal2020, Poddar2021b, Douglas1955, Peaceman1955}. These traditional methods can become computationally demanding, particularly when resolving complex boundary interactions and transient concentration gradients. Recently, Physics-Informed Neural Networks (PINNs) have emerged as a powerful, mesh-free deep learning framework capable of solving nonlinear partial differential equations by embedding physical laws directly into the network's loss function \citep{Raissi2019}.

Early research mainly showed that PINNs could deduce unknown transport parameters from sparse data and solve forward and inverse advection-diffusion equations. For instance, \cite{He2021} demonstrated that PINNs can accurately estimate unknown coefficients from sparse data and reliably recover solutions of advection-dispersion equations over a broad range of P\'eclet numbers. The numerical accuracy and convergence of advection-diffusion-reaction simulations were then improved by the orthogonal-grid PINN framework of \cite{Hou2022}, which incorporated derivative restrictions and structured collocation techniques during training. In recent years, PINNs have been utilised to solve complex environmental transport problems, including coupled surface flow and solute transport \citep{Niu2023}, multispecies contaminant migration with spatially varying transport parameters \citep{Hou2025}, adaptive Runge--Kutta PINNs for stiff multicomponent reactive transport \citep{shu2026}, surrogate modelling of reactive nitrate transport in groundwater \citep{arab2026}, and passive scalar emission \citep{Rawden2026}. In inverse advection-diffusion-reaction problems, where unknown reaction rates and transport coefficients are obtained concurrently with the concentration field, PINNs have also been investigated \citep{mamud2023}. Furthermore, advanced PINN architectures have been increasingly deployed for complex, coupled systems, such as reactive solute transport in parameterised groundwater models \citep{Jiao2026}, solute concentration distributions in continuous crystallizers \citep{Cui2026}, and water and nitrogen transport in unsaturated soils \citep{Kamil2025}. Related machine-learning techniques have been utilised to infer flow fields around active colloidal particles \citep{Mohapatra2025}, explore soliton collisions \citep{Qiu2025}, and discover multiple PDE solutions using deep ensembles \citep{Zou2025}. Most studies of solute dispersion adopt a Dirac delta initial condition to model instantaneous tracer release, whereas only a few consider Gaussian pulses \citep{Teng2023}. The singularity associated with the Dirac delta source poses a significant challenge for PINNs, which rely on smooth neural-network approximations and automatic differentiation.

Motivated by these computational challenges and the recent successes of deep learning in fluid mechanics, this study introduces a physics-informed neural network (PINN) framework specifically tailored for two-dimensional reactive solute dispersion in canonical shear flows. While traditional grid-based solvers demand high-resolution meshing to accurately capture sharp concentration gradients near reactive boundaries, our mesh-free approach embeds the governing convection--diffusion equation, localised initial source distributions, and reactive Robin boundary constraints directly into a unified composite loss function. To rigorously establish the physical fidelity of this data-driven approach, the learned PINN solutions are systematically validated against a classical alternating-direction implicit (ADI) finite-difference benchmark \citep{Douglas1955, Peaceman1955, Mondal2006}. However, the novelty of the present work lies not merely in reconstructing the spatiotemporal concentration field with high accuracy, but in leveraging the continuous, fully differentiable nature of the trained PINN surrogate to extract complex transport diagnostics. By circumventing the discrete approximations required by classical numerical methods, we directly compute the time-dependent effective dispersion coefficient, total surviving mass, and localised wall fluxes across Couette, Poiseuille, and Couette--Poiseuille flows. This continuous representation enables a precise evaluation of reaction-induced transverse asymmetry, cumulative wall-removal dynamics, and the streamwise organisation of reactive uptake. Ultimately, this study demonstrates that PINNs can serve not just as an alternative PDE solver but as a highly interpretable, data-efficient tool for decoding the coupled hydrodynamic and chemical mechanisms governing solute transport.

The remainder of this paper is organised as follows. Section \ref{sec:math-formulation} details the mathematical formulation of the two-dimensional reactive transport model, including the governing equations, initial source configurations, and the mean-centred canonical shear flows. Section \ref{sec:pinn-method} outlines the physics-informed neural network methodology, detailing the composite loss construction, training strategy, physical interpretation of the learned solution, and the ADI finite-difference benchmark. Section \ref{sec:results-discussion} presents the results and discussion, beginning with a rigorous validation of the PINN framework against the ADI benchmark, followed by an in-depth analysis of wall-resolved reactive dispersion, reaction-induced transverse asymmetry, and cumulative wall-removal dynamics. Finally, the main conclusions of this study are summarised in Section \ref{sec:conclusions}.

\section{Mathematical formulation}
\label{sec:math-formulation}

\subsection{Geometry and variables}
We consider two-dimensional solute transport in a laminar shear flow between two parallel plates located at $y^*=\pm H$. The coordinates $x^*$ and $y^*$ denote the streamwise and wall-normal directions, respectively, and $t^*$ denotes time. The solute concentration is denoted by $C^*(x^*,y^*,t^*)$. The dimensional flow domain is
\[
\Omega^*=\{(x^*,y^*): x^*\in\mathbb{R},\; -H\leq y^*\leq H\}.
\]

\subsection{Dimensional convection--diffusion model}
The solute is transported by a prescribed unidirectional shear flow $u^*(y^*)$ and diffuses isotropically with molecular diffusivity $D$. The governing dimensional convection--diffusion equation is
\begin{equation}
\frac{\partial C^*}{\partial t^*}
+
u^*(y^*)\frac{\partial C^*}{\partial x^*}
=
D\left(
\frac{\partial^2 C^*}{\partial x^{*2}}
+
\frac{\partial^2 C^*}{\partial y^{*2}}
\right),
\qquad (x^*,y^*)\in\Omega^*,\quad t^*>0.
\label{eq:dim-conv-diff}
\end{equation}
Here $u^*(y^*)$ represents a canonical parallel-plate shear profile, specified below.

At the plates, the solute undergoes first-order uptake. With reaction rates \(\kappa_1\) and \(\kappa_2\) at the upper wall \(y^*=H\) and lower wall \(y^*=-H\), respectively, the dimensional Robin conditions are
\begin{equation}
-D\frac{\partial C^*}{\partial y^*}-\kappa_1 C^*=0
\quad \text{at } y^*=H,
\qquad
D\frac{\partial C^*}{\partial y^*}-\kappa_2 C^*=0
\quad \text{at } y^*=-H.
\label{eq:dim-robin}
\end{equation}
The case $\kappa_i=0$ corresponds to an impermeable wall, while increasing $\kappa_i$ represents stronger wall absorption. These first-order Robin conditions are widely used to model irreversible absorption, reversible phase exchange, and retention kinetics at the boundaries of confined flows \cite{Purnama1988, Ng2008, Poddar2021}.

A localised solute release is prescribed at $t^*=0$:
\begin{equation}
C^*(x^*,y^*,0)=C_0^*(x^*,y^*).
\label{eq:dim-ic}
\end{equation}
\subsection{Nondimensionalization}
We introduce the dimensionless variables
\[
x=\frac{x^*-\overline{u}^*t^*}{H},\qquad
y=\frac{y^*}{H},\qquad
t=\frac{D t^*}{H^2},\qquad
C=\frac{C^*}{C_{\mathrm{ref}}},\qquad
u(y)=\frac{u^*(y^*)}{U_{\mathrm{ref}}},\quad \overline{u}=\frac{\overline{u}^*}{U_{\mathrm{ref}}}
\]
where \(U_{\mathrm{ref}}\) is the characteristic velocity scale associated with the imposed shear flow and the overbar denotes the corresponding cross-sectional average. The corresponding P\'eclet number and wall-reactivity parameters are
\[
\mathrm{Pe}=\frac{U_{\mathrm{ref}}H}{D},
\qquad
\beta_i=\frac{\kappa_i H}{D},\qquad i=1,2.
\]
The dimensionless convection--diffusion equation becomes
\begin{equation}
\frac{\partial C}{\partial t}
+
\mathrm{Pe}\,\left(u(y)-\overline{u}\right)\frac{\partial C}{\partial x}
=
\frac{\partial^2 C}{\partial x^2}
+
\frac{\partial^2 C}{\partial y^2},
\qquad x\in\mathbb{R},\quad -1<y<1,\quad t>0.
\label{eq:nd-conv-diff}
\end{equation}
In the computations below, this mean-deviated velocity is denoted directly by \(u^{(s)}(y)\) for each canonical shear profile.
The dimensionless reactive wall conditions are
\begin{equation}
-\frac{\partial C}{\partial y}-\beta_1 C=0
\quad \text{at } y=1,
\qquad
\frac{\partial C}{\partial y}-\beta_2 C=0
\quad \text{at } y=-1.
\label{eq:nd-robin}
\end{equation}
The associated local dimensionless uptake fluxes at the upper and lower plates are
\begin{equation}
J_+(x,t)=\beta_1 C(x,1,t),
\qquad
J_-(x,t)=\beta_2 C(x,-1,t).
\label{eq:wall-uptake-flux}
\end{equation}
Here \(J_+\) and \(J_-\) denote the uptake fluxes at the upper and lower plates, respectively. These fluxes provide a local physical measure of how each imposed shear flow transports solute toward the reactive walls.

\subsection{Initial source configurations}
To test the sensitivity of reactive dispersion to the form of the initial release, we consider two smooth Gaussian source configurations. The first is a streamwise-localised line-like source,
\begin{equation}
C_0^{\mathrm{L}}(x,y)=\exp(-x^2),
\qquad x\in[x_0,x_1],\quad -1\leq y\leq 1.
\label{eq:ic-line-source}
\end{equation}
This source is localised in the streamwise direction and uniform across the channel height.

The second is a point-like two-dimensional Gaussian source,
\begin{equation}
C_0^{\mathrm{P}}(x,y)=\exp(-x^2-y^2),
\qquad x\in[x_0,x_1],\quad -1\leq y\leq 1.
\label{eq:ic-point-source}
\end{equation}
This source is localised in both the streamwise and wall-normal directions and provides a smooth approximation to a point release centred at the channel midplane.

Both choices avoid directly imposing a singular Dirac delta distribution. A Dirac delta is not a classical continuous function and is therefore not compatible with the pointwise mean-square initial-condition loss used in the present PINN formulation. The Gaussian sources may be interpreted as mollified releases; singular-source formulations are left for future work through weak-form PINNs, Green 's-function-based formulations or distributional residuals.

\subsection{Mean-centred canonical shear-flow choices}
In \eqref{eq:nd-conv-diff}, the imposed velocity is written as
\[
u(y)=u^{(s)}(y),
\qquad
s\in\mathcal{S}:=\{\mathrm{C},\mathrm{Po},\mathrm{CP}\},
\]
where \(\mathrm{C}\), \(\mathrm{Po}\) and \(\mathrm{CP}\) denote Couette, Poiseuille and Couette--Poiseuille flow, respectively. The same source configuration, wall reactivity and axial boundary conditions are imposed for each \(s\in\mathcal{S}\); hence, differences in the computed concentration fields are attributable to the imposed shear profile.

The nondimensionalization fixes the velocity scale through \(U_{\mathrm{ref}}\). The following centring is therefore not a rescaling but a choice of frame. Let
\[
\langle f\rangle_y=\frac{1}{2}\int_{-1}^{1} f(y)\mathrm{d}y
\]
denote the cross-sectional average. For each canonical flow, we write
\begin{equation}
u^{(s)}(y)
=
\widetilde{u}^{(s)}(y)
-
\left\langle \widetilde{u}^{(s)} \right\rangle_y,
\qquad
\left\langle u^{(s)} \right\rangle_y=0.
\label{eq:mean-centred-velocity}
\end{equation}
Here \(\widetilde{u}^{(s)}(y)\) denotes the usual dimensionless shear profile, while \(u^{(s)}(y)\) is the mean-centred profile used in the transport equation. This is equivalent to observing the solute cloud in a frame moving with the bulk velocity. It removes uniform streamwise translation while retaining the shear responsible for dispersion, wall contact and reactive uptake.

For the dimensionless channel \(-1\leq y\leq 1\), the three mean-centred profiles are:
\begin{enumerate}[label=\textbf{(F\arabic*)},leftmargin=*,itemsep=2pt]
\item \textbf{Couette flow.}
The standard one-moving-wall Couette profile is
\[
\widetilde{u}^{(\mathrm{C})}(y)=\frac{1+y}{2},
\qquad
\left\langle \widetilde{u}^{(\mathrm{C})}\right\rangle_y=\frac12.
\]
Hence, the mean-deviated Couette profile is
\begin{equation}
u^{(\mathrm{C})}(y)
=
\frac{1+y}{2}-\frac12
=
\frac{y}{2},
\qquad -1\leq y\leq 1.
\label{eq:u-couette}
\end{equation}

\item \textbf{Poiseuille flow.}
The standard pressure-driven parabolic profile is
\[
\widetilde{u}^{(\mathrm{Po})}(y)=\frac12(1-y^2),
\qquad
\left\langle \widetilde{u}^{(\mathrm{Po})}\right\rangle_y=\frac13.
\]
Therefore, the mean-deviated Poiseuille profile is
\begin{equation}
u^{(\mathrm{Po})}(y)
=
\frac12(1-y^2)-\frac13,
\qquad -1\leq y\leq 1.
\label{eq:u-poiseuille}
\end{equation}

\item \textbf{Couette--Poiseuille flow.}
The combined uncentred profile is
\[
\widetilde{u}^{(\mathrm{CP})}(y)
=
\frac{1+y}{2}
+
\frac12(1-y^2),
\qquad
\left\langle \widetilde{u}^{(\mathrm{CP})}\right\rangle_y=\frac56.
\]
Thus, the mean-deviated Couette--Poiseuille profile is
\begin{equation}
u^{(\mathrm{CP})}(y)
=
\frac{1+y}{2}
+
\frac12(1-y^2)
-
\frac56,
\qquad -1\leq y\leq 1.
\label{eq:u-couette-poiseuille}
\end{equation}
Equivalently,
\[
u^{(\mathrm{CP})}(y)
=
u^{(\mathrm{C})}(y)+u^{(\mathrm{Po})}(y).
\]
\end{enumerate}
The Couette--Poiseuille configuration is a vital model for investigating transport in which pressure-driven and boundary-driven shear flows interact, a scenario common in complex hydrodynamics and microfluidic applications \cite{Paul2008, Barik2019, Poddar2023}.

Consequently, for each shear-flow case \(s\in\mathcal{S}\) and source configuration \(q\in\mathcal{Q}:=\{\mathrm{L},\mathrm{P}\}\), the concentration field \(C^{(s,q)}(x,y,t)\) satisfies
\begin{equation}
\frac{\partial C^{(s,q)}}{\partial t}
+
\mathrm{Pe}\,u^{(s)}(y)
\frac{\partial C^{(s,q)}}{\partial x}
=
\frac{\partial^2 C^{(s,q)}}{\partial x^2}
+
\frac{\partial^2 C^{(s,q)}}{\partial y^2},
\qquad x\in\mathbb{R},\quad -1<y<1,\quad t>0 .
\label{eq:nd-conv-diff-shear}
\end{equation}

\subsection{Axial boundary modelling on a finite window}
For the PINN and finite-difference computations, the infinite streamwise direction is truncated to a finite interval
\[
x\in[x_0,x_1],
\qquad
x_0=-\frac{L_x}{2},\quad x_1=\frac{L_x}{2}.
\]
The computations are performed over the dimensionless time interval \(0\leq t\leq T\), where \(T\) denotes the final observation time. The computational length \(L_x\) is chosen sufficiently large so that the localised solute cloud remains away from the artificial axial boundaries over the time interval considered. Since the initial release is localised near the interior of the domain, we impose homogeneous zero-gradient conditions at the two axial boundaries,
\begin{equation}
\frac{\partial C}{\partial x}\left(-\frac{L_x}{2},y,t\right)=0,
\qquad
\frac{\partial C}{\partial x}\left(\frac{L_x}{2},y,t\right)=0,
\qquad -1\leq y\leq 1,\quad t>0 .
\label{eq:axial-neumann}
\end{equation}
These conditions close the finite computational problem without imposing artificial concentration loss through the streamwise boundaries. The same axial boundary treatment is used in the PINN loss and in the finite-difference benchmark.

\section{Methodology}
\label{sec:pinn-method}

\subsection{Neural-network approximation as a function-space representation}
For each shear-flow case \(s\in\mathcal{S}:=\{\mathrm{C},\mathrm{Po},\mathrm{CP}\}\) and each source configuration \(q\in\mathcal{Q}:=\{\mathrm{L},\mathrm{P}\}\), the concentration field
\[
C^{(s,q)}:(x,y,t)\mapsto C^{(s,q)}(x,y,t)
\]
is approximated by a feed-forward neural network
\[
C^{(s,q)}(x,y,t)\approx C_{\theta}^{(s,q)}(x,y,t)
:=\mathcal{N}_{\theta}^{(s,q)}(x,y,t),
\label{eq:pinn-approx}
\]
where \(\theta\) denotes the trainable weights and biases. Here \(s\) identifies the imposed shear flow, while \(q\) identifies the initial source configuration. The neural network may be viewed as a finite-dimensional nonlinear approximation space; training selects a member of this function class that satisfies the governing convection--diffusion equation, the prescribed initial source and the boundary constraints in a least-squares residual sense.

The input layer receives the space--time variables \((x,y,t)\), and the output layer returns the scalar concentration approximation \(C_{\theta}^{(s,q)}(x,y,t)\). In the computations reported here, we use a fully connected network with four hidden layers, each containing \(64\) neurons with \(\tanh\) activation. The choice of a smooth activation function is important because the residual of the convection--diffusion equation involves first- and second-order derivatives of \(C_{\theta}^{(s,q)}\) with respect to \(x\), \(y\) and \(t\). These derivatives are evaluated by automatic differentiation.

\begin{figure}
    \centering
    \begin{subfigure}[t]{0.6\textwidth}
        \centering
        \includegraphics[width=\textwidth]{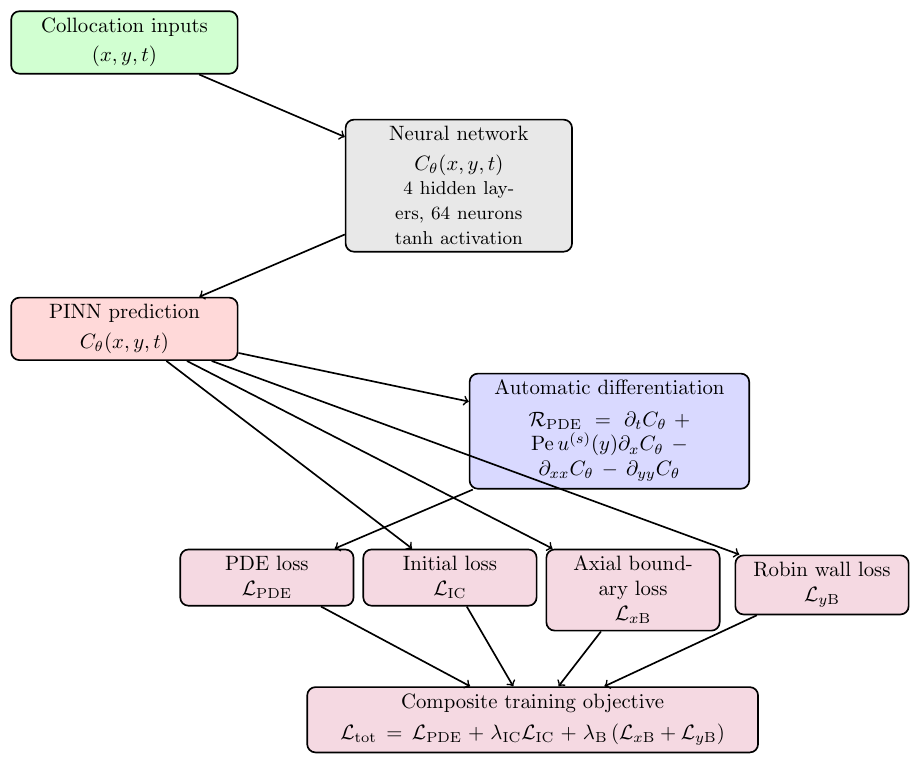}
        \caption{Physics-informed workflow and composite loss construction.}
        \label{fig:pinn-workflow}
    \end{subfigure}
    \hfill
    \begin{subfigure}[t]{0.38\textwidth}
        \centering
        \includegraphics[width=\textwidth]{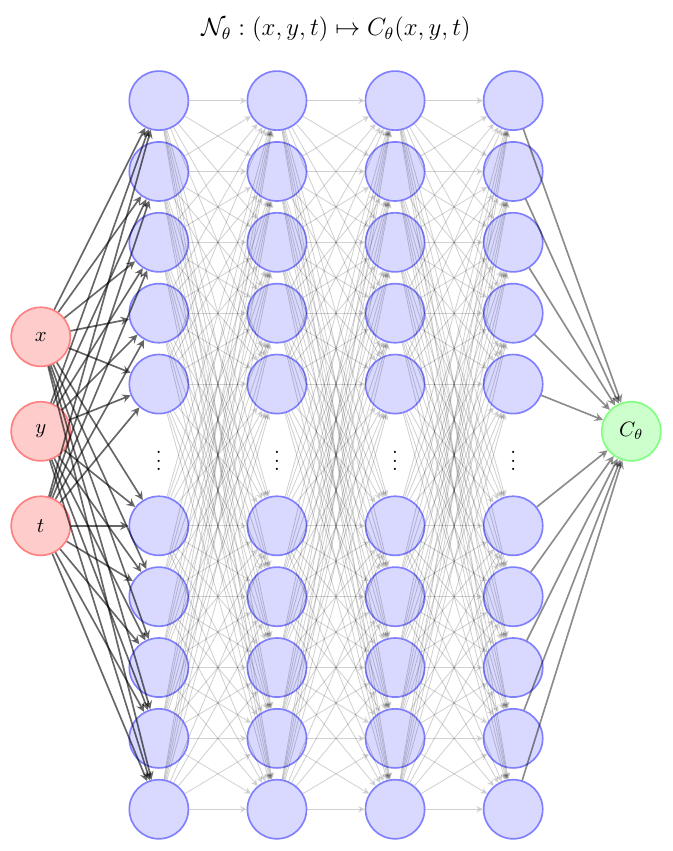}
        \caption{Detailed fully connected neural-network architecture.}
        \label{fig:pinn-detailed-network}
    \end{subfigure}

    \vspace{0.35cm}

    \begin{subfigure}[t]{1\textwidth}
        \centering
        \includegraphics[width=\textwidth]{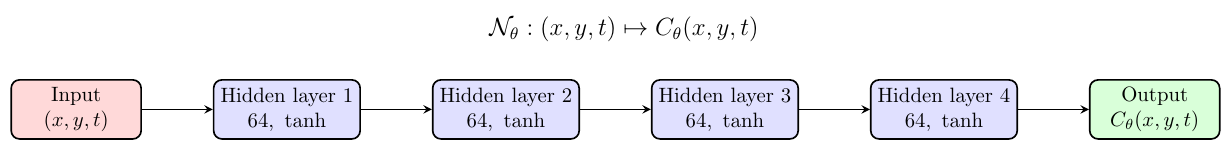}
        \caption{Compact layer-wise representation of the PINN architecture.}
        \label{fig:pinn-compact-network}
    \end{subfigure}

    \caption{
    Schematic of the physics-informed neural-network framework.}
    \label{fig:pinn-framework}
\end{figure}

The overall PINN workflow is summarised in figure~\ref{fig:pinn-framework}. The schematic shows how the space--time inputs are mapped to the concentration prediction, how automatic differentiation is used to construct the PDE and boundary residuals, and how these residuals are combined into the composite training objective.

\subsection{Physics-informed residual}
For a prescribed shear profile \(u^{(s)}(y)\), the dimensionless convection--diffusion equation is enforced through the residual
\begin{equation}
\mathcal{R}_{\theta}^{(s,q)}(x,y,t)
=
\frac{\partial C_{\theta}^{(s,q)}}{\partial t}
+
\mathrm{Pe}\,u^{(s)}(y)
\frac{\partial C_{\theta}^{(s,q)}}{\partial x}
-
\frac{\partial^2 C_{\theta}^{(s,q)}}{\partial x^2}
-
\frac{\partial^2 C_{\theta}^{(s,q)}}{\partial y^2}.
\label{eq:pinn-pde-residual}
\end{equation}
The exact solution satisfies \(\mathcal{R}_{\theta}^{(s,q)}=0\) throughout the interior of the space--time domain. Thus, the PDE loss is defined as
\begin{equation}
\mathcal{L}_{\mathrm{PDE}}^{(s,q)}
=
\frac{1}{N_f}
\sum_{j=1}^{N_f}
\left|
\mathcal{R}_{\theta}^{(s,q)}
\left(x_f^j,y_f^j,t_f^j\right)
\right|^2,
\label{eq:pde-loss}
\end{equation}
where \(\{(x_f^j,y_f^j,t_f^j)\}_{j=1}^{N_f}\) are interior collocation points sampled from
\[
(x,y,t)\in (x_0,x_1)\times(-1,1)\times(0,T).
\]
This formulation, in which the governing physical laws serve as soft penalty constraints during network optimisation, provides the mathematical foundation of the PINN framework \cite{Raissi2019, He2021}.

\subsection{Initial-condition loss}
For each shear-flow case \(s\in\mathcal{S}:=\{\mathrm{C},\mathrm{Po},\mathrm{CP}\}\) and source configuration
\(q\in\mathcal{Q}:=\{\mathrm{L},\mathrm{P}\}\), the initial concentration is prescribed as
\[
C(x,y,0)=C_0^{(q)}(x,y).
\]
Here, the superscript \(s\) identifies the imposed shear flow, while \(q\) identifies the initial source type. The two source configurations are defined in equations \eqref{eq:ic-line-source} and \eqref{eq:ic-point-source}. The source \(C_0^{(\mathrm{L})}\) is a line-like Gaussian release \cite{Teng2023}, localised in the streamwise direction and uniform across the channel height. The source \(C_0^{(\mathrm{P})}\) is a point-like Gaussian release, localised in both the streamwise and wall-normal directions. Thus, the line-like source isolates shear-driven longitudinal spreading, whereas the point-like source also probes transverse diffusion and wall contact.

The corresponding initial-condition loss is
\begin{equation}
\mathcal{L}_{\mathrm{IC}}^{(s,q)}
=
\frac{1}{N_0}
\sum_{j=1}^{N_0}
\left|
C_{\theta}^{(s,q)}(x_0^j,y_0^j,0)
-
C_0^{(q)}(x_0^j,y_0^j)
\right|^2 ,
\label{eq:ic-loss}
\end{equation}
where \(\{(x_0^j,y_0^j)\}_{j=1}^{N_0}\) are sampled over the computational domain at \(t=0\).

A true point-source release would formally be represented by a Dirac delta distribution. However, the Dirac delta is not a classical continuous function and is therefore incompatible with a pointwise mean-square initial-condition loss in the standard PINN setting. Since the present network represents a smooth trial function \(C_{\theta}^{(s,q)}\), imposing a singular distribution directly through pointwise residual minimisation is not well posed. The Gaussian source \(C_0^{(\mathrm{P})}\) may instead be interpreted as a smooth mollified approximation to a point release, while \(C_0^{(\mathrm{L})}\) represents the corresponding line-release configuration. Future extensions may incorporate singular source data more directly using weak-form PINNs, Green 's-function-based representations, adaptive mollification, or distributional residual formulations. Resolving these highly localised initial distributions is critical, as the initial transient stage of dispersion is highly sensitive to the source configuration before approaching the asymptotic Gaussian regime \cite{Jiang2022, Rawden2026}.

\subsection{Boundary-condition losses}
The axial boundaries of the truncated computational domain are assigned homogeneous zero-gradient conditions. The corresponding axial boundary loss is
\begin{equation}
\mathcal{L}_{x\mathrm{B}}^{(s,q)}
=
\frac{1}{N_x}
\sum_{j=1}^{N_x}
\left[
\left|
\frac{\partial C_{\theta}^{(s,q)}}{\partial x}
(x_0,y_x^j,t_x^j)
\right|^2
+
\left|
\frac{\partial C_{\theta}^{(s,q)}}{\partial x}
(x_1,y_x^j,t_x^j)
\right|^2
\right],
\label{eq:xb-loss}
\end{equation}
where \(\{(y_x^j,t_x^j)\}_{j=1}^{N_x}\) are sampled on the two axial boundaries.

At the reactive walls, the Robin boundary conditions are enforced through
\[
\frac{\partial C}{\partial y}-\beta_2 C=0
\quad \text{at } y=-1,
\qquad
-\frac{\partial C}{\partial y}-\beta_1 C=0
\quad \text{at } y=1.
\]
Equivalently, the wall residuals are
\begin{equation}
\mathcal{R}_{-,\theta}^{(s,q)}
=
\frac{\partial C_{\theta}^{(s,q)}}{\partial y}
-
\beta_2 C_{\theta}^{(s,q)}
\quad \text{at } y=-1,
\label{eq:bottom-robin-residual}
\end{equation}
and
\begin{equation}
\mathcal{R}_{+,\theta}^{(s,q)}
=
-\frac{\partial C_{\theta}^{(s,q)}}{\partial y}
-
\beta_1 C_{\theta}^{(s,q)}
\quad \text{at } y=1.
\label{eq:top-robin-residual}
\end{equation}
The Robin wall loss is therefore
\begin{equation}
\mathcal{L}_{y\mathrm{B}}^{(s,q)}
=
\frac{1}{N_y}
\sum_{j=1}^{N_y}
\left|
\mathcal{R}_{-,\theta}^{(s,q)}(x_-^j,-1,t_-^j)
\right|^2
+
\frac{1}{N_y}
\sum_{j=1}^{N_y}
\left|
\mathcal{R}_{+,\theta}^{(s,q)}(x_+^j,1,t_+^j)
\right|^2 .
\label{eq:yb-loss}
\end{equation}
The wall loss controls the reactive solute exchange at the plates, while the axial loss closes the finite computational domain without imposing artificial concentration loss across the streamwise boundaries.

\subsection{Composite training objective}
The trainable parameters \(\theta\) are obtained by minimising the composite physics-informed loss
\begin{equation}
\mathcal{L}_{\mathrm{tot}}^{(s,q)}(\theta)
=
\mathcal{L}_{\mathrm{PDE}}^{(s,q)}
+
\lambda_{\mathrm{IC}}\mathcal{L}_{\mathrm{IC}}^{(s,q)}
+
\lambda_{\mathrm{B}}
\left(
\mathcal{L}_{x\mathrm{B}}^{(s,q)}
+
\mathcal{L}_{y\mathrm{B}}^{(s,q)}
\right).
\label{eq:total-loss}
\end{equation}
Here \(\lambda_{\mathrm{IC}}\) and \(\lambda_{\mathrm{B}}\) are positive weights that balance the initial-condition and boundary constraints relative to the interior PDE residual. In the computations reported here, these constraints are weighted more strongly to ensure accurate imposition of the localised source, the axial zero-gradient condition, and the reactive-wall conditions.

The minimisation problem is therefore
\begin{equation}
\theta_{\star}^{(s,q)}
=
\operatorname*{arg\,min}_{\theta}\,
\mathcal{L}_{\mathrm{tot}}^{(s,q)}(\theta),
\qquad
s\in\mathcal{S},\quad q\in\mathcal{Q}.
\label{eq:pinn-minimization}
\end{equation}
Once trained, the PINN solution is evaluated as
\[
C_{\mathrm{PINN}}^{(s,q)}(x,y,t)
=
C_{\theta_{\star}}^{(s,q)}(x,y,t).
\]
The same network architecture, sampling strategy and loss structure are used across the shear-flow cases and source configurations, so that differences in the predicted transport can be attributed to \(u^{(s)}(y)\), \(C_0^{(q)}\), and the imposed wall reactivity.

\subsection{Collocation strategy and training}
The collocation points are sampled separately for the interior PDE residual, the initial condition, the axial boundary condition and the Robin wall conditions. This separation is essential because the initial and boundary constraints lie on lower-dimensional subsets of the space--time domain and would otherwise be underrepresented by purely uniform sampling in the full domain. Thus, at each training epoch, independent point sets are drawn from the interior domain, the initial plane, the two axial boundaries and the two reactive walls.

The trainable parameters are optimised using the Adam algorithm with an exponentially decaying learning rate. At each epoch, the loss components are evaluated over the corresponding collocation sets, and the network parameters are updated to reduce the composite physics-informed objective. This sampling strategy ensures that the PDE residual, the selected initial source configuration, the artificial axial boundaries, and the reactive wall conditions all contribute explicitly to the training process.

All spatial and temporal derivatives appearing in \eqref{eq:pinn-pde-residual}, \eqref{eq:bottom-robin-residual} and \eqref{eq:top-robin-residual} are computed by automatic differentiation with respect to the neural-network representation \(C_\theta^{(s,q)}(x,y,t)\). This avoids finite-difference approximation of derivatives inside the PINN loss and enables mesh-free enforcement of the dimensionless convection--diffusion equation and Robin wall constraints. By leveraging exact automatic differentiation rather than discrete approximations, the network avoids the classical truncation errors inherent in grid-based derivative calculations, which is a primary advantage of deep learning solvers in fluid mechanics \cite{Raissi2019, Zou2025}.

\subsection{Computational setup}
\label{subsec:computational-setup}

All computations are performed on the finite domain
\[
x\in[-5,5],\qquad y\in[-1,1],\qquad 0\leq t\leq 1,
\]
with \(\mathrm{Pe}=10\). The PINN uses the network architecture described above, namely four hidden layers with $64$ neurons in each layer and $\tanh$ activation. For each training epoch, collocation points are sampled separately from the interior domain, the initial plane, the two axial boundaries and the two reactive walls. Unless otherwise stated, the numbers of points used in the PINN loss are \[
N_f=8000,\qquad N_0=1000,\qquad N_{x\mathrm{B}}=800,\qquad N_{y\mathrm{B}}=400.
\] Here $N_f$ denotes interior residual points, $N_0$ denotes initial-condition points, $N_{x\mathrm{B}}$ denotes axial-boundary points, and $N_{y\mathrm{B}}$ denotes points on each reactive wall. A sufficiently large finite interval is used to truncate the infinite longitudinal domain. Because the concentration may stay non-zero at the artificial longitudinal boundaries throughout the simulation, homogeneous Neumann boundary conditions are required there. The Neumann condition lessens the impact of domain truncation on the solution of interest by avoiding unnecessarily restricting the concentration field, in contrast to homogeneous Dirichlet conditions. The axial-boundary collocation points are used to impose the homogeneous zero-gradient condition $\partial C_\theta/\partial x=0$ at $x=-5$ and $x=5$.

The trainable parameters are optimised using the Adam algorithm with an initial learning rate $10^{-3}$. The learning rate is multiplied by a factor \(0.95\) every \(500\) epochs, allowing rapid initial optimisation followed by progressively smaller parameter updates. The composite loss is evaluated using stronger weights on the initial and boundary constraints, $\lambda_{\mathrm{IC}}=30, \lambda_{\mathrm{B}}=30,$
so that the localised initial release and the boundary conditions remain accurately imposed during training. Most reported validation cases are trained for $5000$ epochs, while selected runs are continued up to $20000$ epochs to monitor longer optimisation behaviour and to generate the epoch-wise loss values reported later.

For reproducibility, the ADI finite-difference benchmark uses the same physical domain as
\[
N_x=101,\qquad N_y=51,\qquad \Delta t=10^{-4},\qquad T=1.
\]
The same initial source profiles, wall-reaction coefficients and shear-flow configurations are used in the PINN and ADI computations. This common setup ensures that the reported PINN--ADI discrepancies measure the approximation accuracy of the learned solution rather than differences in the underlying physical problem.

\subsection{Finite-difference benchmark}
To assess the accuracy of the PINN predictions, we compute reference solutions using a finite-difference alternating-direction implicit (ADI) scheme on the same truncated domain. Originally developed for parabolic partial differential equations \cite{Peaceman1955, Douglas1955}, the ADI method provides an unconditionally stable and computationally efficient framework for multi-dimensional transport. In our implementation, the dimensionless convection--diffusion equation is advanced in time by an alternating-direction splitting of the diffusive terms, with the advective contribution evaluated on the same Cartesian grid. The initial source configuration, axial zero-gradient condition and Robin wall conditions are imposed consistently with the PINN formulation.

The ADI solution is used only as a benchmark for validation. Comparisons are performed using full concentration contours and cross-sectionally averaged concentration profiles. The latter are defined by
\begin{equation}
\langle C \rangle(x,t)
=
\langle C\rangle_y
=
\frac{1}{2}\int_{-1}^{1}C(x,y,t)\mathrm{d}y .
\label{eq:y-averaged-concentration}
\end{equation}
For visualisation, the averaged profiles are plotted in the centroid-shifted coordinate \(x-x_g(t)\), where
\begin{equation}
x_g(t)
=
\frac{\displaystyle\int_{x_0}^{x_1}x\,\langle C \rangle(x,t)\mathrm{d}x}
{\displaystyle\int_{x_0}^{x_1}\langle C \rangle(x,t)\mathrm{d}x}.
\label{eq:axial-centroid-xg}
\end{equation}
The agreement between PINN and ADI solutions is quantified using \(L_2\)-type errors at selected times, providing a direct validation of the learned solution against a classical grid-based method.

In addition to concentration-based validation, we also use the ADI benchmark to validate the integrated wall-removal diagnostics. Let
\[
C_{\mathrm{ADI}}^{(s,q)}(x,y,t)
\]
denote the ADI solution corresponding to the same shear profile, source configuration and reactive-wall parameters as the PINN solution. The ADI wall uptake fluxes are evaluated analogously as
\[
J_{+,\mathrm{ADI}}^{(s,q)}(x,t)
=
\beta_1 C_{\mathrm{ADI}}^{(s,q)}(x,1,t),
\qquad
J_{-,\mathrm{ADI}}^{(s,q)}(x,t)
=
\beta_2 C_{\mathrm{ADI}}^{(s,q)}(x,-1,t).
\]
The corresponding integrated wall uptake rates are
\[
\mathcal{J}_{+,\mathrm{ADI}}^{(s,q)}(t)
=
\int_{x_0}^{x_1}
J_{+,\mathrm{ADI}}^{(s,q)}(x,t)\mathrm{d}x,
\qquad
\mathcal{J}_{-,\mathrm{ADI}}^{(s,q)}(t)
=
\int_{x_0}^{x_1}
J_{-,\mathrm{ADI}}^{(s,q)}(x,t)\mathrm{d}x.
\]
The cumulative wall uptakes are then computed as
\[
\mathcal{U}_{+,\mathrm{ADI}}^{(s,q)}(t)
=
\int_0^t
\mathcal{J}_{+,\mathrm{ADI}}^{(s,q)}(\tau)\,d\tau,
\qquad
\mathcal{U}_{-,\mathrm{ADI}}^{(s,q)}(t)
=
\int_0^t
\mathcal{J}_{-,\mathrm{ADI}}^{(s,q)}(\tau)\,d\tau,
\]
where \(\tau\) is a dummy time-integration variable. Thus,
\[
\mathcal{U}_{\mathrm{tot},\mathrm{ADI}}^{(s,q)}(t)
=
\mathcal{U}_{+,\mathrm{ADI}}^{(s,q)}(t)
+
\mathcal{U}_{-,\mathrm{ADI}}^{(s,q)}(t).
\]
The corresponding PINN quantity is denoted by
\[
\mathcal{U}_{\mathrm{tot},\mathrm{PINN}}^{(s,q)}(t).
\]
The final-time relative error in cumulative wall uptake is defined as
\begin{equation}
E_{\mathcal{U}}^{(s,q)}
=
\frac{
\left|
\mathcal{U}_{\mathrm{tot},\mathrm{PINN}}^{(s,q)}(T)
-
\mathcal{U}_{\mathrm{tot},\mathrm{ADI}}^{(s,q)}(T)
\right|
}
{
\left|
\mathcal{U}_{\mathrm{tot},\mathrm{ADI}}^{(s,q)}(T)
\right|
}
\times 100.
\label{eq:utot-relative-error}
\end{equation}
This metric directly assesses whether the learned concentration field reproduces the cumulative wall-removal dynamics obtained from the ADI benchmark.

\subsection{Physical interpretation of the learned solution}

The trained network provides a differentiable surrogate for the concentration field,
\[
C_{\mathrm{PINN}}^{(s,q)}(x,y,t)
=
C_{\theta_\star}^{(s,q)}(x,y,t),
\qquad s\in\mathcal{S},\quad q\in\mathcal{Q}.
\]
Beyond pointwise concentration values, this representation allows direct evaluation of physically relevant wall-uptake quantities. In particular, the local dimensionless uptake fluxes at the upper and lower walls are computed as
\begin{equation}
J_+^{(s,q)}(x,t)
=
\beta_1 C_{\theta_\star}^{(s,q)}(x,1,t),
\qquad
J_-^{(s,q)}(x,t)
=
\beta_2 C_{\theta_\star}^{(s,q)}(x,-1,t).
\label{eq:pinn-wall-flux}
\end{equation}
Here \(J_+^{(s,q)}\) and \(J_-^{(s,q)}\) denote the uptake fluxes at the upper and lower plates, respectively. The corresponding integrated wall uptake rates are
\begin{equation}
\mathcal{J}_+^{(s,q)}(t)
=
\int_{x_0}^{x_1}J_+^{(s,q)}(x,t)\mathrm{d}x,
\qquad
\mathcal{J}_-^{(s,q)}(t)
=
\int_{x_0}^{x_1}J_-^{(s,q)}(x,t)\mathrm{d}x,
\label{eq:pinn-integrated-wall-flux}
\end{equation}
and the total integrated uptake rate is
\begin{equation}
\mathcal{J}_{\mathrm{tot}}^{(s,q)}(t)
=
\mathcal{J}_+^{(s,q)}(t)
+
\mathcal{J}_-^{(s,q)}(t).
\label{eq:pinn-total-wall-flux}
\end{equation}
These quantities measure the instantaneous rate at which solute is removed by the reactive plates.

The cumulative wall uptakes are then defined by
\begin{equation}
\mathcal{U}_+^{(s,q)}(t)
=
\int_0^t \mathcal{J}_+^{(s,q)}(\tau)\,d\tau,
\qquad
\mathcal{U}_-^{(s,q)}(t)
=
\int_0^t \mathcal{J}_-^{(s,q)}(\tau)\,d\tau,
\label{eq:pinn-cumulative-wall-uptake}
\end{equation}
where \(\tau\) is a dummy time-integration variable. The total cumulative wall uptake is
\begin{equation}
\mathcal{U}_{\mathrm{tot}}^{(s,q)}(t)
=
\mathcal{U}_+^{(s,q)}(t)
+
\mathcal{U}_-^{(s,q)}(t).
\label{eq:pinn-total-cumulative-uptake}
\end{equation}
Thus, \(\mathcal{J}_{\pm}^{(s,q)}(t)\) describe instantaneous wall-removal rates, whereas \(\mathcal{U}_{\pm}^{(s,q)}(t)\) describe the accumulated amount of solute removed by each wall up to time \(t\).

To quantify the final partition of removal between the two plates, we define the lower-wall cumulative uptake fraction
\begin{equation}
\Phi_-^{(s,q)}(T)
=
\frac{
\mathcal{U}_-^{(s,q)}(T)
}
{
\mathcal{U}_{\mathrm{tot}}^{(s,q)}(T)
}.
\label{eq:lower-wall-uptake-fraction}
\end{equation}
For the non-reactive case, \(\mathcal{U}_{\mathrm{tot}}^{(s,q)}(T)=0\), so this fraction is not interpreted and is used only for reactive configurations. Values \(\Phi_-^{(s,q)}(T)\approx 1/2\) indicate balanced upper- and lower-wall uptake, whereas \(\Phi_-^{(s,q)}(T)>1/2\) and \(\Phi_-^{(s,q)}(T)<1/2\) indicate lower- and upper-wall-dominated removal, respectively.

We also introduce a cumulative wall-dominance index
\begin{equation}
D_w^{(s,q)}(t)
=
\frac{
\mathcal{U}_-^{(s,q)}(t)
-
\mathcal{U}_+^{(s,q)}(t)
}
{
\mathcal{U}_{\mathrm{tot}}^{(s,q)}(t)
}.
\label{eq:wall-dominance-index}
\end{equation}
This index is interpreted for times at which \(\mathcal{U}_{\mathrm{tot}}^{(s,q)}(t)>0\). Thus \(D_w^{(s,q)}(t)=0\) corresponds to balanced wall removal, \(D_w^{(s,q)}(t)>0\) indicates lower-wall-dominated uptake, and \(D_w^{(s,q)}(t)<0\) indicates upper-wall-dominated uptake. This bounded diagnostic provides a compact measure of the time-dependent asymmetry in cumulative wall removal.

To characterise the streamwise organisation of wall uptake, we define flux-weighted uptake centroids
\begin{equation}
x_{J_+}^{(s,q)}(t)
=
\frac{
\displaystyle\int_{x_0}^{x_1}x\,J_+^{(s,q)}(x,t)\mathrm{d}x
}
{
\displaystyle\int_{x_0}^{x_1}J_+^{(s,q)}(x,t)\mathrm{d}x
},
\qquad
x_{J_-}^{(s,q)}(t)
=
\frac{
\displaystyle\int_{x_0}^{x_1}x\,J_-^{(s,q)}(x,t)\mathrm{d}x
}
{
\displaystyle\int_{x_0}^{x_1}J_-^{(s,q)}(x,t)\mathrm{d}x
}.
\label{eq:wall-flux-centroids}
\end{equation}
These centroids are evaluated only when the corresponding integrated wall uptake rate is non-zero. The corresponding flux-weighted spreads are
\begin{equation}
\sigma_{J_+}^{(s,q)}(t)
=
\left[
\frac{
\displaystyle\int_{x_0}^{x_1}
\left(x-x_{J_+}^{(s,q)}(t)\right)^2
J_+^{(s,q)}(x,t)\mathrm{d}x
}
{
\displaystyle\int_{x_0}^{x_1}
J_+^{(s,q)}(x,t)\mathrm{d}x
}
\right]^{1/2},
\label{eq:upper-wall-flux-spread}
\end{equation}
and
\begin{equation}
\sigma_{J_-}^{(s,q)}(t)
=
\left[
\frac{
\displaystyle\int_{x_0}^{x_1}
\left(x-x_{J_-}^{(s,q)}(t)\right)^2
J_-^{(s,q)}(x,t)\mathrm{d}x
}
{
\displaystyle\int_{x_0}^{x_1}
J_-^{(s,q)}(x,t)\mathrm{d}x
}
\right]^{1/2}.
\label{eq:lower-wall-flux-spread}
\end{equation}
The quantities \(x_{J_\pm}^{(s,q)}(t)\) and \(\sigma_{J_\pm}^{(s,q)}(t)\) identify where along the channel the wall uptake is concentrated and how broadly the wall-flux distribution is spread. They therefore reveal the spatial organisation of wall removal that may not be visible from total uptake alone.

In the reactive-wall analysis, we focus on three representative configurations,
\[
(\beta_1,\beta_2)=(1,1),\qquad
(\beta_1,\beta_2)=(0.2,2),\qquad
(\beta_1,\beta_2)=(2,0.2).
\]
The first corresponds to symmetric wall uptake, while the latter two represent asymmetric absorption with dominant lower- and upper-wall reactivity, respectively. The non-reactive case \((\beta_1,\beta_2)=(0,0)\) is used as a validation baseline.

The same wall-uptake diagnostics are also evaluated from the ADI finite-difference benchmark solution to assess the accuracy of the PINN-predicted cumulative wall removal.

\section{Results and discussion}
\label{sec:results-discussion}

The results are organised to separate solution validation from physical interpretation. We first assess the PINN approximation using concentration-field errors, cross-sectionally averaged concentration profiles and epoch-wise training histories. We then validate the derived transport quantities, including the apparent axial dispersion coefficient, the total surviving solute mass, the axial variance, and the cumulative wall uptake. After these validation steps, the trained PINN solution is used to examine wall-resolved reactive dispersion through uptake rates, cumulative removal, wall-selective asymmetry and the streamwise organisation of local wall fluxes. The emphasis is therefore not only on reconstructing \(C(x,y,t)\), but also on extracting interpretable diagnostics that describe how shear-driven dispersion and wall absorption jointly control solute removal.

\subsection{Concentration-field validation across source and flow configurations}
\label{subsec:concentration-validation}

To assess the local spatial accuracy of the learned concentration field, we use the blockwise absolute error
\begin{equation}
B_{\mathrm{err}}(x,y,t)=\left|C_{\mathrm{PINN}}(x,y,t)-C_{\mathrm{ADI}}(x,y,t)\right|.
\label{eq:blockwise-absolute-error-results}
\end{equation}
Here $C_{\mathrm{PINN}}(x,y,t)$ denotes the concentration field predicted by the trained physics-informed neural network, while $C_{\mathrm{ADI}}(x,y,t)$ denotes the corresponding ADI finite-difference benchmark solution. The quantity $B_{\mathrm{err}}(x,y,t)$ therefore measures the local pointwise discrepancy between the PINN and ADI solutions at each selected space--time location.

We first use the non-reactive case, $\beta_1=\beta_2=0$, as a baseline to assess the advection--diffusion component of the learned solution. This baseline isolates the PINN's ability to reproduce shear-driven spreading before wall absorption is introduced. The two initial source configurations are the line-like Gaussian source \eqref{eq:ic-line-source} and the point-like Gaussian source \eqref{eq:ic-point-source}. The line-like source is initially uniform across the channel height, and primarily tests shear-induced axial spreading, whereas the point-like source is localised in both the streamwise and wall-normal directions and therefore also tests transverse diffusion and near-wall concentration evolution.

\begin{table}
    \centering
     \caption{
    \(L_2\) errors between PINN predictions and ADI finite-difference solutions at selected times for the three shear-flow configurations with \(\mathrm{Pe}=10\) and \(\beta_1=\beta_2=0\). Results are reported for both line-like and point-like source configurations.
    }
    \begin{tabular}{llllllll}
    \toprule
        Flow type & Reaction rate & Source type 
        & \(t=0.001\) & \(t=0.01\) & \(t=0.1\) & \(t=0.5\) & \(t=1\) \\
    \midrule
        Poiseuille & \(\beta_1=\beta_2=0\) & Line-like 
        & 0.0240489 & 0.0241400 & 0.0251867 & 0.0269552 & 0.0267071 \\
        Poiseuille & \(\beta_1=\beta_2=0\) & Point-like 
        & 0.0108551 & 0.0119165 & 0.0141633 & 0.0174312 & 0.0201506 \\
        Couette & \(\beta_1=\beta_2=0\) & Line-like 
        & 0.0081305 & 0.0087978 & 0.0137522 & 0.0398162 & 0.0414874 \\
        Couette & \(\beta_1=\beta_2=0\) & Point-like 
        & 0.0123710 & 0.0129826 & 0.0143972 & 0.0262930 & 0.0277750 \\
        Couette--Poiseuille & \(\beta_1=\beta_2=0\) & Line-like 
        & 0.0201866 & 0.0206141 & 0.0295364 & 0.0544209 & 0.0720882 \\
        Couette--Poiseuille & \(\beta_1=\beta_2=0\) & Point-like 
        & 0.0124994 & 0.0130910 & 0.0170186 & 0.0170186 & 0.0391011 \\
    \bottomrule
    \end{tabular}
    \label{tab:l2-error-three-flows-sources}
\end{table}

The error values in table~\ref{tab:l2-error-three-flows-sources} provide a quantitative baseline validation across the three mean-centred shear profiles. The corresponding non-reactive blockwise-error plots for all source--flow combinations are reported in the supplementary material. This keeps the main manuscript focused on the reactive cases while retaining the complete concentration-field validation.

\begin{figure}
\centering
\includegraphics[width=0.92\linewidth]{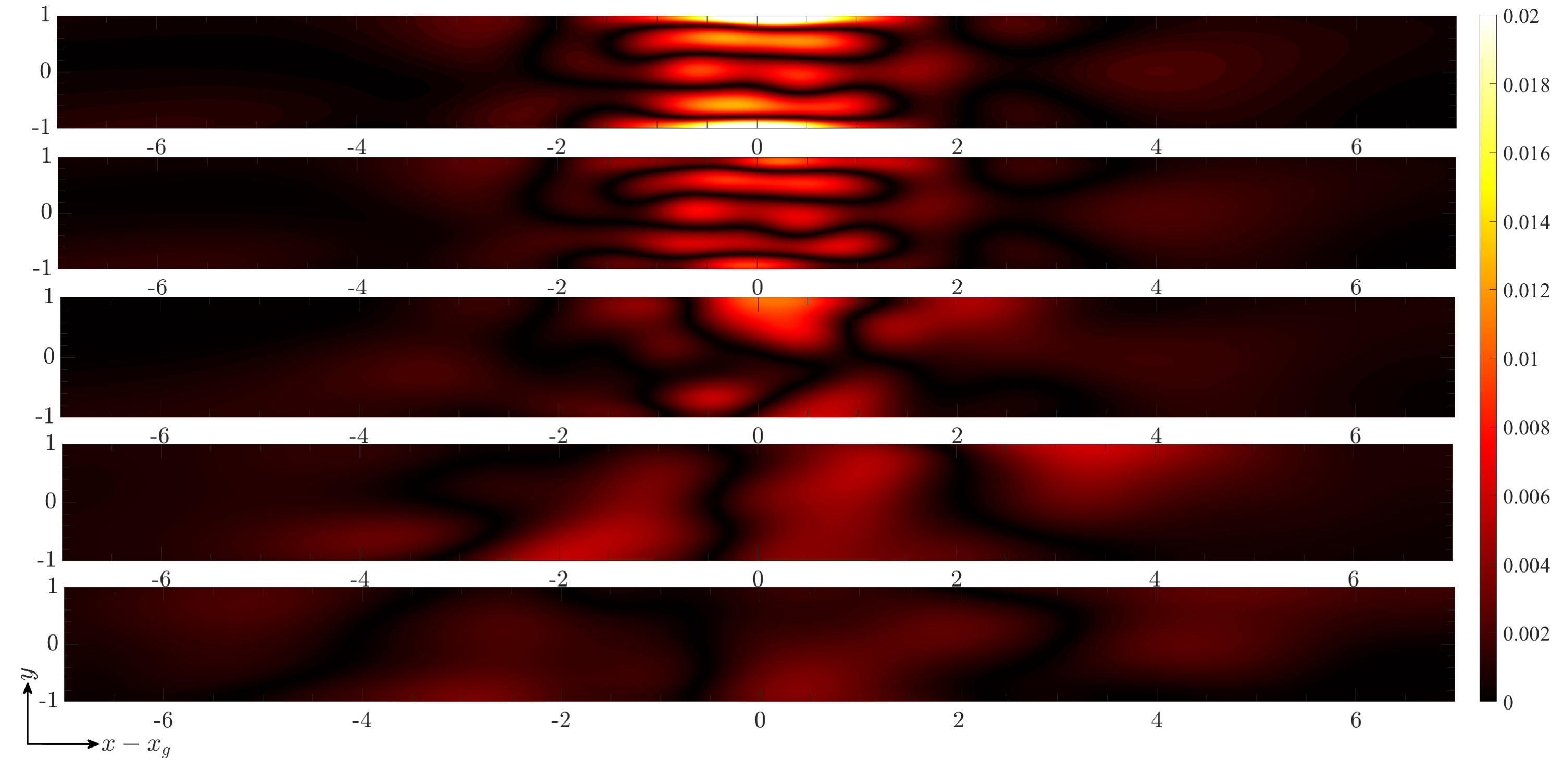}
\caption{
Blockwise absolute error $B_{\mathrm{err}}=|C_{\mathrm{PINN}}-C_{\mathrm{ADI}}|$ for the point-like source in Couette flow with symmetric wall absorption $(\beta_1,\beta_2)=(1,1)$.
}
\label{fig:berr-c-s2-beta11}
\end{figure}

\begin{figure}
\centering
\includegraphics[width=0.92\linewidth]{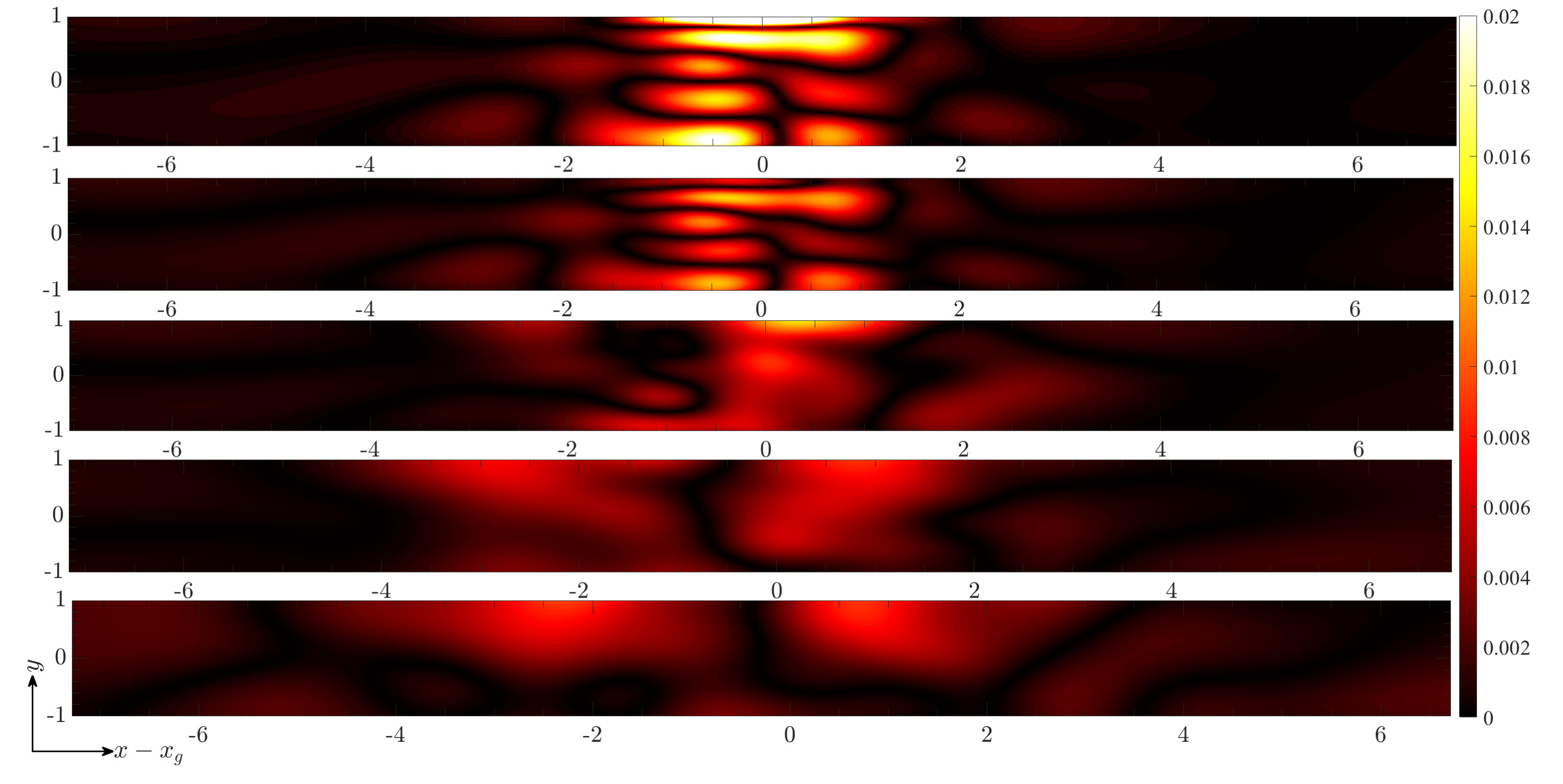}
\caption{
Blockwise absolute error $B_{\mathrm{err}}=|C_{\mathrm{PINN}}-C_{\mathrm{ADI}}|$ for the point-like source in Poiseuille flow with lower-wall-dominated absorption $(\beta_1,\beta_2)=(0.2,2)$.
}
\label{fig:berr-p-s2-beta022}
\end{figure}

\begin{figure}
\centering
\includegraphics[width=0.92\linewidth]{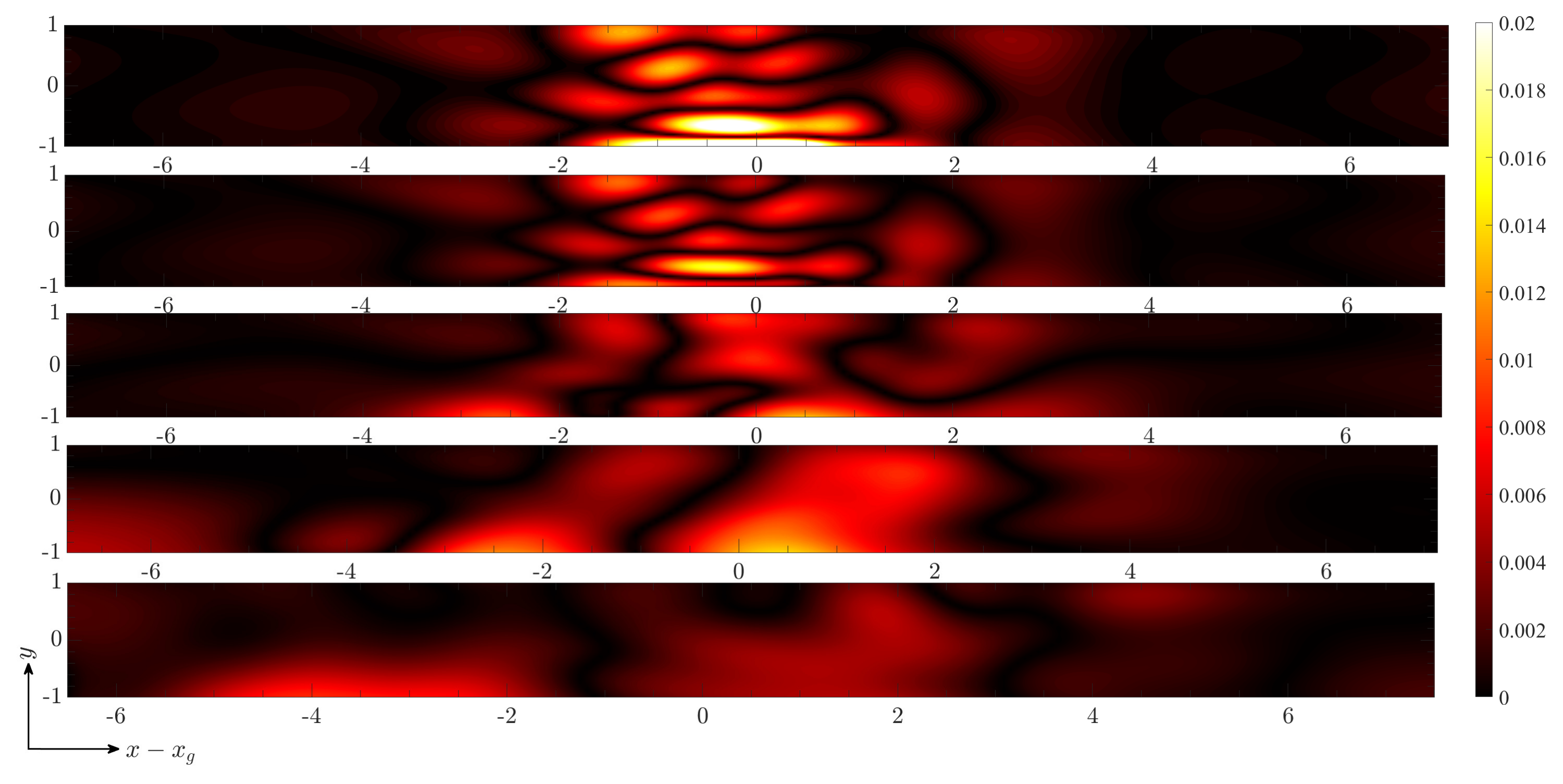}
\caption{
Blockwise absolute error $B_{\mathrm{err}}=|C_{\mathrm{PINN}}-C_{\mathrm{ADI}}|$ for the point-like source in Couette--Poiseuille flow with upper-wall-dominated absorption $(\beta_1,\beta_2)=(2,0.2)$.
}
\label{fig:berr-cp-s2-beta202}
\end{figure}

We next show representative reactive blockwise-error results in the main manuscript, since wall absorption is the central feature of the present study. The point-like source is used here because it is the source configuration used later for wall-resolved uptake diagnostics. The three cases shown in figures~\ref{fig:berr-c-s2-beta11}--\ref{fig:berr-cp-s2-beta202} span symmetric wall absorption, lower-wall-dominated absorption, upper-wall-dominated absorption, and the three canonical shear profiles.

Figures~\ref{fig:berr-c-s2-beta11}--\ref{fig:berr-cp-s2-beta202} demonstrate that the learned concentration field remains close to the ADI benchmark after reactive wall absorption is introduced. These representative cases are deliberately selected to cover different shear structures and different transverse absorption biases. Thus, the validation is not restricted to the non-reactive limit, but directly supports the reactive wall-uptake analysis developed in the following subsections. The localised error structures are physically expected because the largest discrepancies occur where the concentration gradients are strongest, namely near the evolving plume front and near reactive boundaries. In these regions, advection, transverse diffusion and Robin wall uptake act simultaneously, making the local concentration field more demanding to approximate than in weak-gradient regions of the domain.

Additional blockwise-error results for the line-like reactive cases are provided in the supplementary material. These results complement the representative point-like reactive validation shown in the main manuscript while avoiding excessive repetition of similar contour plots. 

\begin{table}
\centering
\caption{PINN--ADI error metrics for the line-like source under different shear-flow configurations and wall-reaction coefficients. Results are reported at selected times for $\mathrm{Pe}=10$.}
\label{tab:pinn-adi-line-source-errors}
\begin{tabular}{llcccccc}
\toprule
\textbf{Flow Regime} & \textbf{Reaction rate} & \textbf{$t=0.001$} & \textbf{$t=0.01$} & \textbf{$t=0.1$} & \textbf{$t=0.5$} & \textbf{$t=1.0$} \\
\midrule
Poiseuille          & $\beta_1=0,\ \beta_2=0$ & 0.0017 & 0.0017 & 0.0052 & 0.0162 & 0.0212 \\
Couette             & $\beta_1=0,\ \beta_2=0$ & 0.0067 & 0.0066 & 0.0088 & 0.0169 & 0.0161 \\
Couette--Poiseuille & $\beta_1=0,\ \beta_2=0$ & 0.0062 & 0.0061 & 0.0112 & 0.0209 & 0.0199 \\
\addlinespace
Poiseuille          & $\beta_1=1,\ \beta_2=1$ & 0.0038 & 0.0033 & 0.0040 & 0.0087 & 0.0070 \\
Poiseuille          & $\beta_1=0.2,\ \beta_2=2$ & 0.0057 & 0.0052 & 0.0088 & 0.0127 & 0.0116 \\
Poiseuille          & $\beta_1=2,\ \beta_2=0.2$ & 0.0073 & 0.0062 & 0.0066 & 0.0101 & 0.0120 \\
\addlinespace
Couette             & $\beta_1=1,\ \beta_2=1$ & 0.0033 & 0.0039 & 0.0087 & 0.0093 & 0.0060 \\
Couette             & $\beta_1=2,\ \beta_2=0.2$ & 0.0077 & 0.0062 & 0.0078 & 0.0116 & 0.0117 \\
Couette             & $\beta_1=0.2,\ \beta_2=2$ & 0.0068 & 0.0059 & 0.0082 & 0.0101 & 0.0095 \\
\addlinespace
Couette--Poiseuille & $\beta_1=1,\ \beta_2=1$ & 0.0032 & 0.0024 & 0.0083 & 0.0120 & 0.0077 \\
Couette--Poiseuille & $\beta_1=2,\ \beta_2=0.2$ & 0.0055 & 0.0044 & 0.0074 & 0.0157 & 0.0097 \\
Couette--Poiseuille & $\beta_1=0.2,\ \beta_2=2$ & 0.0046 & 0.0054 & 0.0104 & 0.0171 & 0.0137 \\
\bottomrule
\end{tabular}
\end{table}

\begin{table}
\centering
\caption{PINN--ADI error metrics for the point-like source under different shear-flow configurations and wall-reaction coefficients. Results are reported at selected times for $\mathrm{Pe}=10$.}
\label{tab:pinn-adi-point-source-errors}
\begin{tabular}{llcccccc}
\toprule
\textbf{Flow Regime} & \textbf{Reaction rate} & \textbf{$t=0.001$} & \textbf{$t=0.01$} & \textbf{$t=0.1$} & \textbf{$t=0.5$} & \textbf{$t=1.0$} \\
\midrule
Poiseuille          & $\beta_1=0,\ \beta_2=0$ & 0.0018 & 0.0028 & 0.0077 & 0.0146 & 0.0181 \\
Couette             & $\beta_1=0,\ \beta_2=0$ & 0.0042 & 0.0055 & 0.0104 & 0.0168 & 0.0177 \\
Couette--Poiseuille & $\beta_1=0,\ \beta_2=0$ & 0.0042 & 0.0054 & 0.0106 & 0.0173 & 0.0147 \\
\addlinespace
Poiseuille          & $\beta_1=1,\ \beta_2=1$ & 0.0067 & 0.0068 & 0.0084 & 0.0077 & 0.0067 \\
Poiseuille          & $\beta_1=0.2,\ \beta_2=2$ & 0.0038 & 0.0042 & 0.0075 & 0.0087 & 0.0076 \\
Poiseuille          & $\beta_1=2,\ \beta_2=0.2$ & 0.0048 & 0.0045 & 0.0006 & 0.0084 & 0.0092 \\
\addlinespace
Couette             & $\beta_1=1,\ \beta_2=1$ & 0.0039 & 0.0035 & 0.0051 & 0.0075 & 0.0052 \\
Couette             & $\beta_1=2,\ \beta_2=0.2$ & 0.0048 & 0.0050 & 0.0067 & 0.0069 & 0.0077 \\
Couette             & $\beta_1=0.2,\ \beta_2=2$ & 0.0052 & 0.0049 & 0.0055 & 0.0092 & 0.0072 \\
\addlinespace
Couette--Poiseuille & $\beta_1=1,\ \beta_2=1$ & 0.0040 & 0.0041 & 0.0076 & 0.0104 & 0.0071 \\
Couette--Poiseuille & $\beta_1=2,\ \beta_2=0.2$ & 0.0058 & 0.0057 & 0.0073 & 0.0120 & 0.0082 \\
Couette--Poiseuille & $\beta_1=0.2,\ \beta_2=2$ & 0.0056 & 0.0062 & 0.0095 & 0.0146 & 0.0104 \\
\bottomrule
\end{tabular}
\end{table}

Tables~\ref{tab:pinn-adi-line-source-errors} and \ref{tab:pinn-adi-point-source-errors} extend the quantitative validation beyond the non-reactive baseline by including both source configurations and the selected symmetric and asymmetric wall-reaction cases. The errors remain small across the reported times, indicating that the trained PINN solutions remain close to the ADI benchmark after reactive wall absorption is introduced.

To further assess the accuracy of the learned streamwise transport, we also compare the cross-sectionally averaged concentration profiles. For each solution, the mean concentration is defined as
\begin{equation}
\langle C \rangle(x,t)
=
\frac{1}{2}\int_{-1}^{1} C(x,y,t)\mathrm{d}y .
\label{eq:mean-concentration-results}
\end{equation}
The corresponding mean-concentration error is
\begin{equation}
E_{\langle C \rangle}(x,t)
=
\left|
\langle C \rangle_{\mathrm{PINN}}(x,t)
-
\langle C \rangle_{\mathrm{ADI}}(x,t)
\right|.
\label{eq:mean-concentration-error-results}
\end{equation}
This diagnostic complements the blockwise two-dimensional error by showing whether the PINN accurately reproduces the streamwise solute distribution after averaging across the channel height.

\begin{figure}
\centering
\includegraphics[width=0.92\linewidth]{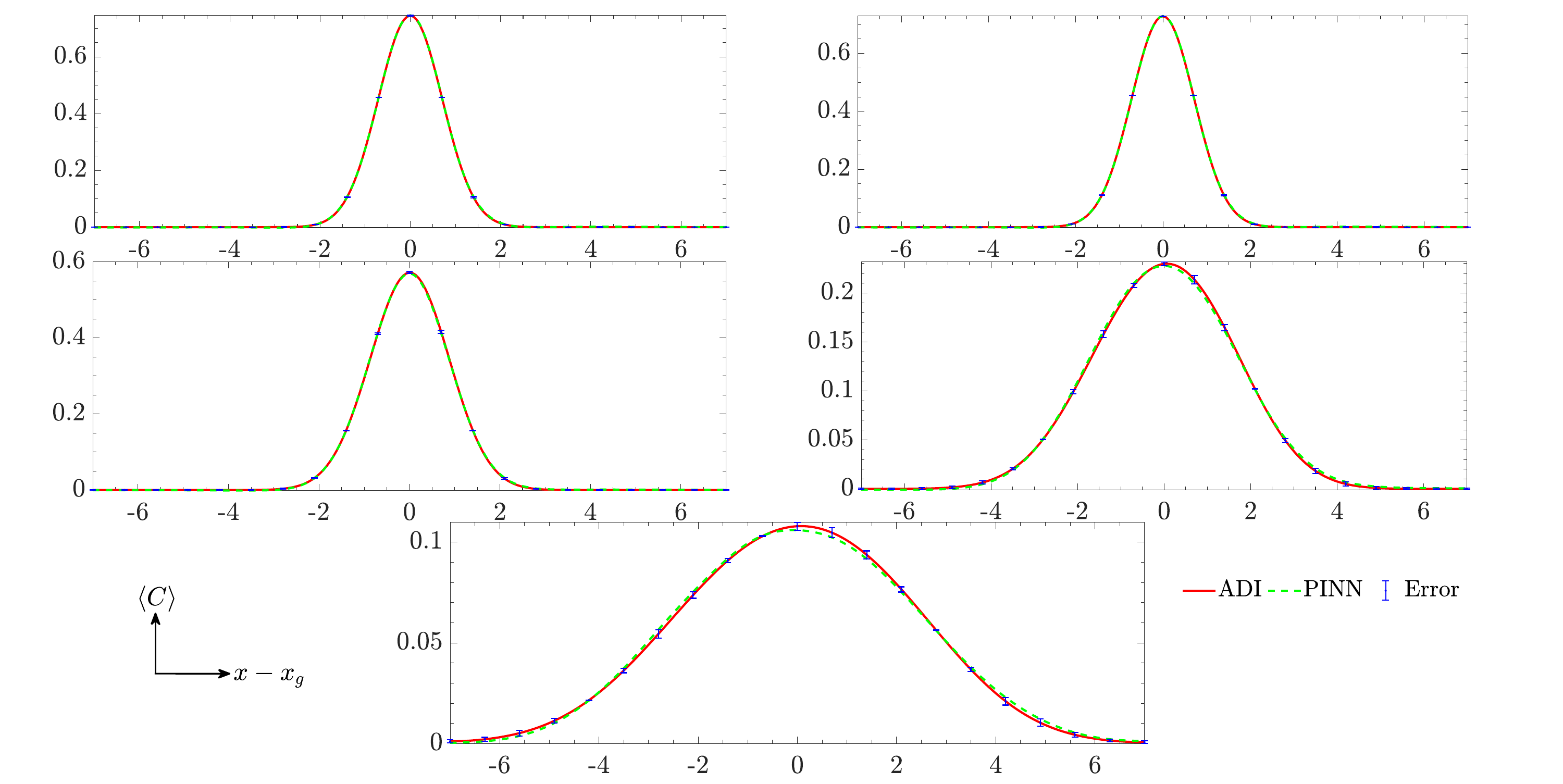}
\caption{
Cross-sectionally averaged concentration comparison and mean-concentration error for the point-like source in Couette flow with symmetric wall absorption $(\beta_1,\beta_2)=(1,1)$. The figure compares $\langle C \rangle_{\mathrm{PINN}}(x,t)$ with $\langle C \rangle_{\mathrm{ADI}}(x,t)$ and reports the corresponding error $E_{\langle C \rangle}(x,t)$.
}
\label{fig:cmwe-c-s2-beta11}
\end{figure}

\begin{figure}
\centering
\includegraphics[width=0.92\linewidth]{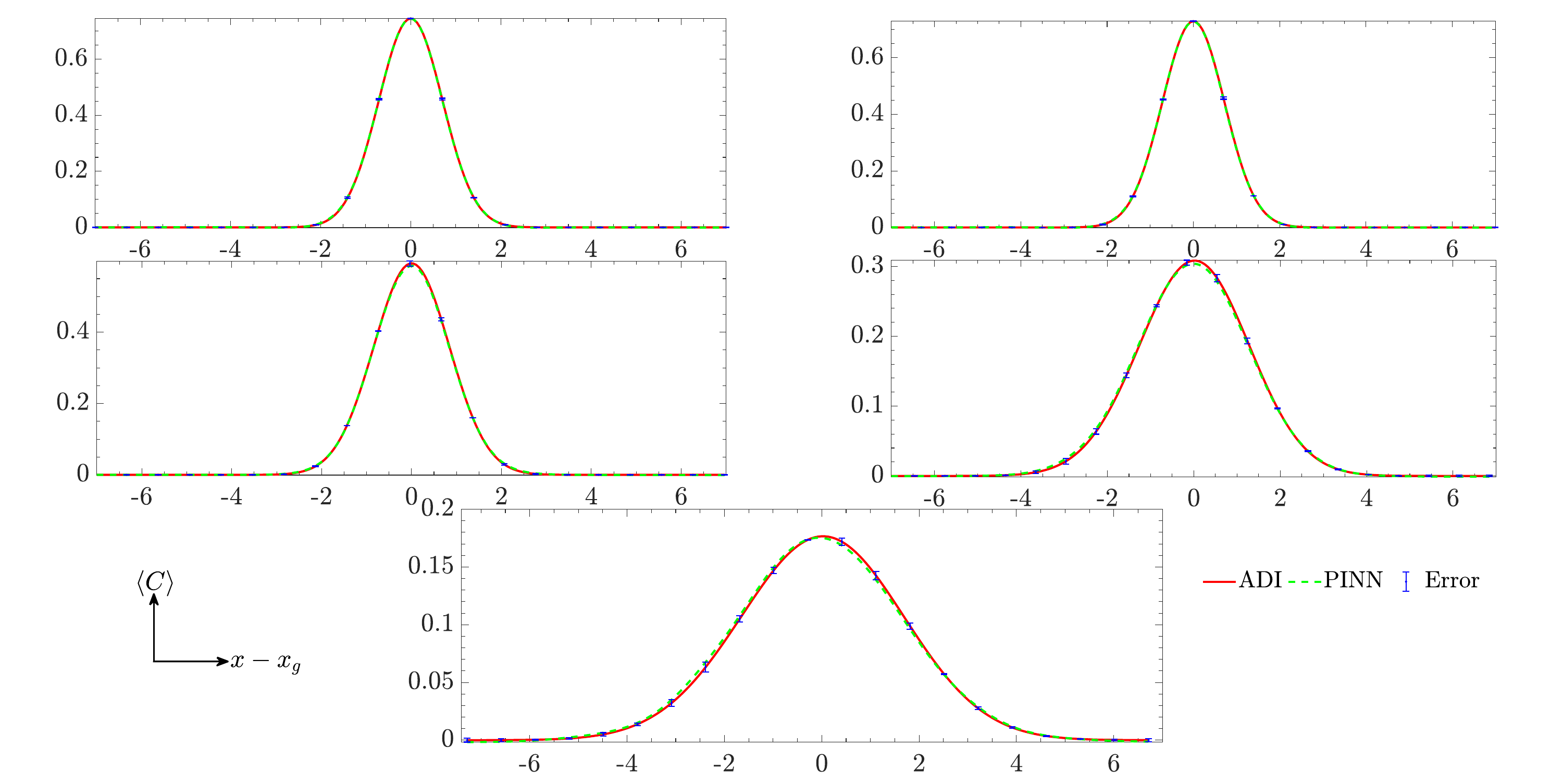}
\caption{
Cross-sectionally averaged concentration comparison and mean-concentration error for the point-like source in Poiseuille flow with lower-wall-dominated absorption $(\beta_1,\beta_2)=(0.2,2)$. The figure compares $\langle C \rangle_{\mathrm{PINN}}(x,t)$ with $\langle C \rangle_{\mathrm{ADI}}(x,t)$ and reports the corresponding error $E_{\langle C \rangle}(x,t)$.
}
\label{fig:cmwe-p-s2-beta022}
\end{figure}

\begin{figure}
\centering
\includegraphics[width=0.92\linewidth]{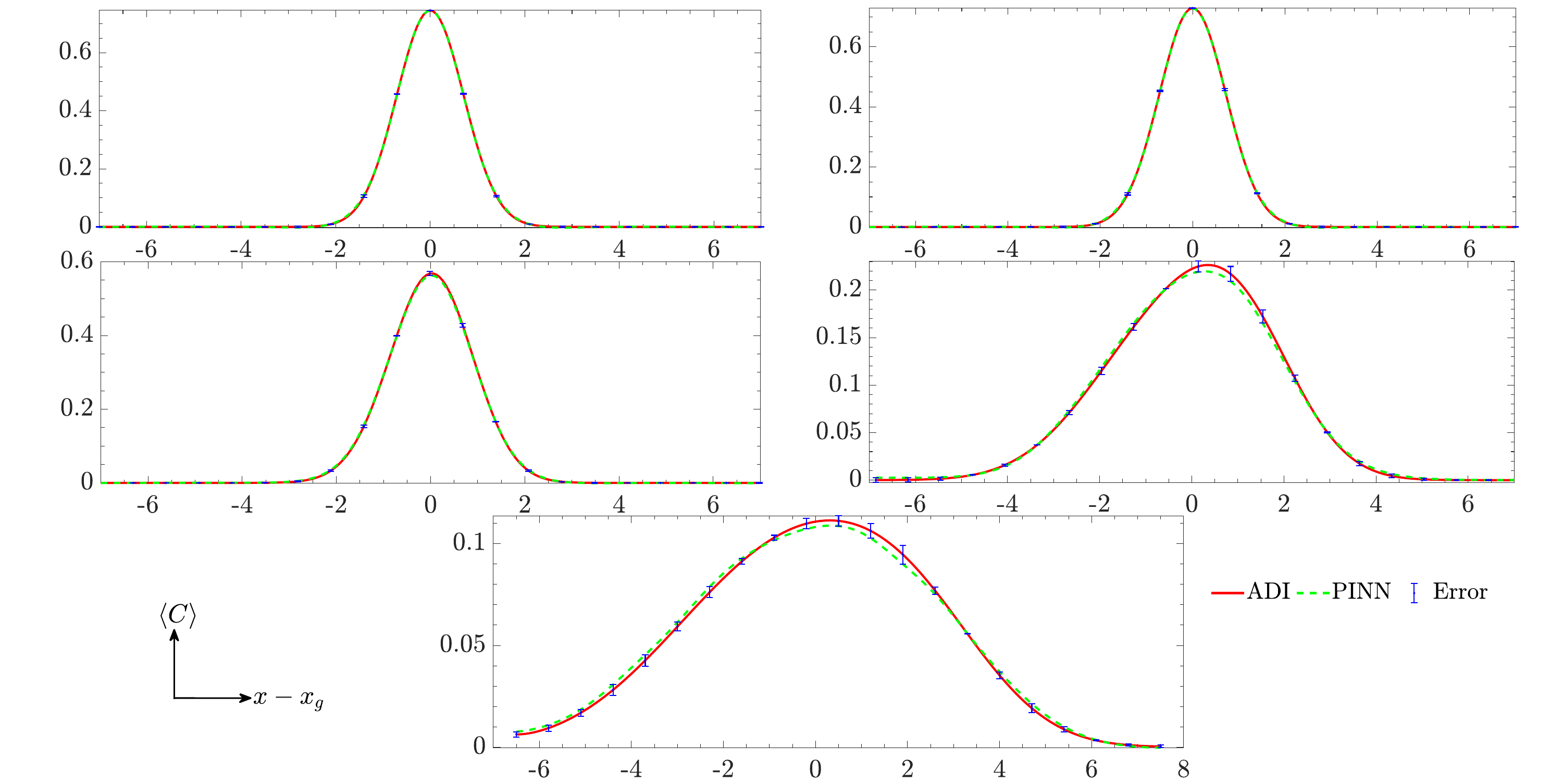}
\caption{
Cross-sectionally averaged concentration comparison and mean-concentration error for the point-like source in Couette--Poiseuille flow with upper-wall-dominated absorption $(\beta_1,\beta_2)=(2,0.2)$. The figure compares $\langle C \rangle_{\mathrm{PINN}}(x,t)$ with $\langle C \rangle_{\mathrm{ADI}}(x,t)$ and reports the corresponding error $E_{\langle C \rangle}(x,t)$.
}
\label{fig:cmwe-cp-s2-beta202}
\end{figure}

Figures~\ref{fig:cmwe-c-s2-beta11}--\ref{fig:cmwe-cp-s2-beta202} show that the PINN also reproduces the cross-sectionally averaged streamwise concentration profiles for representative reactive cases. Thus, the validation includes both local two-dimensional concentration accuracy and averaged streamwise transport accuracy. Additional mean-concentration comparison figures for the line-like reactive cases are reported in the supplementary material. Because cross-sectional averaging removes local wall-normal fluctuations, agreement in \(\langle C\rangle(x,t)\) confirms that the PINN captures the net streamwise redistribution of solute produced by the imposed shear flow and wall absorption.

\subsection{Training convergence of the PINN models}
\label{subsec:training-convergence}

Before extracting wall-resolved reactive-dispersion diagnostics, we examine the training histories of the PINN models. This check is included to verify the numerical optimisation behaviour of the physics-informed objective, whereas the ADI comparisons provide solution-level validation. The monitored quantity is the composite loss $\mathcal{L}_{\mathrm{tot}}^{(s,q)}(\theta)$ defined in \eqref{eq:total-loss}. The selected cases are arranged according to source configuration, shear profile, wall reactivity and activation function, so that the convergence behaviour can be assessed without mixing physically different comparisons in the same panel.

\begin{figure}
\centering

\begin{subfigure}[t]{0.48\linewidth}
\centering
\includegraphics[width=\linewidth,height=5.2cm]{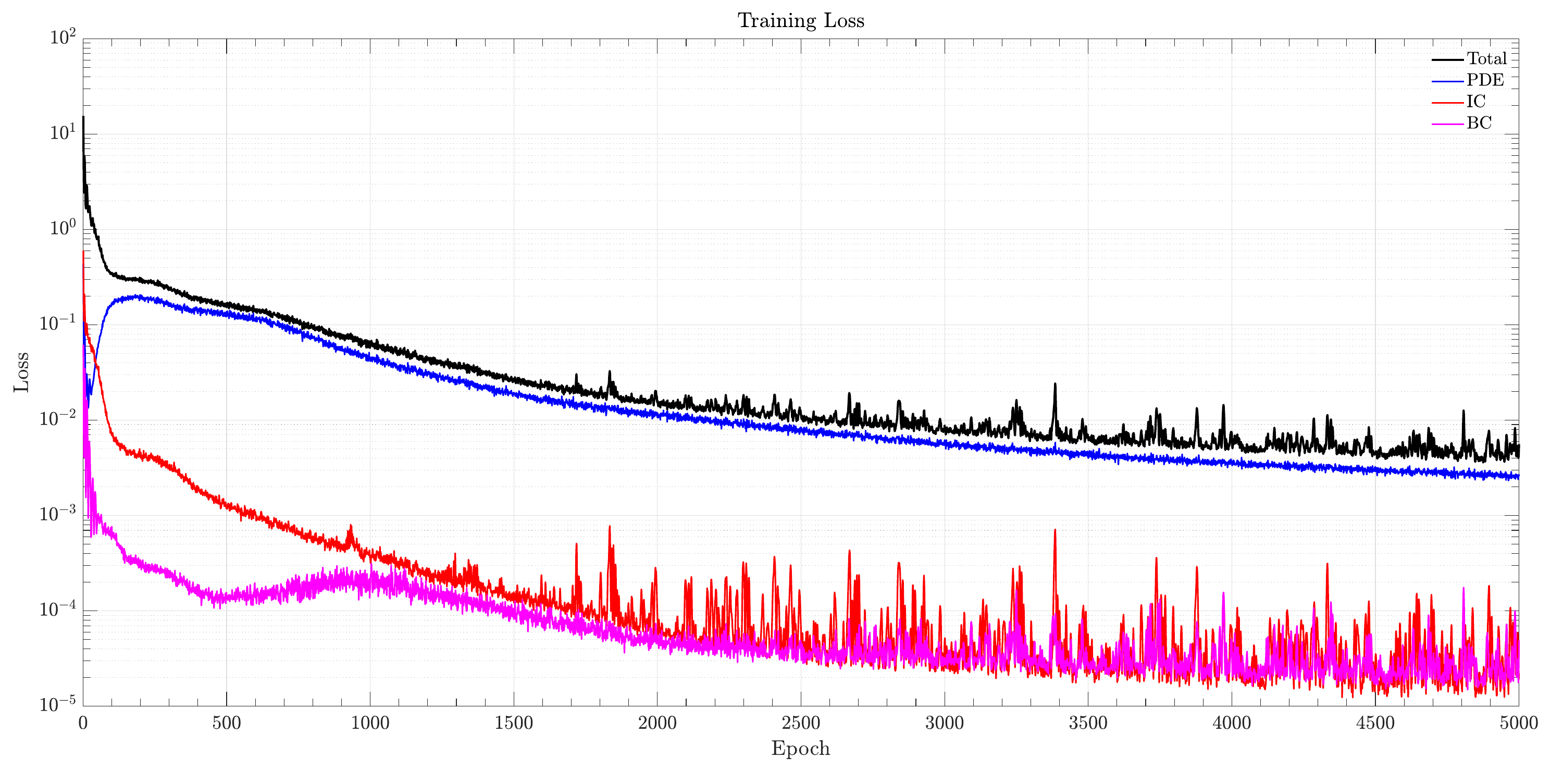}
\caption{$s=\mathrm{C}$, $(\beta_1,\beta_2)=(0,0)$.}
\label{fig:loss-c-beta00-s1}
\end{subfigure}
\hfill
\begin{subfigure}[t]{0.48\linewidth}
\centering
\includegraphics[width=\linewidth,height=5.2cm]{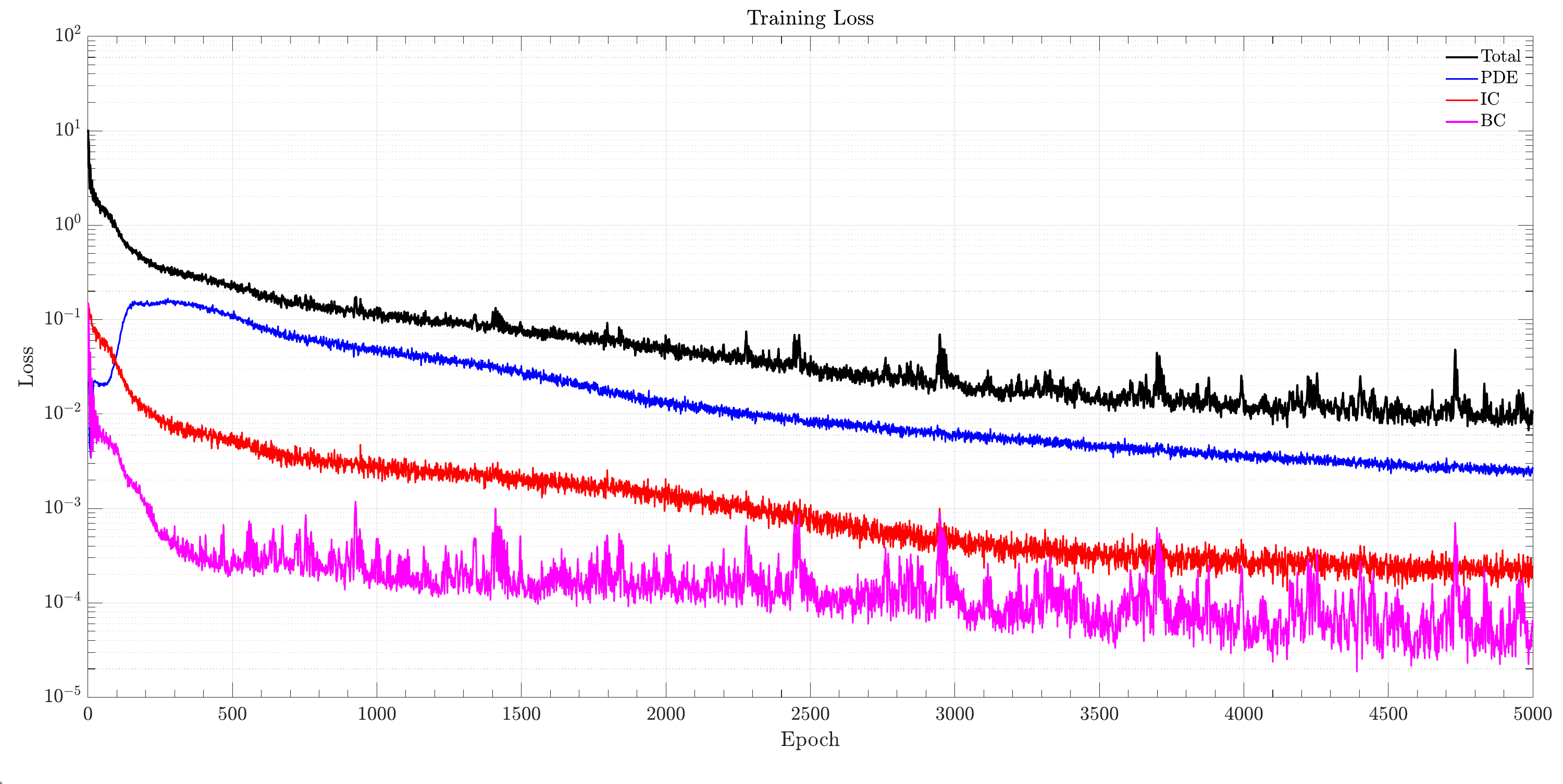}
\caption{$s=\mathrm{C}$, $(\beta_1,\beta_2)=(1,1)$.}
\label{fig:loss-c-beta11-s1}
\end{subfigure}

\vspace{0.25cm}

\begin{subfigure}[t]{0.48\linewidth}
\centering
\includegraphics[width=\linewidth,height=5.2cm]{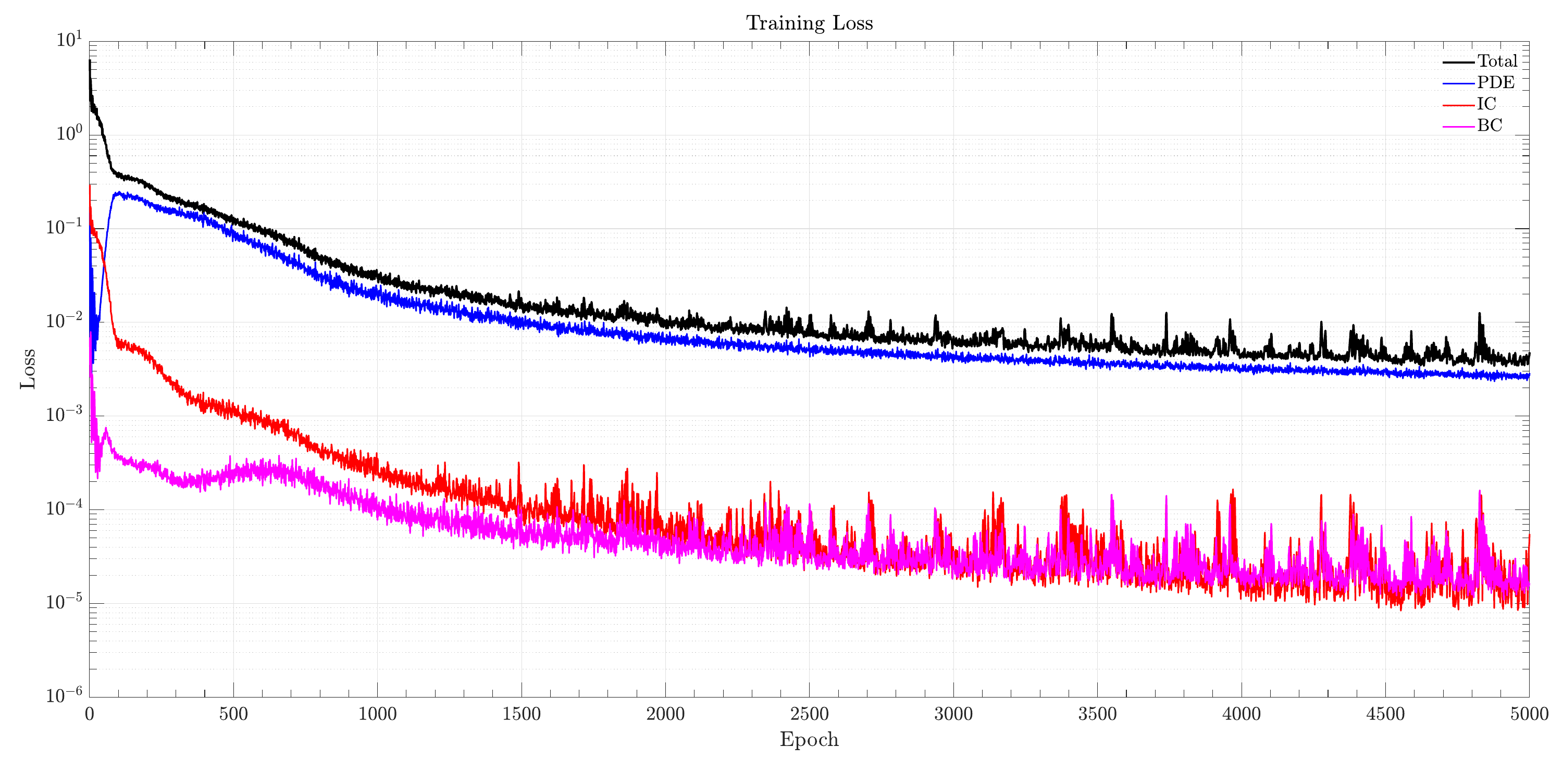}
\caption{$s=\mathrm{CP}$, $(\beta_1,\beta_2)=(0,0)$.}
\label{fig:loss-cp-beta00-s1}
\end{subfigure}
\hfill
\begin{subfigure}[t]{0.48\linewidth}
\centering
\includegraphics[width=\linewidth,height=5.2cm]{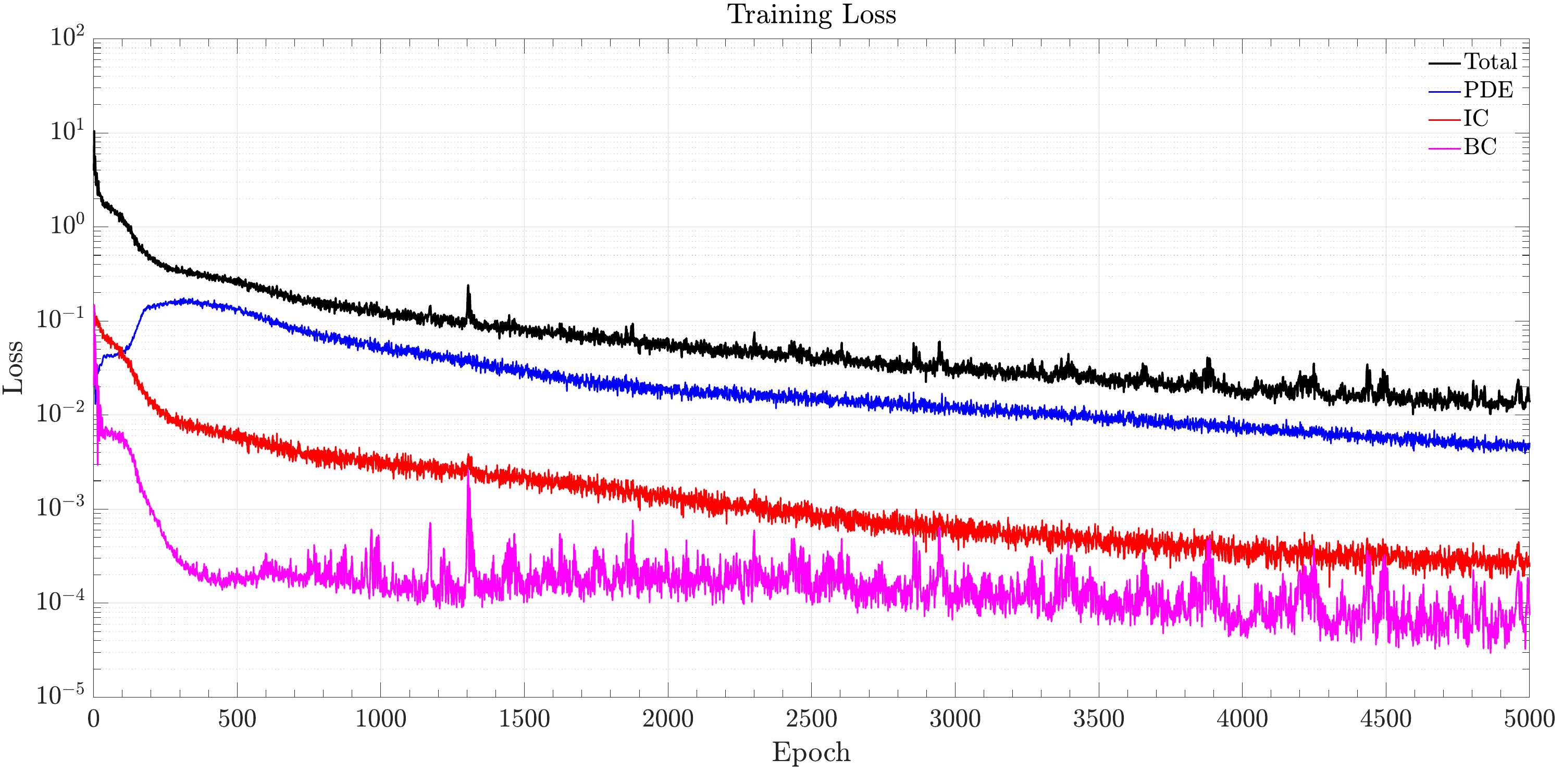}
\caption{$s=\mathrm{CP}$, $(\beta_1,\beta_2)=(1,1)$.}
\label{fig:loss-cp-beta11-s1}
\end{subfigure}

\vspace{0.25cm}

\begin{subfigure}[t]{0.48\linewidth}
\centering
\includegraphics[width=\linewidth,height=5.2cm]{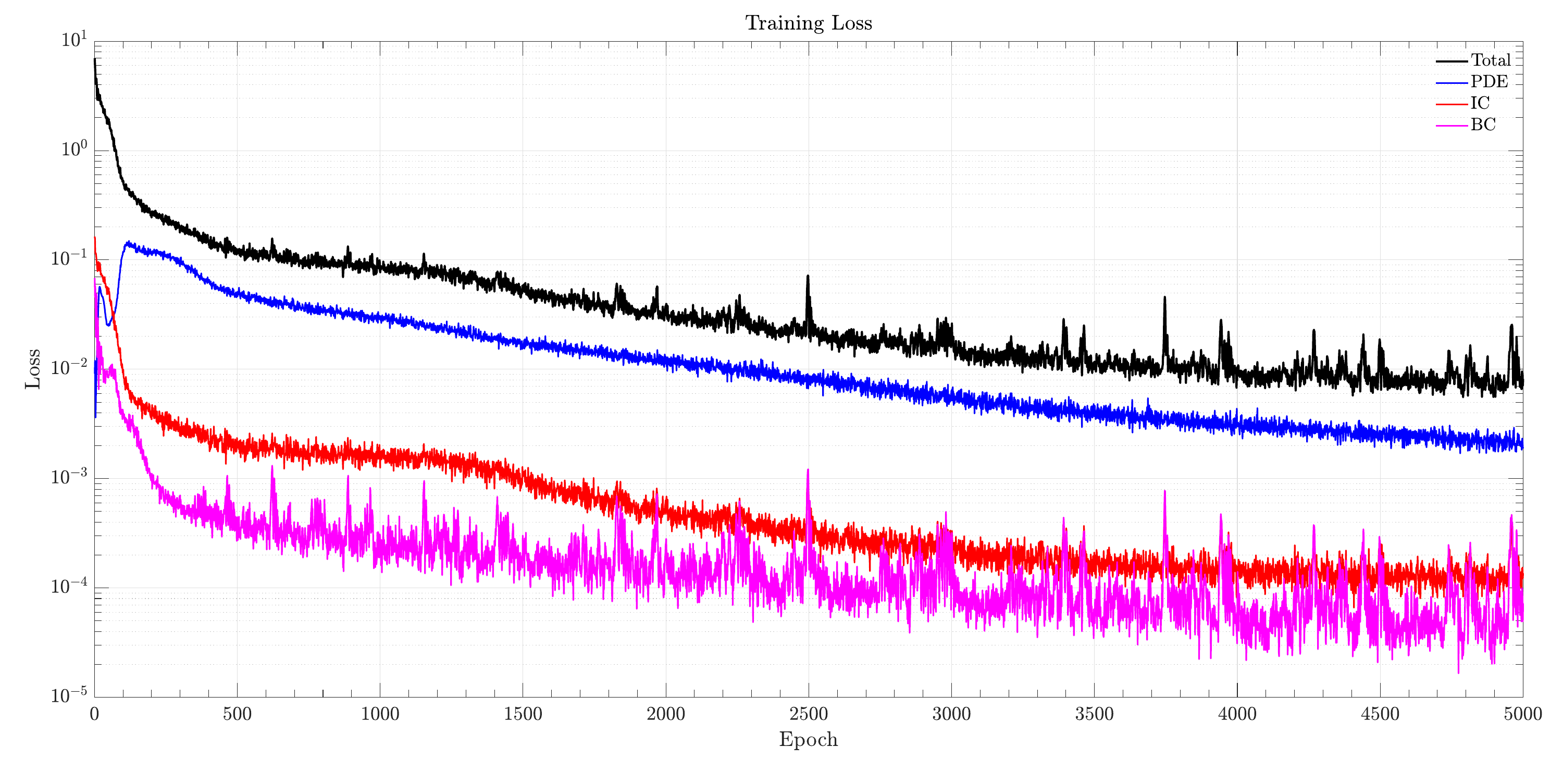}
\caption{$s=\mathrm{Po}$, $(\beta_1,\beta_2)=(1,1)$.}
\label{fig:loss-p-beta11-s1}
\end{subfigure}
\hfill
\begin{subfigure}[t]{0.48\linewidth}
\centering
\includegraphics[width=\linewidth,height=5.2cm]{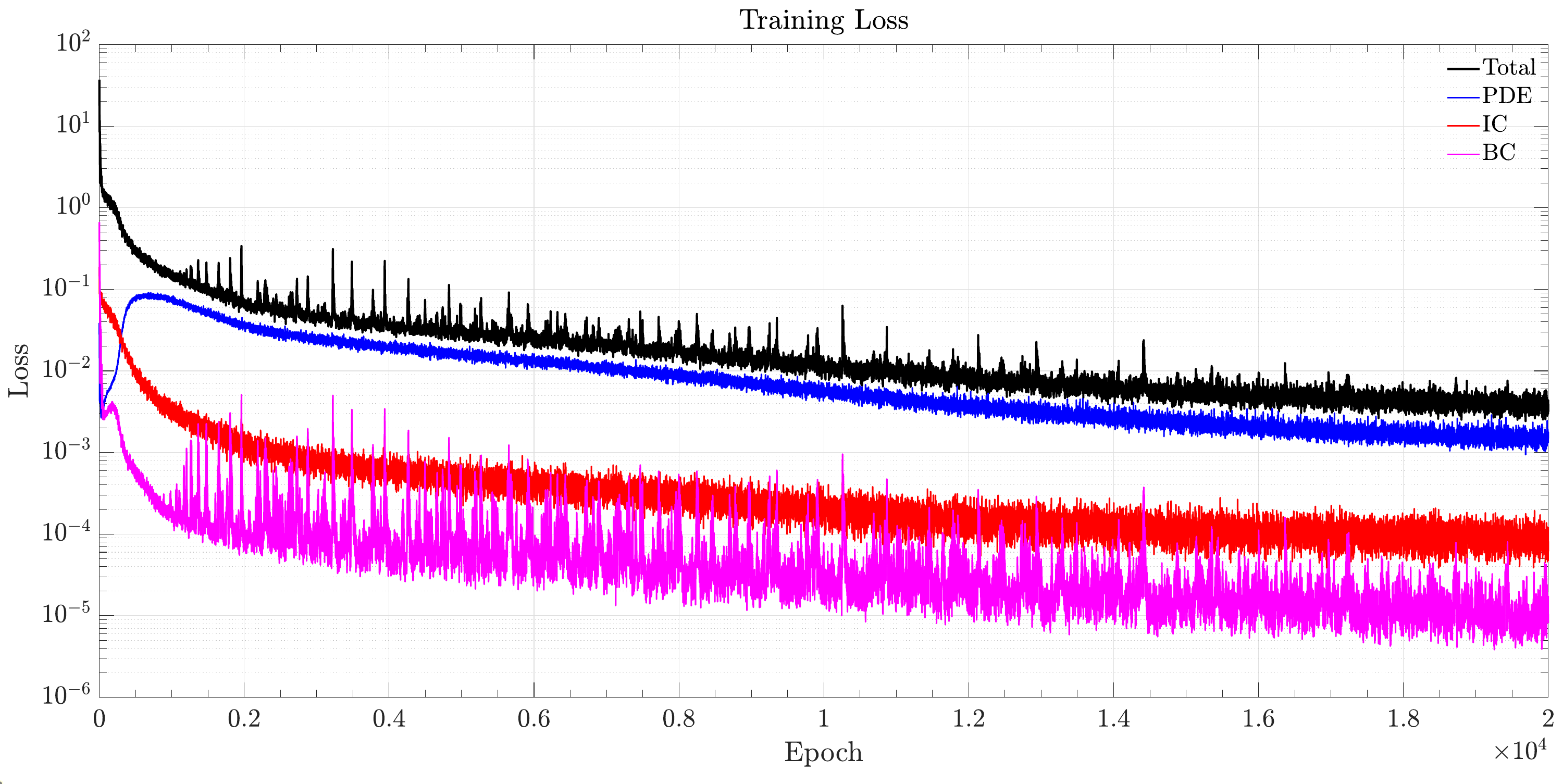}
\caption{$s=\mathrm{Po}$, $(\beta_1,\beta_2)=(2,0.2)$.}
\label{fig:loss-p-beta202-s1}
\end{subfigure}

\caption{
Training loss histories for the line-like source $q=\mathrm{L}$ using the default activation $\phi(z)=\tanh z$. The panels cover Couette, Poiseuille and Couette--Poiseuille flows under non-reactive, symmetric-reactive and asymmetric-reactive wall conditions. The decrease of $\mathcal{L}_{\mathrm{tot}}^{(s,q)}(\theta)$ indicates stable minimisation of the physics-informed objective for initially cross-sectionally distributed solute fields.
}
\label{fig:loss-line-source}
\end{figure}

Figure~\ref{fig:loss-line-source} reports training histories for the line-like source $q=\mathrm{L}$. These cases test whether the PINN optimisation remains stable when the initial concentration is distributed across the channel height, and the wall-reactivity configuration is varied across non-reactive, symmetric and asymmetric regimes. It shows that the optimisation remains stable for the line-like source across different velocity profiles and reaction strengths. This is important because the line-like source interacts immediately with the full transverse structure of the velocity field and the reactive walls. The comparable decay of $\mathcal{L}_{\mathrm{tot}}^{(s,q)}(\theta)$ across these cases supports the robustness of the training procedure for distributed initial data.

\begin{figure}
\centering

\begin{subfigure}[t]{0.48\linewidth}
\centering
\includegraphics[width=\linewidth,height=5.2cm]{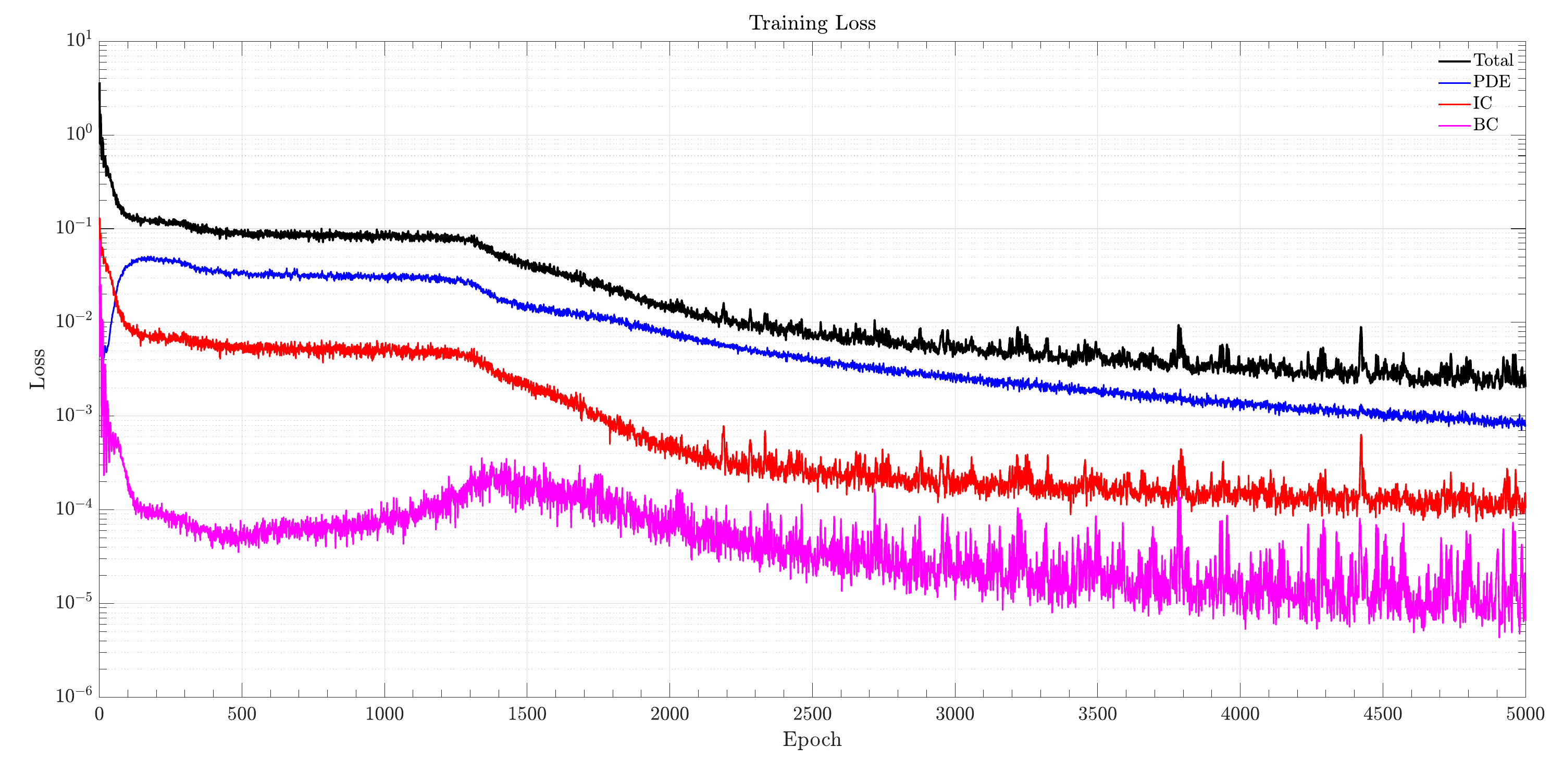}
\caption{$\phi(z)=\tanh z$, $(\beta_1,\beta_2)=(0,0)$.}
\label{fig:loss-p-beta00-s2}
\end{subfigure}
\hfill
\begin{subfigure}[t]{0.48\linewidth}
\centering
\includegraphics[width=\linewidth,height=5.2cm]{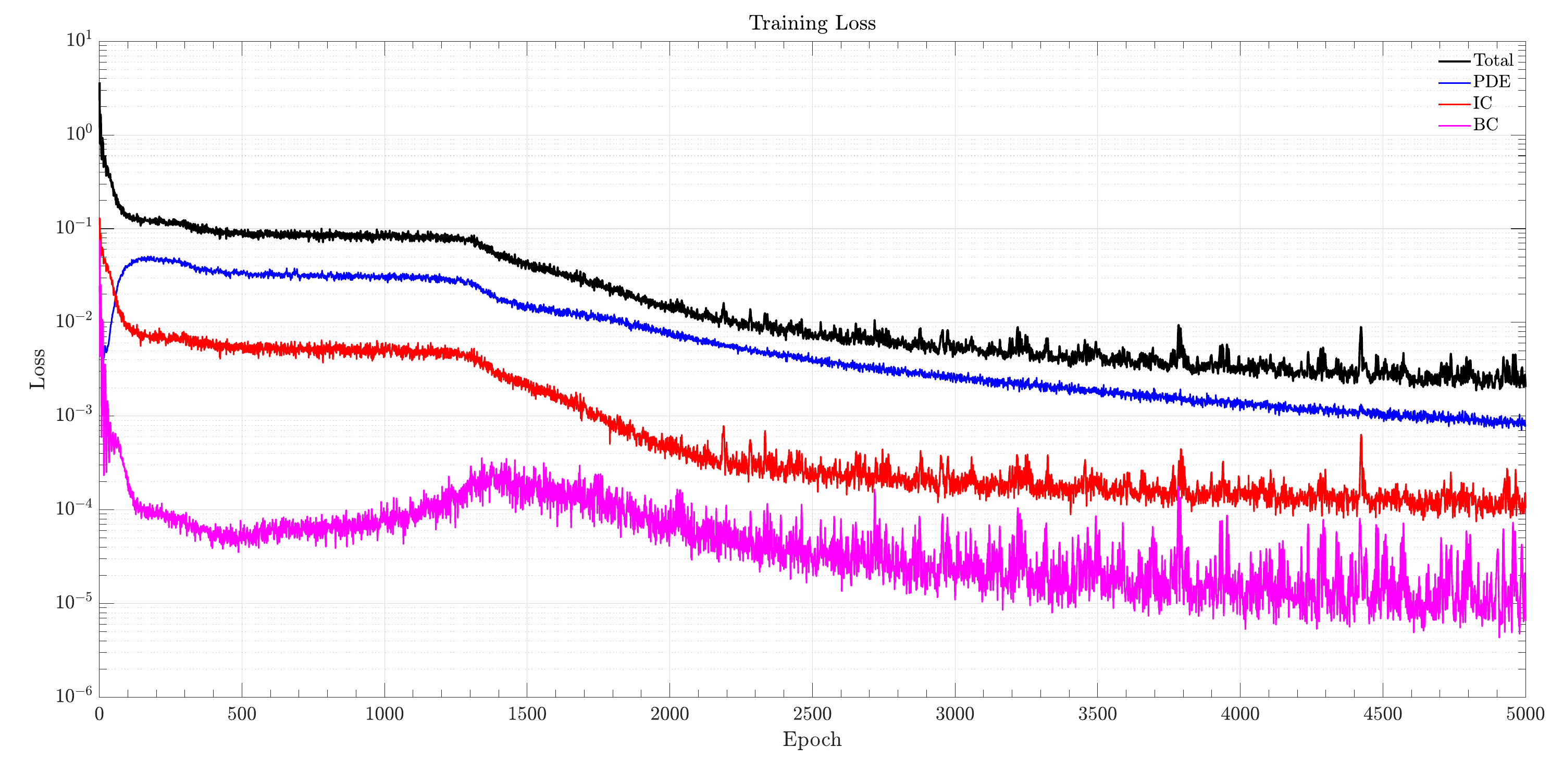}
\caption{$\phi(z)=\tanh z$, $(\beta_1,\beta_2)=(1,1)$.}
\label{fig:loss-p-beta11-s2}
\end{subfigure}
\hfill
\begin{subfigure}[t]{0.48\linewidth}
\centering
\includegraphics[width=\linewidth,height=5.2cm]{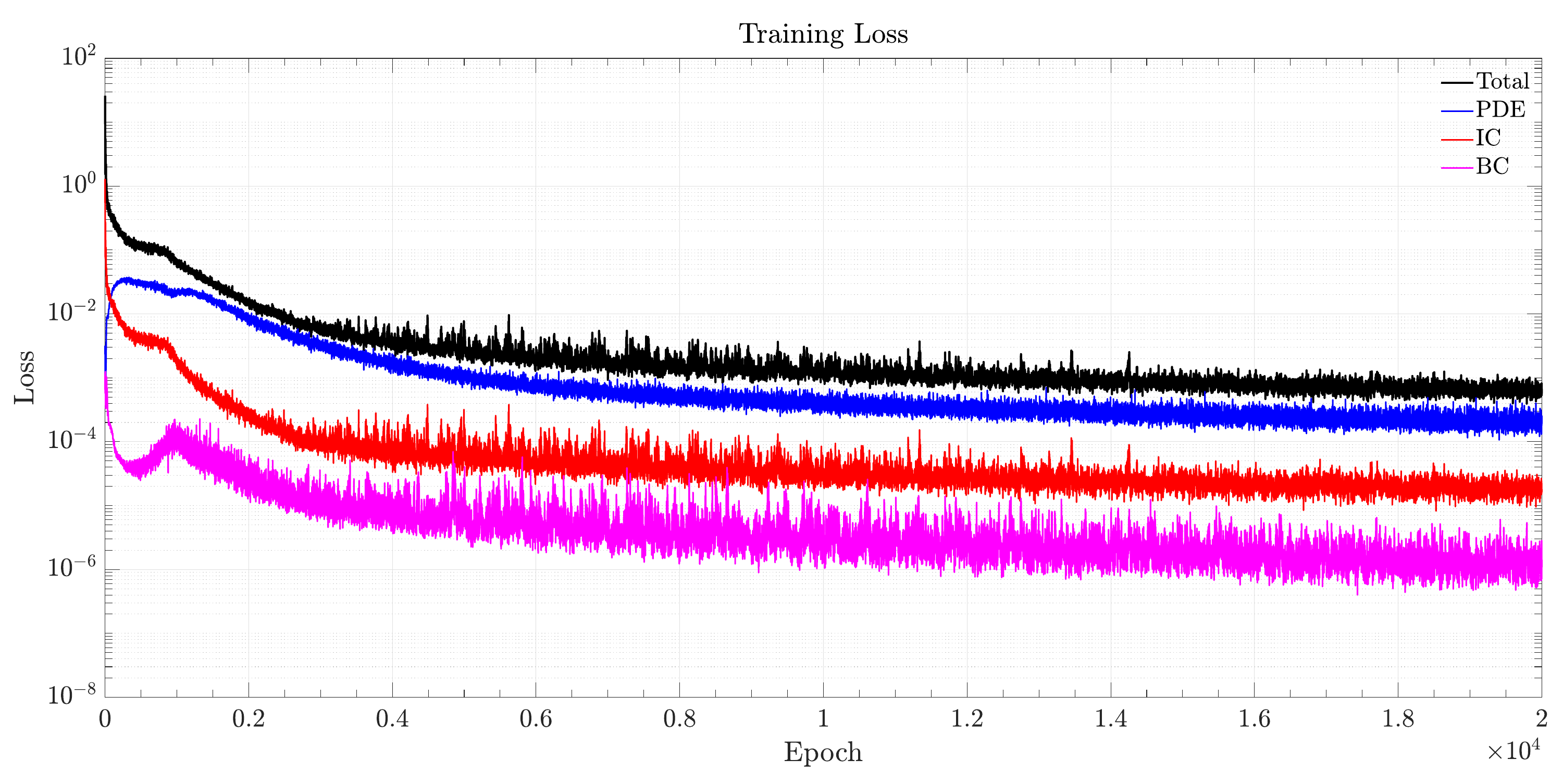}
\caption{$\phi_G(z)=\exp(-z^2)$, $(\beta_1,\beta_2)=(0,0)$.}
\label{fig:loss-p-beta00-s2-gaussian}
\end{subfigure}

\caption{
Training loss histories for the point-like source $q=\mathrm{P}$ in Poiseuille flow, $s=\mathrm{Po}$. Panels $(a)$ and $(b)$ use the default activation $\phi(z)=\tanh z$, while panel $(c)$ uses the Gaussian activation $\phi_G(z)=\exp(-z^2)$. The comparison between panels $(a)$ and $(c)$ isolates the effect of the activation function for the same non-reactive case, whereas panel $(b)$ shows the corresponding behaviour after symmetric wall absorption is introduced.
}
\label{fig:loss-point-poiseuille-activation}
\end{figure}

Figure~\ref{fig:loss-point-poiseuille-activation} focuses on the point-like source in Poiseuille flow. This group is useful because it provides a direct activation comparison for the same non-reactive case and includes the corresponding symmetric-reactive case trained with the default $\tanh$ activation. It separates the activation-sensitivity check from the broader flow--reaction comparisons. The point-like source is a stringent case because the initial solute field is localised in both the streamwise and transverse directions. The comparison shows that the default $\tanh$-activated PINN gives stable convergence for both non-reactive and reactive Poiseuille cases, while the Gaussian-activation case provides an additional check that the observed training behaviour is not tied to a single activation choice.

\begin{figure}
\centering

\begin{subfigure}[t]{0.48\linewidth}
\centering
\includegraphics[width=\linewidth,height=5.2cm]{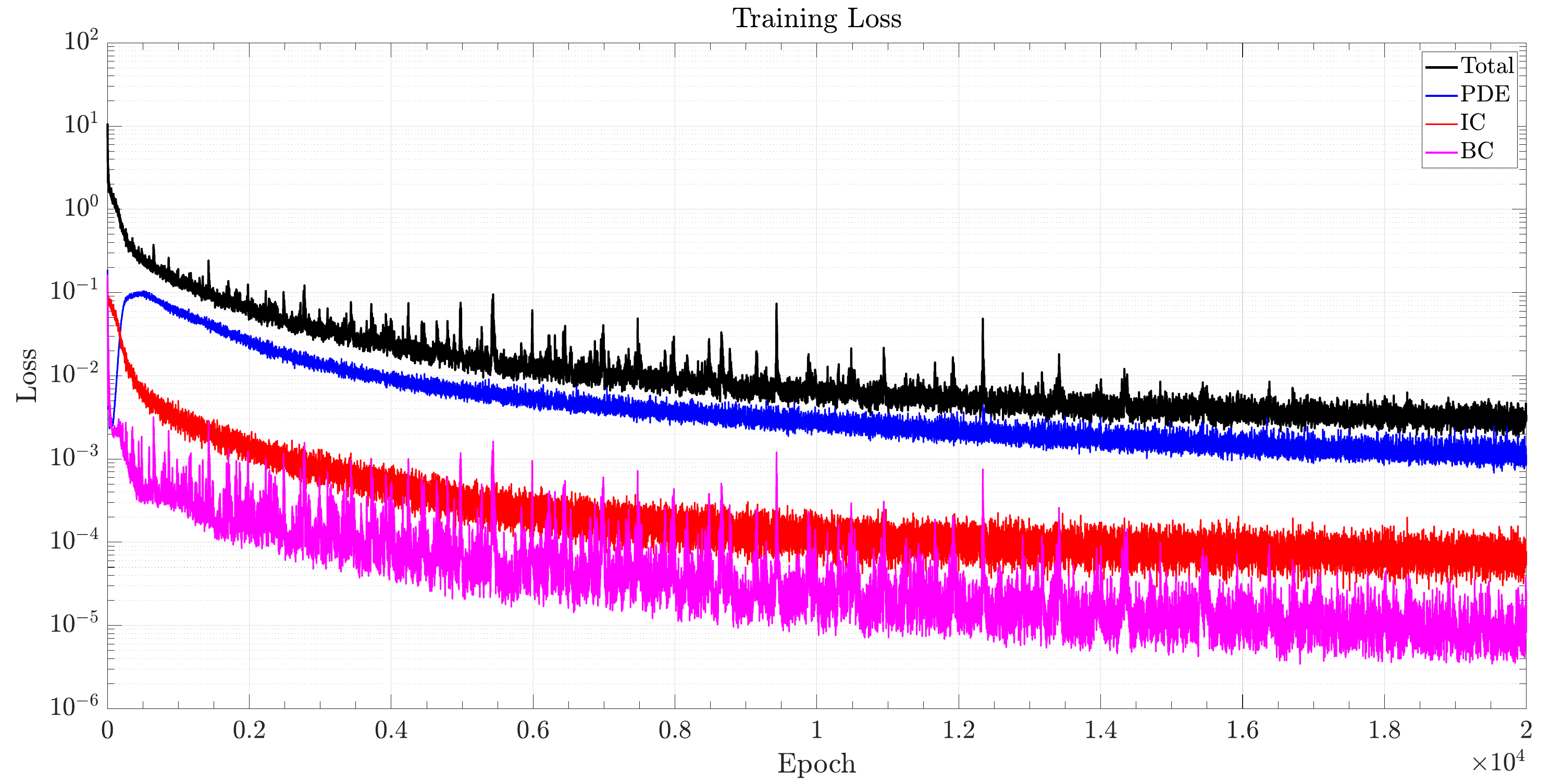}
\caption{$q=\mathrm{L}$ and $(\beta_1,\beta_2)=(0.2,2)$.}
\label{fig:loss-cp-beta022-s1}
\end{subfigure}
\hfill
\begin{subfigure}[t]{0.48\linewidth}
\centering
\includegraphics[width=\linewidth,height=5.2cm]{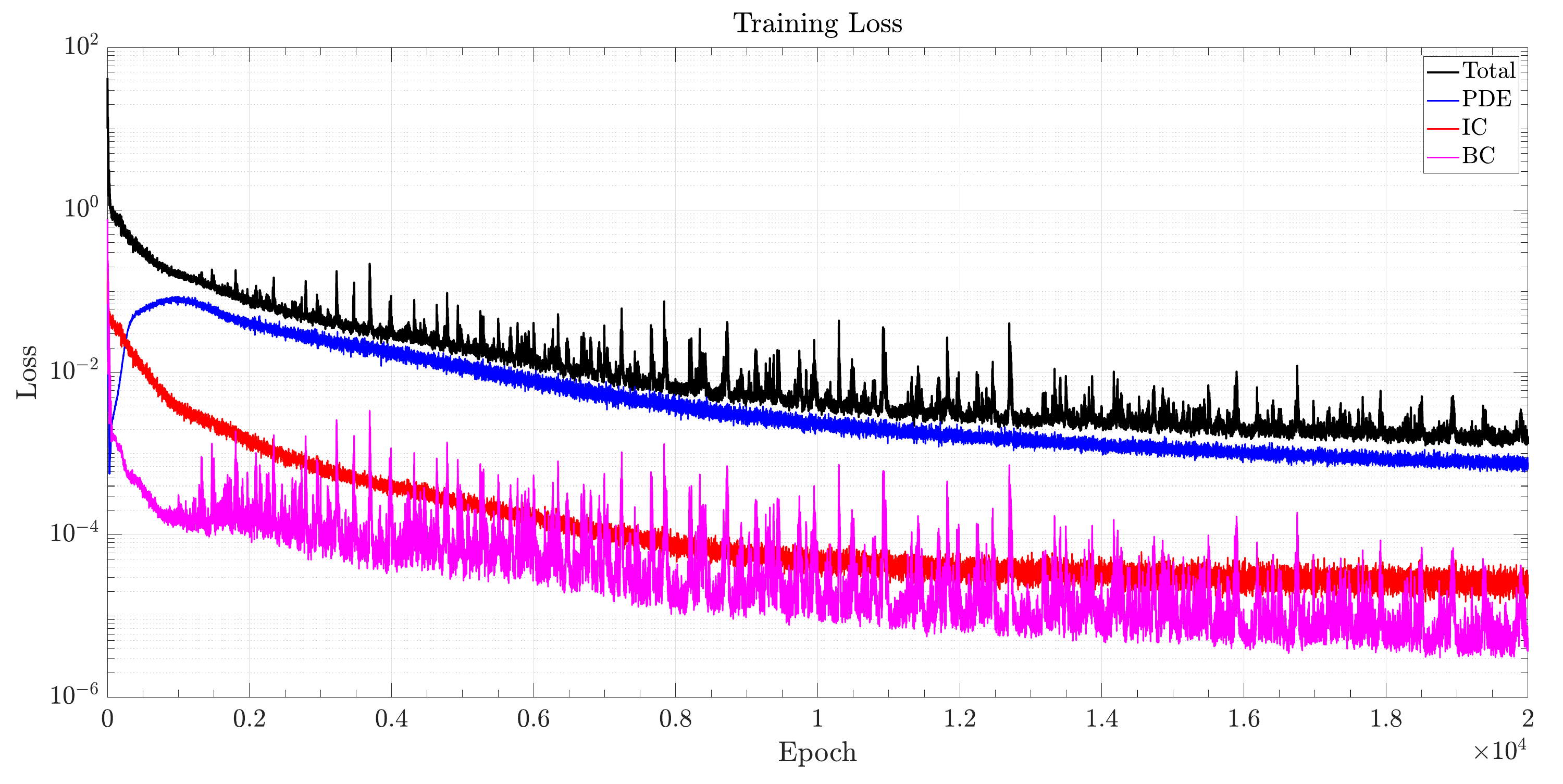}
\caption{$q=\mathrm{P}$ and $(\beta_1,\beta_2)=(2,0.2)$.}
\label{fig:loss-cp-beta202-s2}
\end{subfigure}

\caption{
Training-loss histories for asymmetric reactive cases in Couette--Poiseuille flow, $s=\mathrm{CP}$, using the default activation $\phi(z)=\tanh z$. Panel $(a)$ corresponds to a line-like source with lower-wall-dominated absorption, while panel $(b)$ corresponds to a point-like source with upper-wall-dominated absorption. These two cases test whether the PINN training remains stable when mixed shear, source localisation, and wall-reactivity asymmetry act together.
}
\label{fig:loss-asymmetric-cp-source}
\end{figure}

Figure~\ref{fig:loss-asymmetric-cp-source} reports two additional Couette--Poiseuille cases with asymmetric wall reactivity. These cases are retained in the main manuscript because they test the training behaviour under mixed shear and source-dependent reactive imbalance. This completes the convergence assessment by including two asymmetric reactive configurations in the Couette--Poiseuille profile. This flow contains both linear and parabolic shear contributions, and the asymmetric wall reactivity preferentially removes solute from one side of the channel. The observed decrease of $\mathcal{L}_{\mathrm{tot}}^{(s,q)}(\theta)$ therefore supports the stability of the optimisation procedure in cases where the concentration field is shaped simultaneously by mixed shear, wall-normal diffusion and selective wall absorption.

Entirely, figures~\ref{fig:loss-line-source}--\ref{fig:loss-asymmetric-cp-source} show that the PINN objective can be minimised reliably across the source configurations, canonical shear profiles, wall-reactivity regimes and activation functions used in the validation study. These training histories are not used as a substitute for ADI-based accuracy assessment; rather, they document the convergence behaviour of the optimisation process underlying the validated PINN solutions.

\begin{table}
	\centering
	\caption{Epoch-wise values of the monitored composite loss \(\mathcal{L}_{\mathrm{tot}}^{(s,q)}(\theta)\) for Poiseuille-flow training cases under different wall-reaction coefficients.}
	\label{tab:poiseuille-epoch-loss-values}
	\begin{tabular}{rcccc}
		\toprule
		\textbf{Case} & \textbf{Epoch 5000} & \textbf{Epoch 10000} & \textbf{Epoch 15000} & \textbf{Epoch 20000} \\
		\midrule
		\multicolumn{5}{l}{\textbf{Panel A: Configuration $\beta_1 = 0, \beta_2 = 0$}} \\
		\midrule
		1 & 0.0405298 & 0.0038469 & 0.0086044 & 0.0016563 \\
		2 & 0.0414216 & 0.0039663 & 0.0085759 & 0.0017034 \\
		3 & 0.0481322 & 0.0062045 & 0.0085097 & 0.0052009 \\
		4 & 0.0425191 & 0.0162327 & 0.0163997 & 0.0162187 \\
		5 & 0.0427171 & 0.0208591 & 0.0199454 & 0.0211767 \\
		\midrule
		\multicolumn{5}{l}{\textbf{Panel B: Configuration $\beta_1 = 1, \beta_2 = 1$}} \\
		\midrule
		1 & 0.0158459 & 0.0102640 & 0.0033206 & 0.0037620 \\
		2 & 0.0131776 & 0.0093736 & 0.0040590 & 0.0033488 \\
		3 & 0.0161532 & 0.0094519 & 0.0063048 & 0.0039626 \\
		4 & 0.0254394 & 0.0145121 & 0.0127275 & 0.0086915 \\
		5 & 0.0228691 & 0.0124900 & 0.0118653 & 0.0070485 \\
		\midrule
		\multicolumn{5}{l}{\textbf{Panel C: Configuration $\beta_1 = 2, \beta_2 = 0.2$}} \\
		\midrule
		1 & 0.0260612 & 0.0117902 & 0.0072850 & 0.0056742 \\
		2 & 0.0233718 & 0.0095431 & 0.0065202 & 0.0051939 \\
		3 & 0.0298999 & 0.0131758 & 0.0108181 & 0.0087885 \\
		4 & 0.0530314 & 0.0189359 & 0.0160859 & 0.0127159 \\
		5 & 0.0343961 & 0.0171918 & 0.0141833 & 0.0115769 \\
		\midrule
		\multicolumn{5}{l}{\textbf{Panel D: Configuration $\beta_1 = 0.2, \beta_2 = 2$}} \\
		\midrule
		1 &0.0264832	&0.0177098&	0.0073796	&0.007279\\
			
2 &0.0228858	&0.0158657	&0.0054081	&0.0062097\\
			
3 &0.0293819	&0.013583	&0.0041329	&0.0065858\\
			
4&0.0261047	&0.014644	&0.0089583	&0.0101251\\
			
5&0.0251966	&0.0154548	&0.0121291	&0.0120438\\

		\bottomrule
	\end{tabular}
\end{table}

To provide a numerical supplement to the graphical loss histories, representative values of the monitored composite loss are reported in table~\ref{tab:poiseuille-epoch-loss-values}. This complements the graphical training-loss histories and should be interpreted as an optimisation diagnostic. The ADI comparisons remain the primary solution-level validation of the trained PINN models.

\subsection{Validation of the time-dependent dispersion coefficient}
\label{subsec:dispersion-coefficient-validation}

The preceding comparisons establish the accuracy of the PINN at the levels of the two-dimensional concentration field and the cross-sectionally averaged concentration. We next examine whether this agreement is retained for the time-dependent axial spreading rate of the solute cloud. This constitutes a more demanding validation because the dispersion coefficient is obtained from the spatially integrated plume structure and the temporal derivative of its second central moment. Consequently, errors in plume displacement, deformation, or spreading can affect this quantity even when the corresponding concentration fields appear visually close.

Using the cross-sectionally averaged concentration $\langle C \rangle(x,t)$ and the streamwise centroid $x_g(t)$ introduced previously, the normalised second central moment is written as
\begin{equation}
\nu_2(t)=\frac{\displaystyle\int_{x_0}^{x_1}\left[x-x_g(t)\right]^2\langle C \rangle(x,t)\,\mathrm{d}x}{\displaystyle\int_{x_0}^{x_1}\langle C \rangle(x,t)\,\mathrm{d}x}.
\label{eq:second-central-moment-dispersion}
\end{equation}
The corresponding apparent axial dispersion coefficient is
\begin{equation}
D_{\mathrm{a}}(t)=\frac{1}{2}\frac{\mathrm{d}\nu_2(t)}{\mathrm{d}t}.
\label{eq:apparent-dispersion-coefficient}
\end{equation}
This moment-based definition captures the transient evolution of the effective spreading rate and connects the plume-width dynamics to the apparent axial dispersion coefficient \cite{Aris1956, Barton1983}.

The absolute PINN--ADI discrepancy is defined by
\begin{equation}
E_D(t)=\left|D_{\mathrm{a}}^{\mathrm{PINN}}(t)-D_{\mathrm{a}}^{\mathrm{ADI}}(t)\right|.
\label{eq:dispersion-coefficient-error}
\end{equation}
Unlike a pointwise concentration error, $E_D(t)$ assesses a derived transport quantity involving spatial integration, normalisation, and temporal differentiation. Agreement in $D_{\mathrm{a}}(t)$ therefore provides evidence that the PINN reproduces the evolving streamwise spreading dynamics generated by advection, molecular diffusion and reactive wall removal.

\begin{figure}
\centering
\includegraphics[width=0.82\linewidth,height=7.2cm,keepaspectratio]{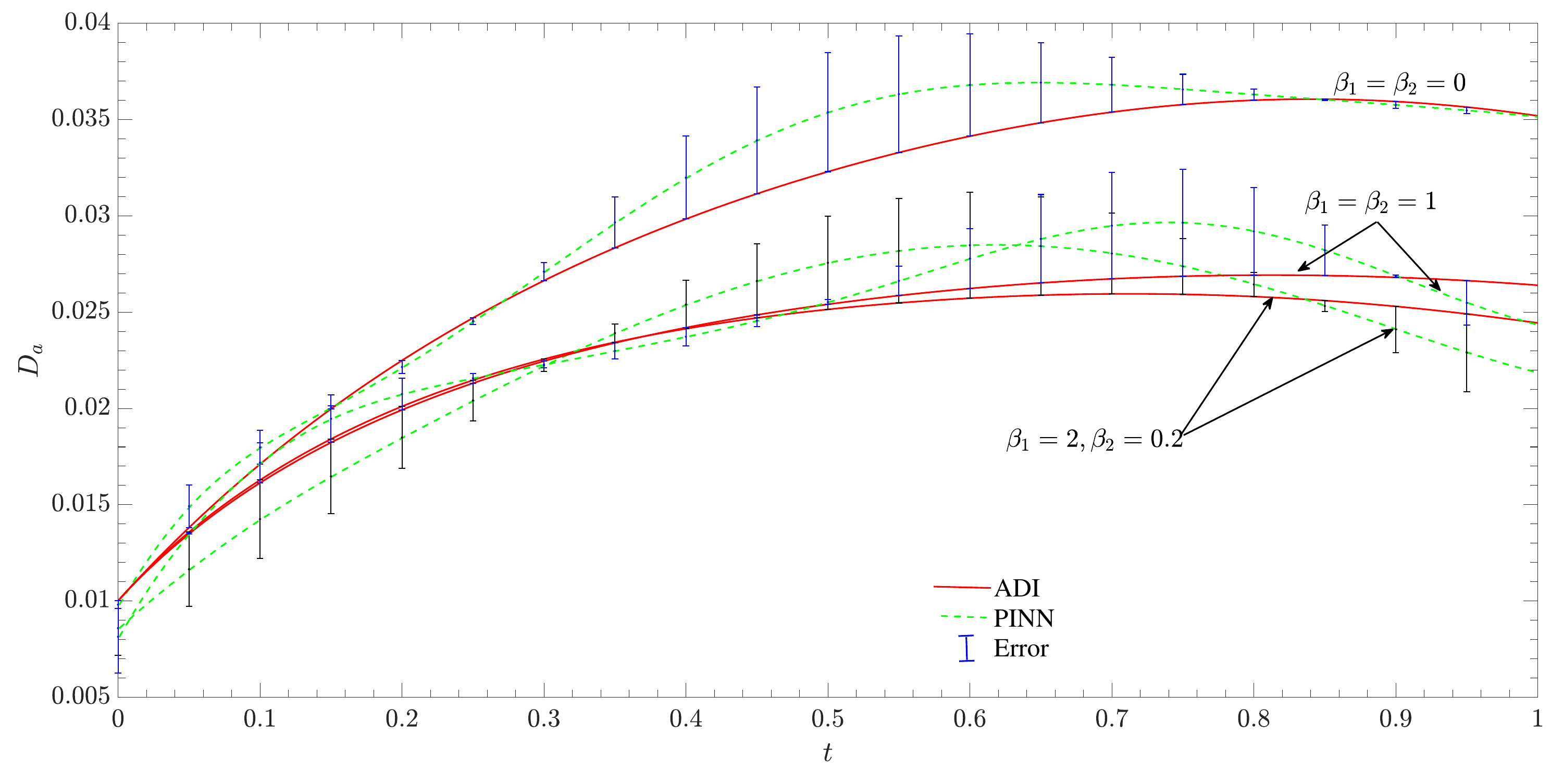}
\caption{
Validation of the effective axial dispersion coefficient for Couette flow with the line-like source, $q=\mathrm{L}$. The internal panels correspond to the wall-reactivity configurations $(\beta_1,\beta_2)=(0,0)$, $(\beta_1,\beta_2)=(1,1)$ and $(\beta_1,\beta_2)=(2,0.2)$. For each configuration, $D_{\mathrm{a}}^{\mathrm{PINN}}(t)$ is compared with $D_{\mathrm{a}}^{\mathrm{ADI}}(t)$, together with the corresponding absolute discrepancy $E_D(t)$.
}
\label{fig:dispersion-validation-line-couette}
\end{figure}

Figure~\ref{fig:dispersion-validation-line-couette} provides a focused assessment of the line-like source under Couette shear. Because the initial concentration is distributed across the transverse direction, the subsequent variation of $D_{\mathrm{a}}(t)$ reflects the combined action of the linear velocity gradient, molecular diffusion and selective wall removal. Presenting the three wall-reactivity cases together permits a direct evaluation of whether the PINN retains its accuracy as the transport mechanism changes from non-reactive dispersion to symmetric and asymmetric reactive dispersion.

Physically, the early-time growth of \(D_{\mathrm{a}}(t)\) is dominated by advective stretching and transverse diffusion, while the later-time tendency toward a plateau over the observed interval reflects the developing balance between shear-induced longitudinal spreading and transverse homogenization. In Poiseuille flow, the symmetric velocity profile restricts the maximal shear to the near-wall regions, whereas Couette flow maintains a constant shear rate across the entire domain. Consequently, the effective dispersion coefficient $D_{\mathrm{a}}(t)$ in Couette flow exhibits a fundamentally different transient growth phase, as the solute is continuously stretched without the homogenising effect of a zero-shear centreline.

\begin{figure}
\centering

\begin{subfigure}[t]{0.48\linewidth}
\centering
\includegraphics[width=\linewidth,height=5.6cm,keepaspectratio]{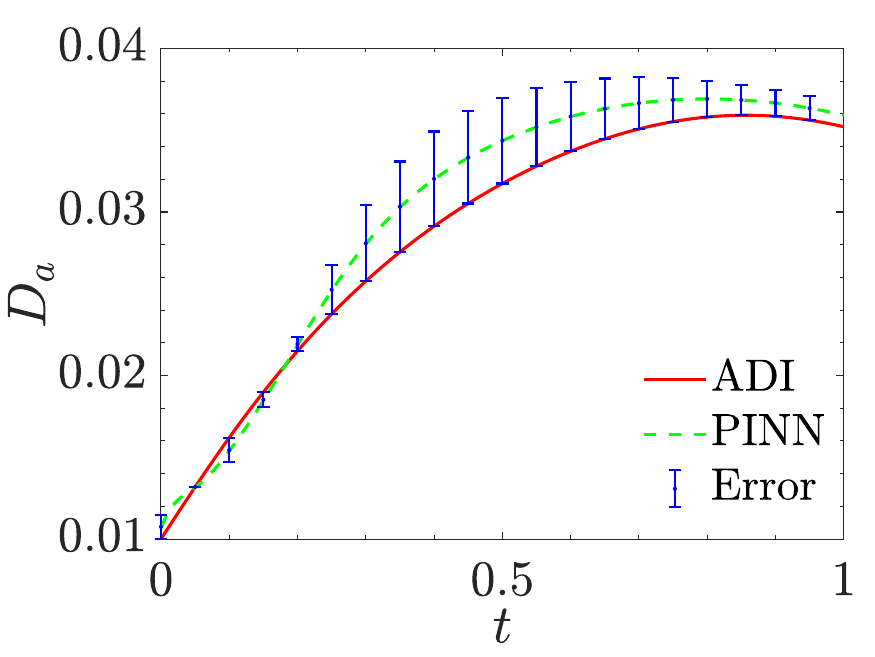}
\caption{Couette flow with $q=\mathrm{P}$ and $(\beta_1,\beta_2)=(0,0)$.}
\label{fig:daterr-s2-c-beta00}
\end{subfigure}
\hfill
\begin{subfigure}[t]{0.48\linewidth}
\centering
\includegraphics[width=\linewidth,height=5.6cm,keepaspectratio]{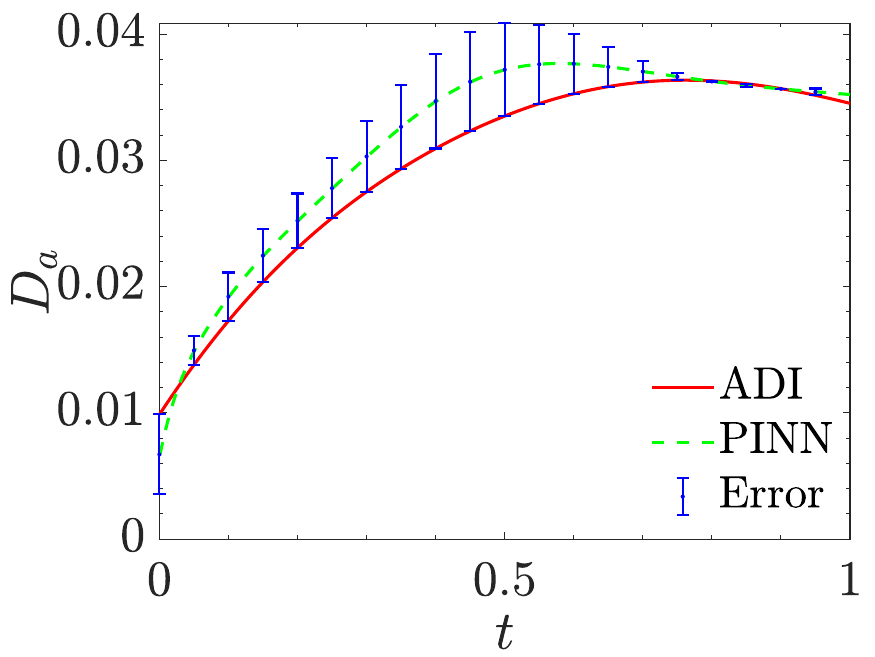}
\caption{Couette--Poiseuille flow with $q=\mathrm{P}$ and $(\beta_1,\beta_2)=(0,0)$.}
\label{fig:daterr-s2-cp-beta00}
\end{subfigure}

\caption{
Validation of the effective axial dispersion coefficient for non-reactive point-like-source configurations. The panels compare $D_{\mathrm{a}}^{\mathrm{PINN}}(t)$ with $D_{\mathrm{a}}^{\mathrm{ADI}}(t)$ and report the corresponding absolute discrepancy $E_D(t)$. The comparison between Couette and Couette--Poiseuille flow tests whether the PINN reproduces the spreading of an initially localised solute cloud under distinct shear structures before reactive wall removal is introduced.
}
\label{fig:dispersion-validation-point-nonreactive}
\end{figure}

Figure~\ref{fig:dispersion-validation-point-nonreactive} establishes the non-reactive point-source baseline. In contrast to the line-like source, the point-like source introduces an initially localized wall-normal distribution. Its effective dispersion coefficient is therefore influenced by the early transverse redistribution of the solute as well as by its subsequent streamwise stretching. The agreement between the two velocity profiles confirms that the learned solution captures these coupled stages of plume evolution.

\begin{figure}
\centering

\begin{subfigure}[t]{0.48\linewidth}
\centering
\includegraphics[width=\linewidth,height=5.6cm,keepaspectratio]{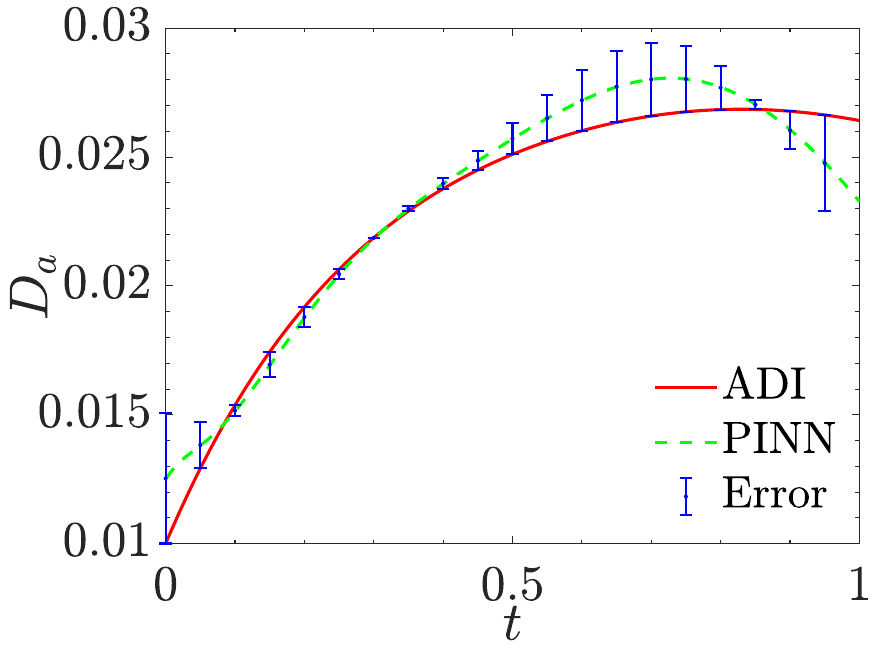}
\caption{Couette flow with $q=\mathrm{P}$ and $(\beta_1,\beta_2)=(1,1)$.}
\label{fig:daterr-s2-c-beta11}
\end{subfigure}
\hfill
\begin{subfigure}[t]{0.48\linewidth}
\centering
\includegraphics[width=\linewidth,height=5.6cm,keepaspectratio]{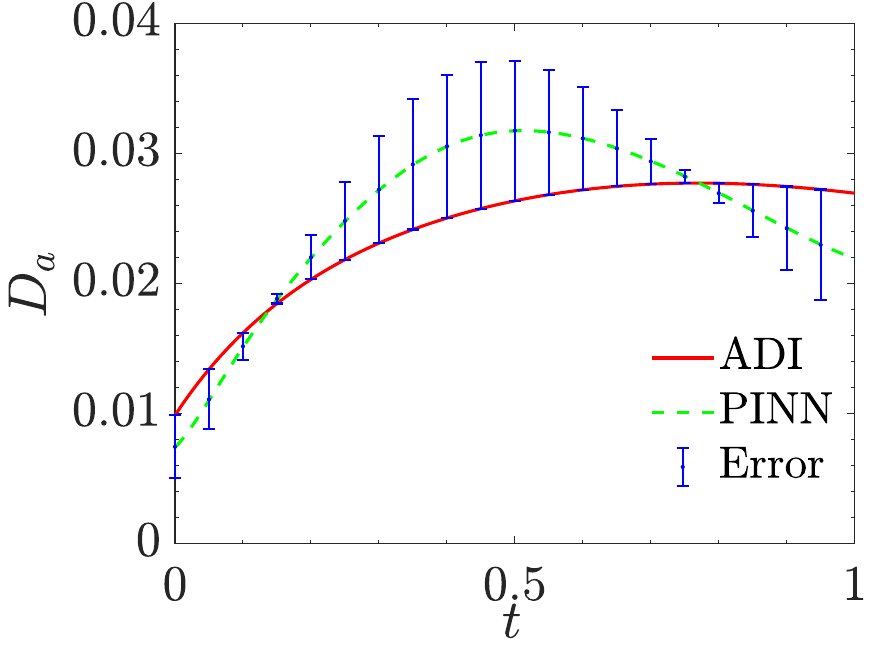}
\caption{Couette--Poiseuille flow with $q=\mathrm{P}$ and $(\beta_1,\beta_2)=(1,1)$.}
\label{fig:daterr-s2-cp-beta11}
\end{subfigure}

\caption{
Validation of the effective axial dispersion coefficient for symmetric wall absorption, $(\beta_1,\beta_2)=(1,1)$. The panels compare $D_{\mathrm{a}}^{\mathrm{PINN}}(t)$ with $D_{\mathrm{a}}^{\mathrm{ADI}}(t)$ for point-like releases in Couette and Couette--Poiseuille flow and show the associated absolute discrepancy $E_D(t)$.
}
\label{fig:dispersion-validation-symmetric}
\end{figure}

Figure~\ref{fig:dispersion-validation-symmetric} extends the comparison to equal wall reactivity. Symmetric absorption reduces the surviving solute mass at both boundaries without imposing a preferential transverse direction of removal. Nevertheless, it can modify the normalised second central moment by selectively removing material that reaches the walls. The comparison, therefore, tests whether the PINN captures the effect of symmetric wall uptake on the evolving axial spreading rate.

\begin{figure}
\centering

\begin{subfigure}[t]{0.48\linewidth}
\centering
\includegraphics[width=\linewidth,height=5.6cm,keepaspectratio]{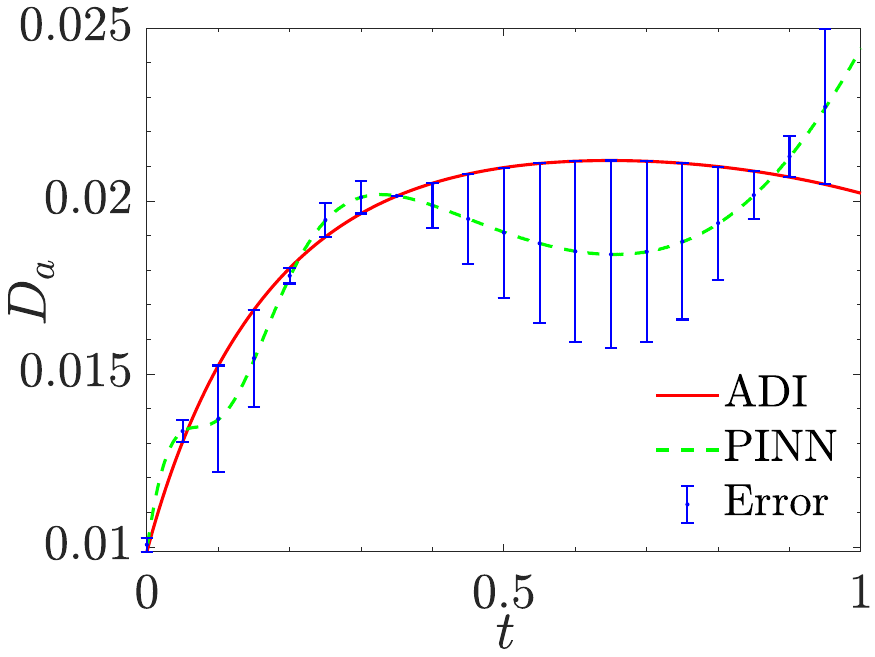}
\caption{Couette--Poiseuille flow with $q=\mathrm{L}$ and $(\beta_1,\beta_2)=(0.2,2)$.}
\label{fig:daterr-s1-cp-beta022}
\end{subfigure}
\hfill
\begin{subfigure}[t]{0.48\linewidth}
\centering
\includegraphics[width=\linewidth,height=5.6cm,keepaspectratio]{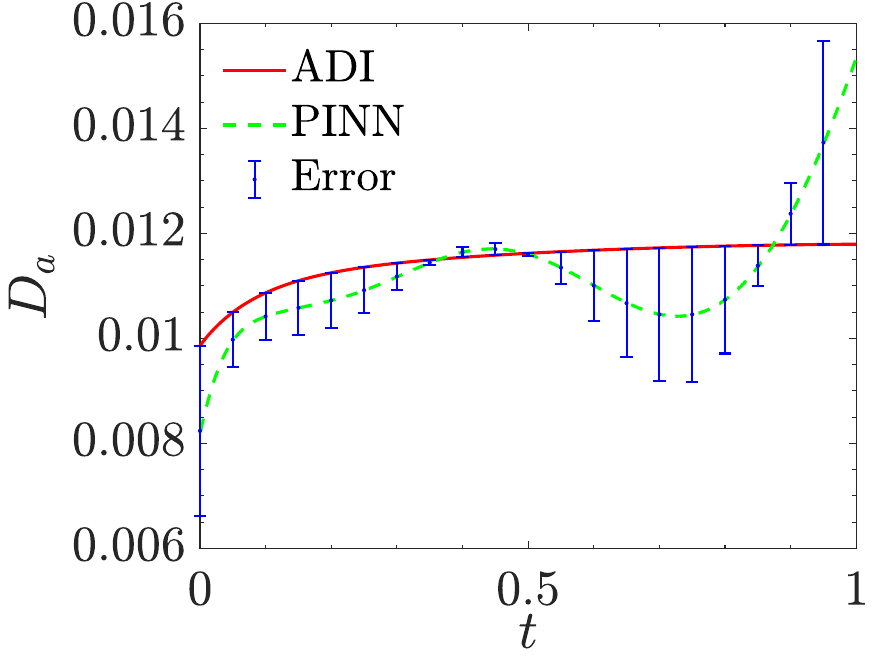}
\caption{Poiseuille flow with $q=\mathrm{P}$ and $(\beta_1,\beta_2)=(0.2,2)$.}
\label{fig:daterr-s2-p-beta022}
\end{subfigure}

\caption{
Validation of the effective axial dispersion coefficient for lower-wall-dominated absorption, $(\beta_1,\beta_2)=(0.2,2)$. The panels compare $D_{\mathrm{a}}^{\mathrm{PINN}}(t)$ with $D_{\mathrm{a}}^{\mathrm{ADI}}(t)$ for line-like and point-like sources under Couette--Poiseuille and Poiseuille flow, respectively, together with the corresponding absolute discrepancy $E_D(t)$.
}
\label{fig:dispersion-validation-lower-dominated}
\end{figure}

Figure~\ref{fig:dispersion-validation-lower-dominated} examines a strongly asymmetric reaction regime in which removal at the lower wall dominates. The unequal values of $\beta_1$ and $\beta_2$ alter the wall-normal distribution of the surviving solute and can consequently change its normalised streamwise variance. Including both source configurations and two different velocity profiles provides a stringent test of whether the PINN reproduces the coupling between shear-induced deformation and selective wall absorption.

\begin{figure}
\centering

\begin{subfigure}[t]{0.48\linewidth}
\centering
\includegraphics[width=\linewidth,height=5.6cm,keepaspectratio]{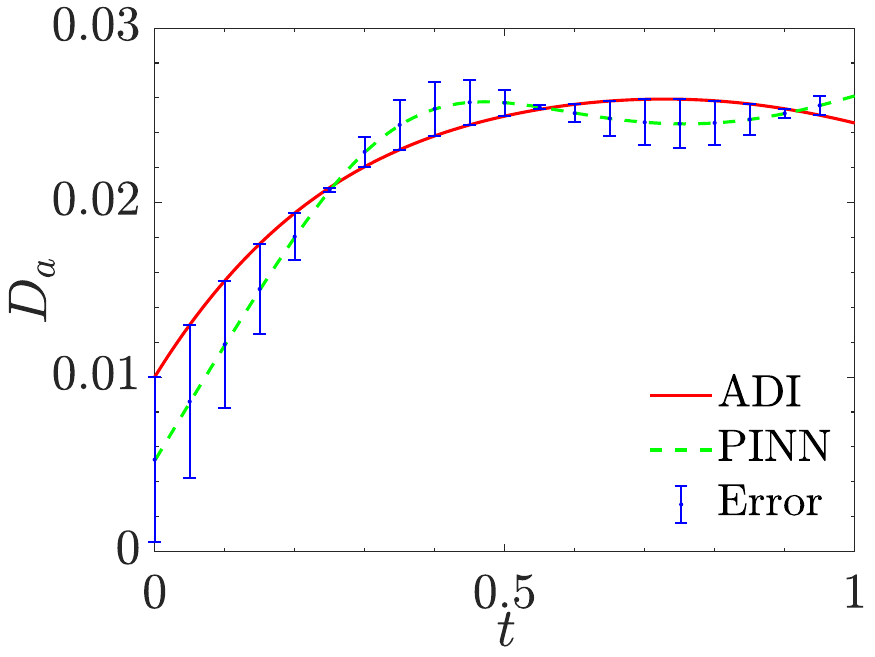}
\caption{Couette flow with $q=\mathrm{P}$ and $(\beta_1,\beta_2)=(2,0.2)$.}
\label{fig:daterr-s2-c-beta202}
\end{subfigure}
\hfill
\begin{subfigure}[t]{0.48\linewidth}
\centering
\includegraphics[width=\linewidth,height=5.6cm,keepaspectratio]{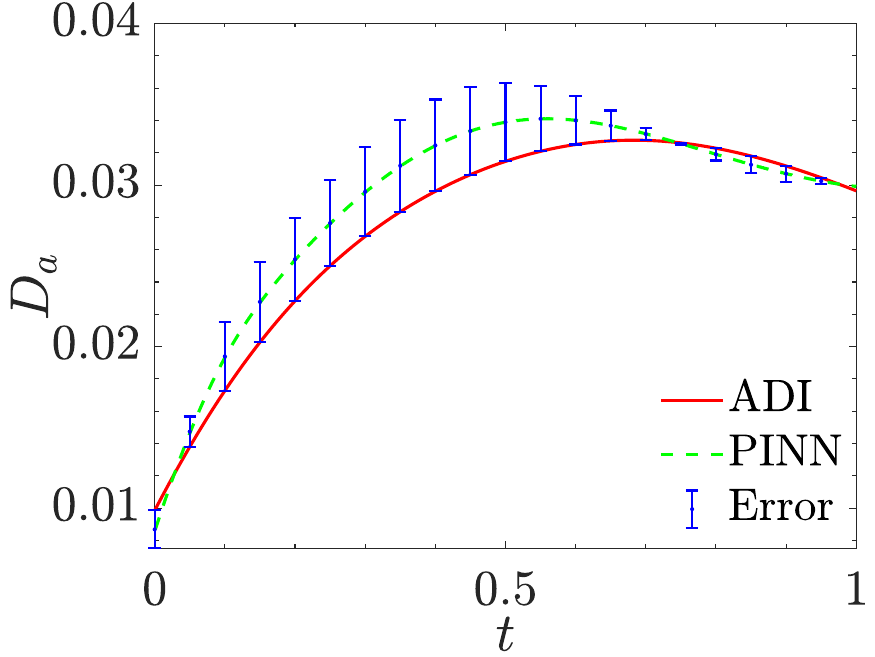}
\caption{Couette--Poiseuille flow with $q=\mathrm{P}$ and $(\beta_1,\beta_2)=(2,0.2)$.}
\label{fig:daterr-s2-cp-beta202}
\end{subfigure}

\caption{
Validation of the effective axial dispersion coefficient for upper-wall-dominated absorption, $(\beta_1,\beta_2)=(2,0.2)$. The panels compare $D_{\mathrm{a}}^{\mathrm{PINN}}(t)$ with $D_{\mathrm{a}}^{\mathrm{ADI}}(t)$ for point-like releases in Couette and Couette--Poiseuille flow and show the corresponding absolute discrepancy $E_D(t)$.
}
\label{fig:dispersion-validation-upper-dominated}
\end{figure}

Figure~\ref{fig:dispersion-validation-upper-dominated} considers the reversed wall-reactivity asymmetry. Transferring the stronger absorption from the lower wall to the upper wall changes which part of the sheared concentration field is preferentially removed. The resulting comparison provides an additional test of the PINN because the effect of the reaction asymmetry depends on the interaction between the transverse concentration distribution and the local streamwise velocity.

Taken together, figures~\ref{fig:dispersion-validation-line-couette}--\ref{fig:dispersion-validation-upper-dominated} establish the accuracy of the time-dependent dispersion coefficient across line-like and point-like sources, Couette, Poiseuille and Couette--Poiseuille flows, and non-reactive, symmetric and asymmetric wall-reactivity regimes. The inclusion of these comparisons in the main manuscript is justified because $D_{\mathrm{a}}(t)$ is a physically interpretable transport measure rather than solely a numerical error diagnostic. Agreement between $D_{\mathrm{a}}^{\mathrm{PINN}}(t)$ and $D_{\mathrm{a}}^{\mathrm{ADI}}(t)$ confirms that the PINN reproduces not only the concentration distribution but also the evolving streamwise spreading rate of the surviving reactive solute.

\subsection{Validation of the total surviving solute mass}
\label{subsec:total-mass-validation}

The local concentration field and cross-sectional profile comparisons are complemented by a global mass-level validation. For both source configurations, the total surviving solute mass is defined as
\begin{equation}
M_0(t)=\int_{x_0}^{x_1}\int_{-1}^{1}C(x,y,t)\mathrm{d}y\,\mathrm{d}x.
\label{eq:total-surviving-mass}
\end{equation}
The corresponding absolute discrepancy between the PINN and ADI predictions is
\begin{equation}
E_{M_0}(t)=\left|M_0^{\mathrm{PINN}}(t)-M_0^{\mathrm{ADI}}(t)\right|.
\label{eq:total-mass-error}
\end{equation}
The quantity $M_0(t)$ provides a global measure of the amount of solute remaining within the channel. It is therefore sensitive to accumulated concentration errors over the full computational domain and, in the reactive cases, to the integrated effect of wall absorption. Agreement in $M_0(t)$ is particularly important because an accurate local concentration field does not by itself guarantee that the total surviving mass or its decay rate is reproduced correctly.

\begin{figure}
\centering

\begin{subfigure}[t]{0.48\linewidth}
\centering
\includegraphics[width=\linewidth,height=5.7cm,keepaspectratio]{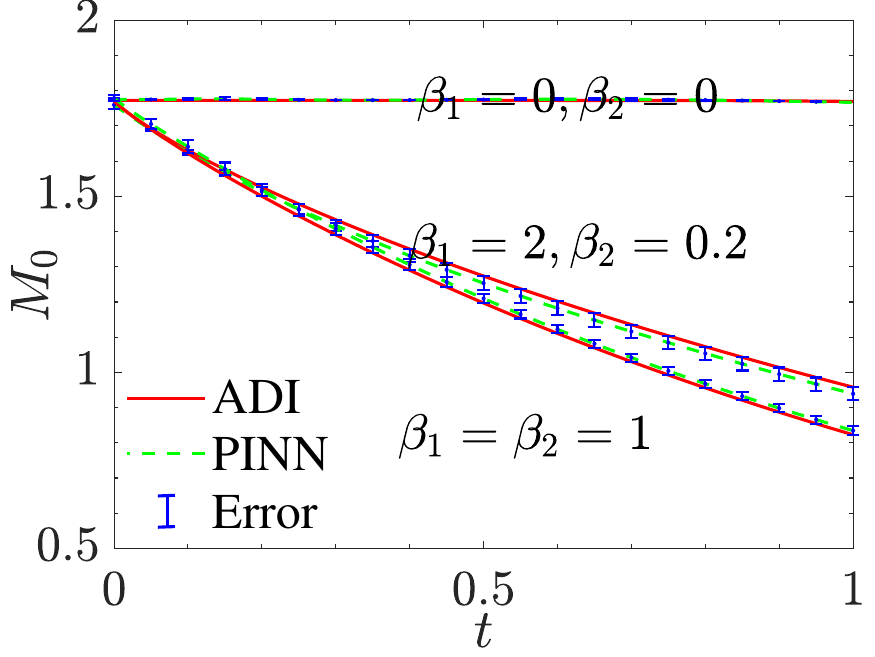}
\caption{Couette flow with the line-like source, $q=\mathrm{L}$.}
\label{fig:m0terr-s1-c}
\end{subfigure}
\hfill
\begin{subfigure}[t]{0.48\linewidth}
\centering
\includegraphics[width=\linewidth,height=5.7cm,keepaspectratio]{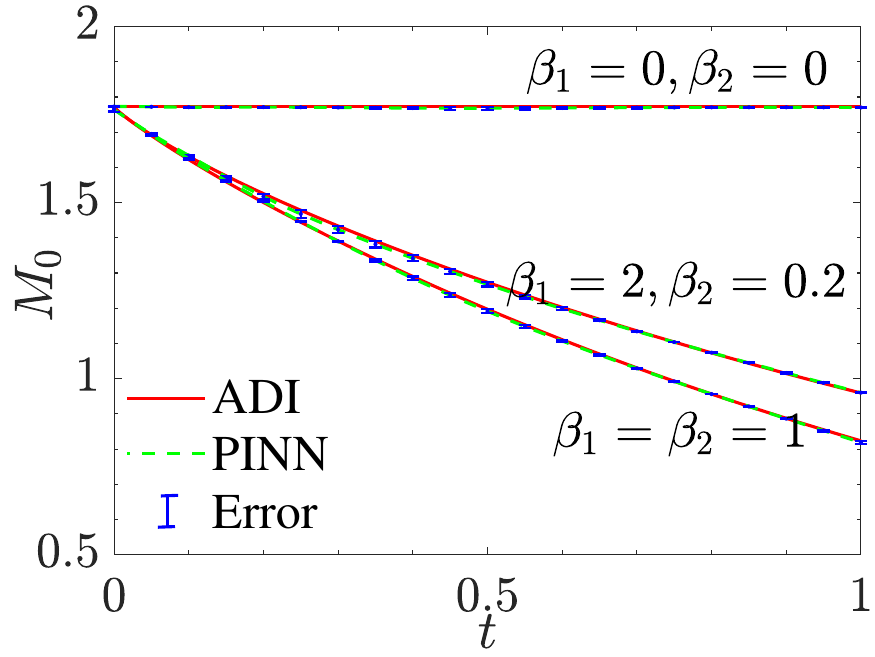}
\caption{Poiseuille flow with the line-like source, $q=\mathrm{L}$.}
\label{fig:m0terr-s1-p}
\end{subfigure}

\vspace{0.25cm}

\begin{subfigure}[t]{0.60\linewidth}
\centering
\includegraphics[width=\linewidth,height=6.2cm,keepaspectratio]{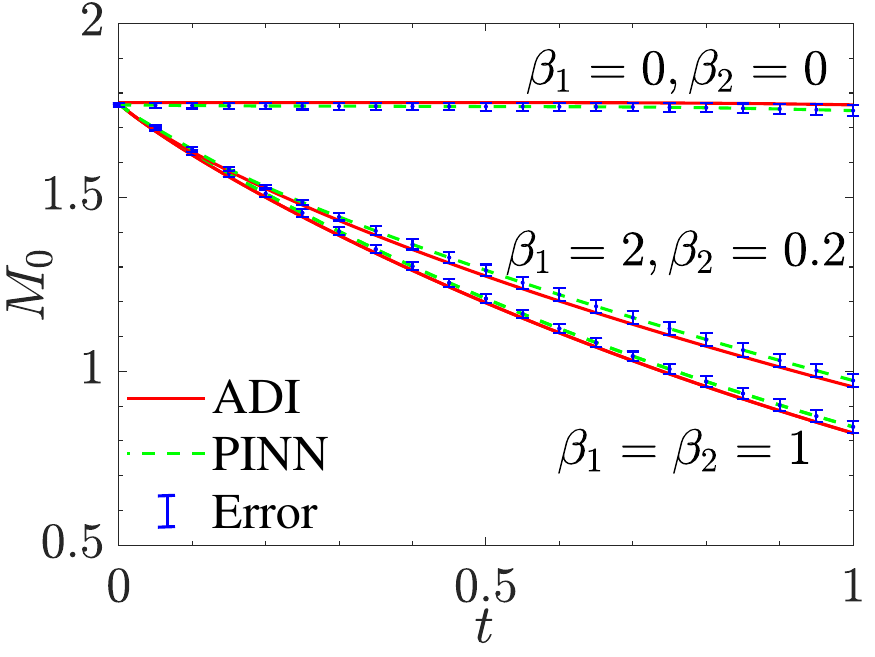}
\caption{Couette--Poiseuille flow with the line-like source, $q=\mathrm{L}$.}
\label{fig:m0terr-s1-cp}
\end{subfigure}

\caption{
Validation of the total surviving solute mass $M_0(t)$ for the line-like source under Couette, Poiseuille and Couette--Poiseuille flows. The wall-reactivity configurations are identified within the individual panels. In each case, $M_0^{\mathrm{PINN}}(t)$ is compared with $M_0^{\mathrm{ADI}}(t)$, together with the corresponding absolute discrepancy $E_{M_0}(t)$.
}
\label{fig:total-mass-validation-line-source}
\end{figure}

Figure~\ref{fig:total-mass-validation-line-source} shows the total-mass validation for the line-like source. Since the initial solute distribution is spread across the channel height, the evolution of $M_0(t)$ reflects the combined effects of transverse diffusion, shear-induced redistribution and reactive removal at the boundaries. The agreement between $M_0^{\mathrm{PINN}}(t)$ and $M_0^{\mathrm{ADI}}(t)$ confirms that the PINN captures the integrated mass evolution for an initially distributed concentration field.

\begin{figure}
\centering

\begin{subfigure}[t]{0.48\linewidth}
\centering
\includegraphics[width=\linewidth,height=5.7cm,keepaspectratio]{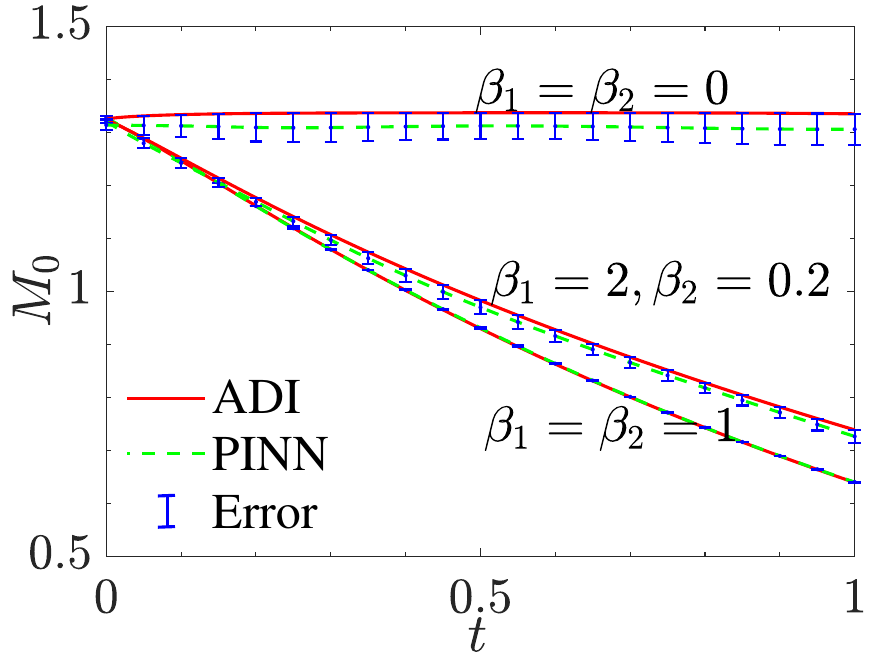}
\caption{Couette flow with the point-like source, $q=\mathrm{P}$.}
\label{fig:m0terr-s2-c}
\end{subfigure}
\hfill
\begin{subfigure}[t]{0.48\linewidth}
\centering
\includegraphics[width=\linewidth,height=5.7cm,keepaspectratio]{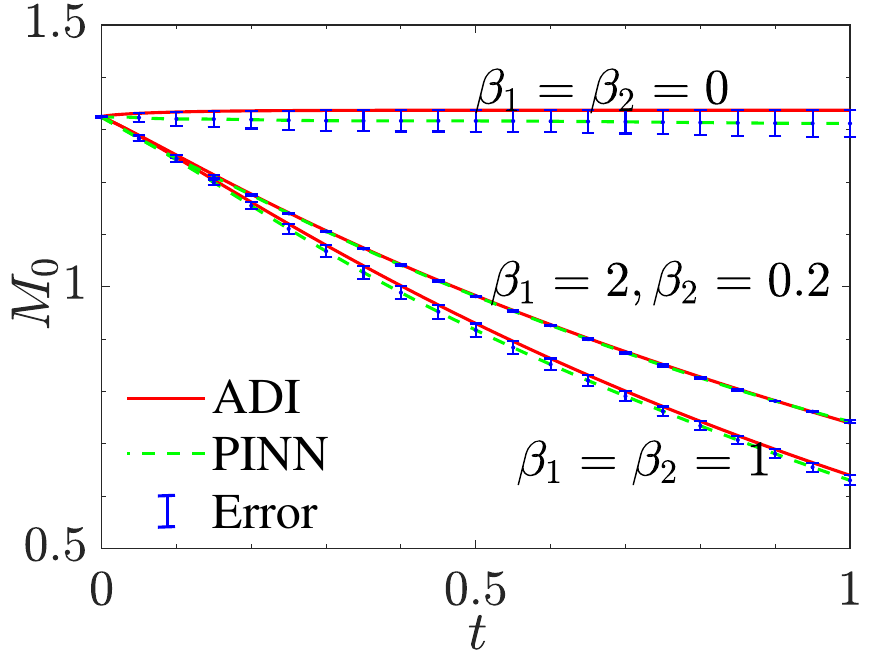}
\caption{Poiseuille flow with the point-like source, $q=\mathrm{P}$.}
\label{fig:m0terr-s2-p}
\end{subfigure}

\vspace{0.25cm}

\begin{subfigure}[t]{0.60\linewidth}
\centering
\includegraphics[width=\linewidth,height=6.2cm,keepaspectratio]{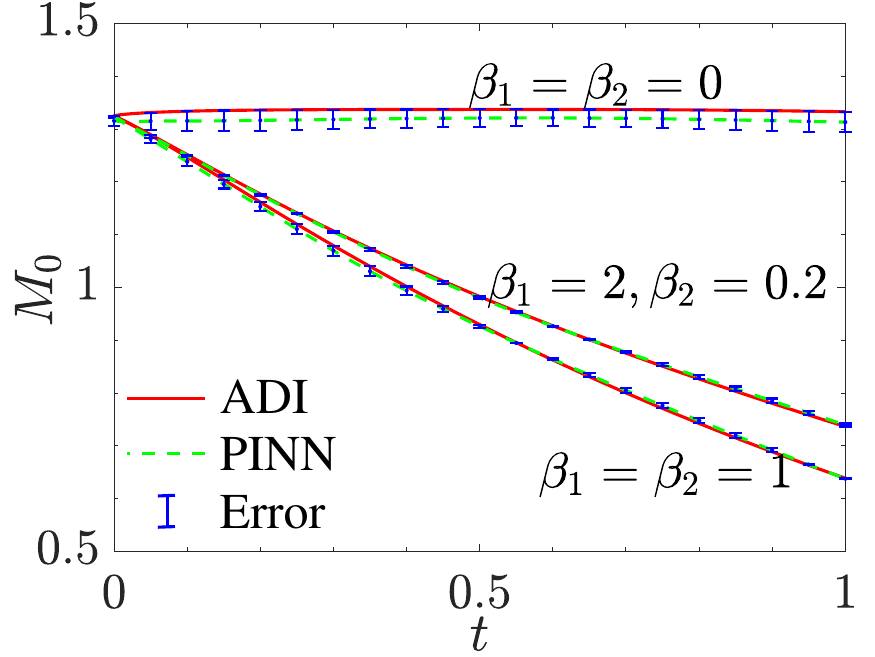}
\caption{Couette--Poiseuille flow with the point-like source, $q=\mathrm{P}$.}
\label{fig:m0terr-s2-cp}
\end{subfigure}

\caption{
Validation of the total surviving solute mass $M_0(t)$ for the point-like source under Couette, Poiseuille and Couette--Poiseuille flows. The wall-reactivity configurations are identified within the individual panels. In each case, $M_0^{\mathrm{PINN}}(t)$ is compared with $M_0^{\mathrm{ADI}}(t)$, together with the corresponding absolute discrepancy $E_{M_0}(t)$.
}
\label{fig:total-mass-validation-point-source}
\end{figure}

Figure~\ref{fig:total-mass-validation-point-source} shows the corresponding validation for the point-like source. In this case, the solute is initially localised, so the subsequent evolution of $M_0(t)$ depends on early transverse redistribution as well as on the transport of solute toward the reactive walls. The agreement between $M_0^{\mathrm{PINN}}(t)$ and $M_0^{\mathrm{ADI}}(t)$ verifies that the PINN captures not only the spatial distribution of concentration but also the integrated mass loss generated by the Robin boundary conditions.

The comparison across Couette, Poiseuille, and Couette--Poiseuille flows is useful because the velocity profile controls how rapidly solute is redistributed within the channel and transported toward regions from which transverse diffusion can deliver it to the reactive walls. The symmetric and asymmetric wall-reactivity cases further test whether the learned solution reproduces changes in the global removal rate as the absorption strengths at the two boundaries vary. Consequently, validating $M_0(t)$ provides an independent mass-balance assessment before considering higher-order spreading measures.

From a hydrodynamic perspective, the disparity in total mass decay rates across the Couette, Poiseuille, and combined flows stems directly from the differing near-wall velocity gradients. These gradients modulate the residence time of the solute within the highly reactive boundary layers. A higher near-wall velocity sweeps the solute past the reactive boundaries more rapidly, reducing local absorption opportunities, whereas a lower near-wall velocity increases the local absorption, thereby accelerating the global mass depletion quantified by $M_0(t)$.

\subsection{Validation of the axial variance}
\label{subsec:axial-variance-validation}

The validation is next extended to the streamwise spreading of the surviving solute, quantified by the axial variance \(\nu_2(t)\). As established in equation \eqref{eq:second-central-moment-dispersion}, \(\nu_2(t)\) represents the normalised second central moment of the cross-sectionally averaged concentration profile. The absolute discrepancy between the PINN and ADI predictions for this quantity is defined as
\begin{equation}
E_{\nu_2}(t)=\left|\nu_2^{\mathrm{PINN}}(t)-\nu_2^{\mathrm{ADI}}(t)\right|.
\label{eq:axial-variance-error}
\end{equation}

The quantity \(\nu_2(t)\) measures the squared streamwise width of the solute distribution about its instantaneous centroid. Unlike the total surviving mass \(M_0(t)\), which quantifies the amount of solute remaining within the channel, \(\nu_2(t)\) characterises the spatial spreading of the surviving solute. The normalisation by \(M_0(t)\) is essential in the reactive problem because the total mass changes continuously owing to wall absorption. Agreement in \(\nu_2(t)\) therefore tests whether the PINN reproduces the evolving plume width independently of the overall reduction in solute mass.

To isolate the influence of the velocity profile, two matched point-like-source cases are considered under the same asymmetric wall-reactivity configuration, $(\beta_1,\beta_2)=(2,0.2)$. The source geometry and wall-reaction parameters are therefore fixed, while the flow is varied between Couette and Couette--Poiseuille profiles.

\begin{figure}
\centering

\begin{subfigure}[t]{0.48\linewidth}
\centering
\includegraphics[width=\linewidth,height=5.8cm,keepaspectratio]{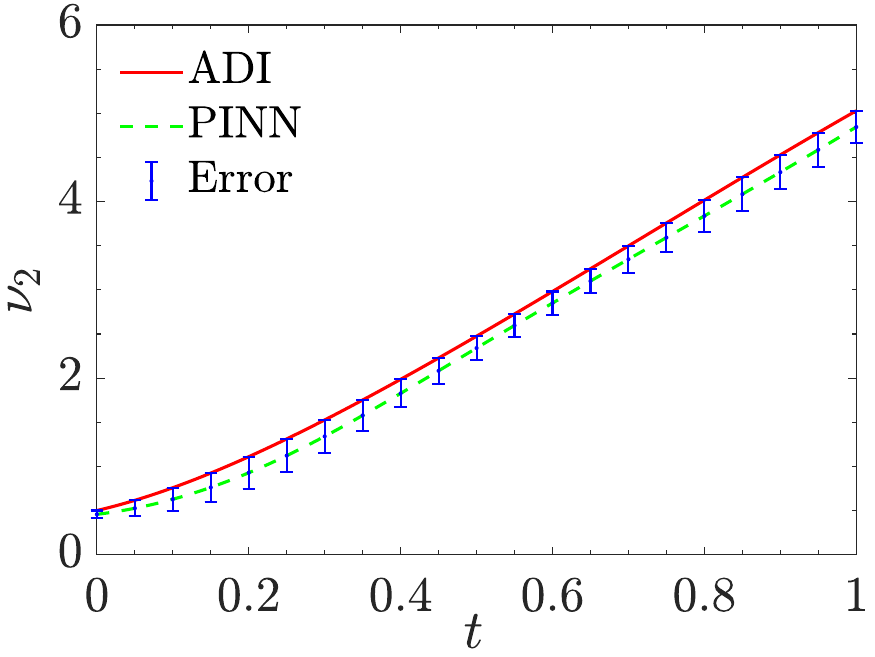}
\caption{Couette flow with $q=\mathrm{P}$ and $(\beta_1,\beta_2)=(2,0.2)$.}
\label{fig:nu2-s2-c-beta202}
\end{subfigure}
\hfill
\begin{subfigure}[t]{0.48\linewidth}
\centering
\includegraphics[width=\linewidth,height=5.8cm,keepaspectratio]{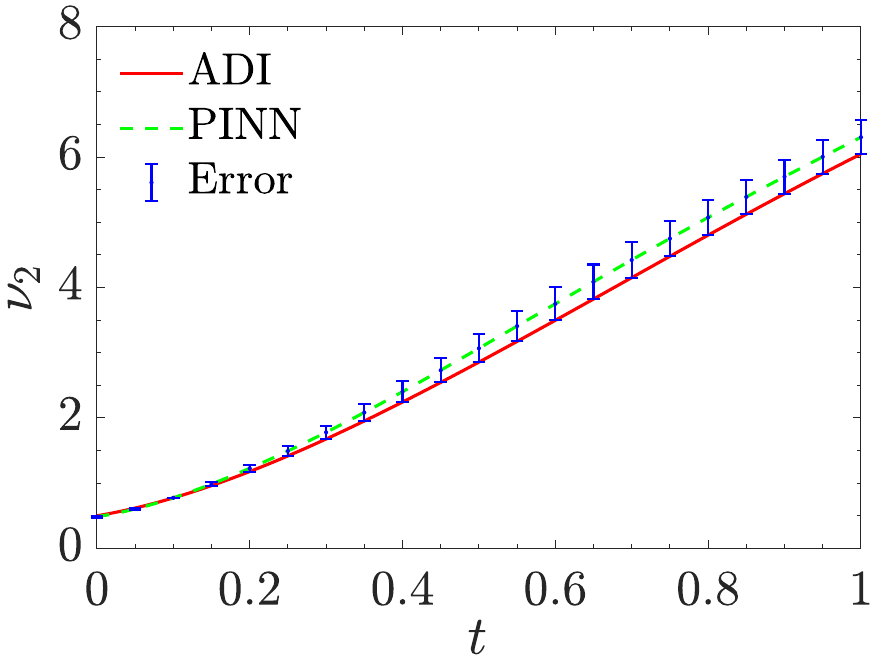}
\caption{Couette--Poiseuille flow with $q=\mathrm{P}$ and $(\beta_1,\beta_2)=(2,0.2)$.}
\label{fig:nu2-s2-cp-beta202}
\end{subfigure}

\caption{
Validation of the axial variance $\nu_2(t)$ for point-like releases under upper-wall-dominated reactivity, $(\beta_1,\beta_2)=(2,0.2)$. Panel $(a)$ corresponds to Couette flow, while panel $(b)$ corresponds to Couette--Poiseuille flow. In each case, $\nu_2^{\mathrm{PINN}}(t)$ is compared with $\nu_2^{\mathrm{ADI}}(t)$, together with the corresponding absolute discrepancy $E_{\nu_2}(t)$.
}
\label{fig:axial-variance-validation}
\end{figure}

Figure~\ref{fig:axial-variance-validation} compares the PINN and ADI predictions of $\nu_2(t)$ for two distinct shear structures under identical source and reaction conditions. In Couette flow, the linear velocity gradient continuously stretches the solute distribution in the streamwise direction. In Couette--Poiseuille flow, the combined linear and parabolic contributions produce a different distribution of streamwise velocities across the channel and therefore a different evolution of the axial plume width.

The asymmetric wall-reactivity pair $(\beta_1,\beta_2)=(2,0.2)$ introduces preferential absorption at one boundary. This selective removal alters the wall-normal composition of the surviving concentration field and, through its interaction with the local streamwise velocity, can modify the axial variance. The matched comparison is therefore more stringent than a comparison involving different reaction parameters because differences between the two panels arise primarily from changes in the imposed velocity profile.

The agreement between $\nu_2^{\mathrm{PINN}}(t)$ and $\nu_2^{\mathrm{ADI}}(t)$ confirms that the PINN reproduces the evolving streamwise width of the surviving solute for both Couette and Couette--Poiseuille flows. The corresponding small values of $E_{\nu_2}(t)$ further support the accuracy of the learned concentration field at the level of the normalised second central moment. Additional axial-variance comparisons for Couette flow with $(\beta_1,\beta_2)=(0.2,2)$ and Couette--Poiseuille flow with $(\beta_1,\beta_2)=(1,1)$ are reported in the supplementary material.


\subsection{Validation of cumulative wall-removal dynamics}
\label{subsec:cumulative-uptake-validation}

\begin{figure}
    \centering

    \begin{subfigure}[t]{0.98\linewidth}
        \centering
        \includegraphics[width=\linewidth]{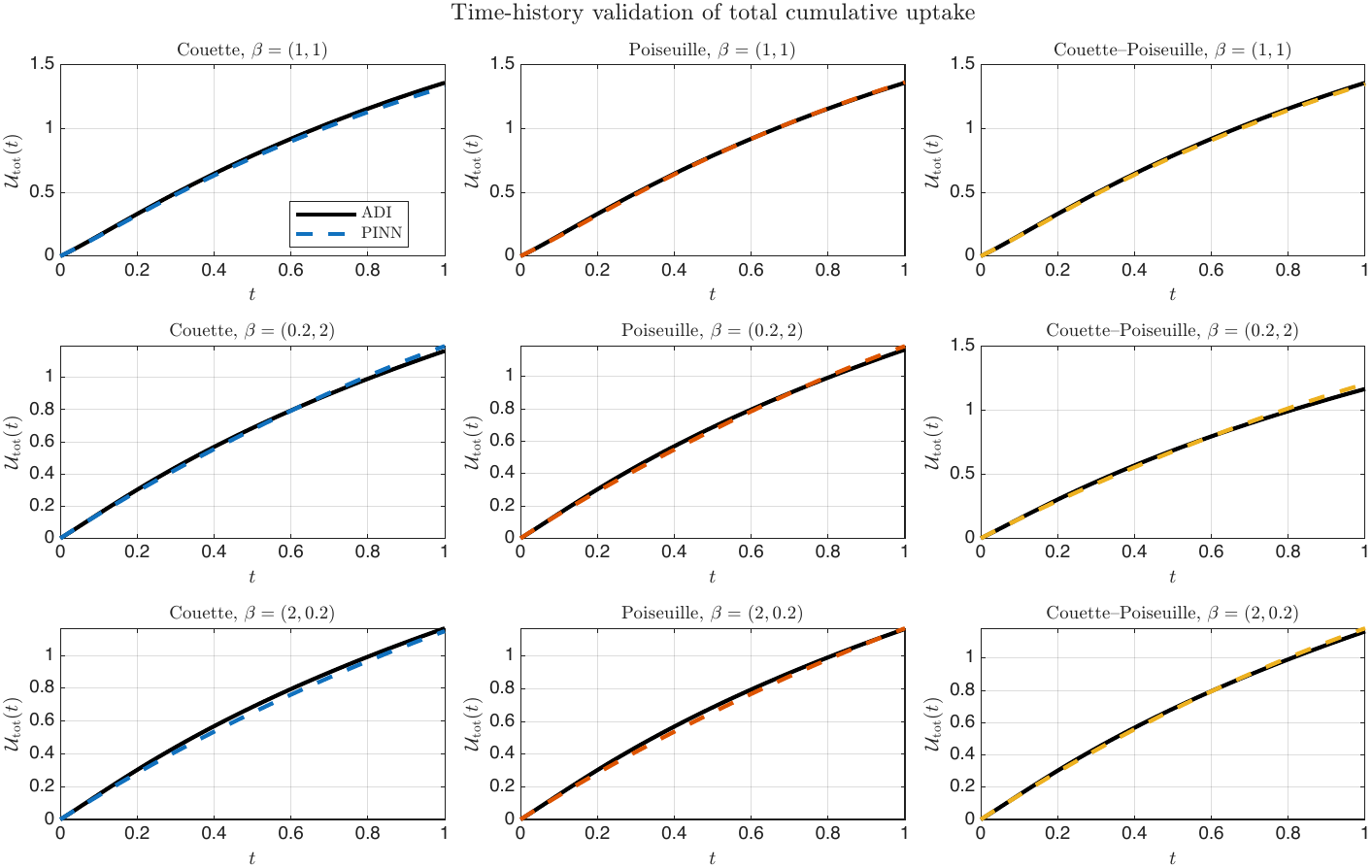}
        \caption{Time-history comparison of \(\mathcal{U}_{\mathrm{tot}}(t)\).}
        \label{fig:utot-time-validation}
    \end{subfigure}

    \vspace{0.25cm}

    \begin{subfigure}[t]{0.48\linewidth}
        \centering
        \includegraphics[width=\linewidth]{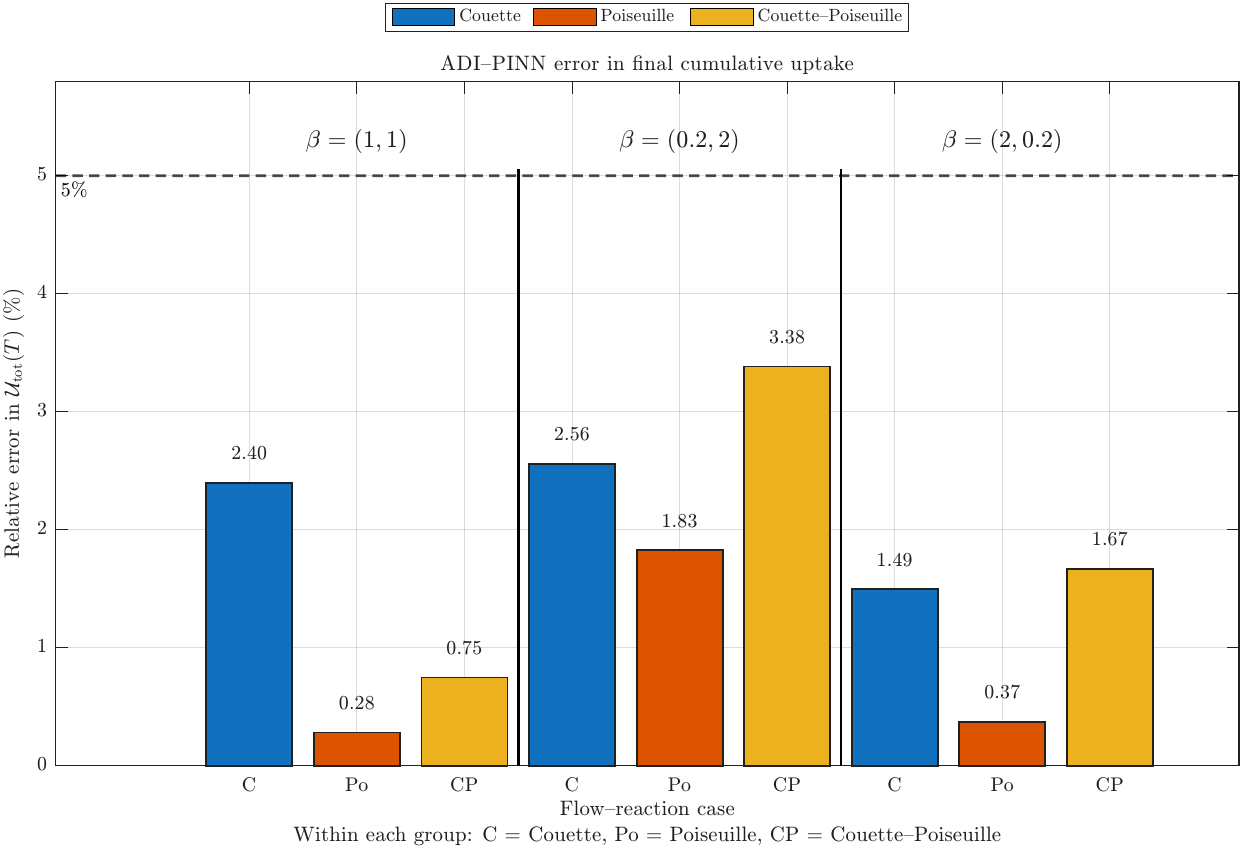}
        \caption{Relative error in \(\mathcal{U}_{\mathrm{tot}}(T)\).}
        \label{fig:utot-relative-error}
    \end{subfigure}
    \hfill
    \begin{subfigure}[t]{0.48\linewidth}
        \centering
        \includegraphics[width=\linewidth]{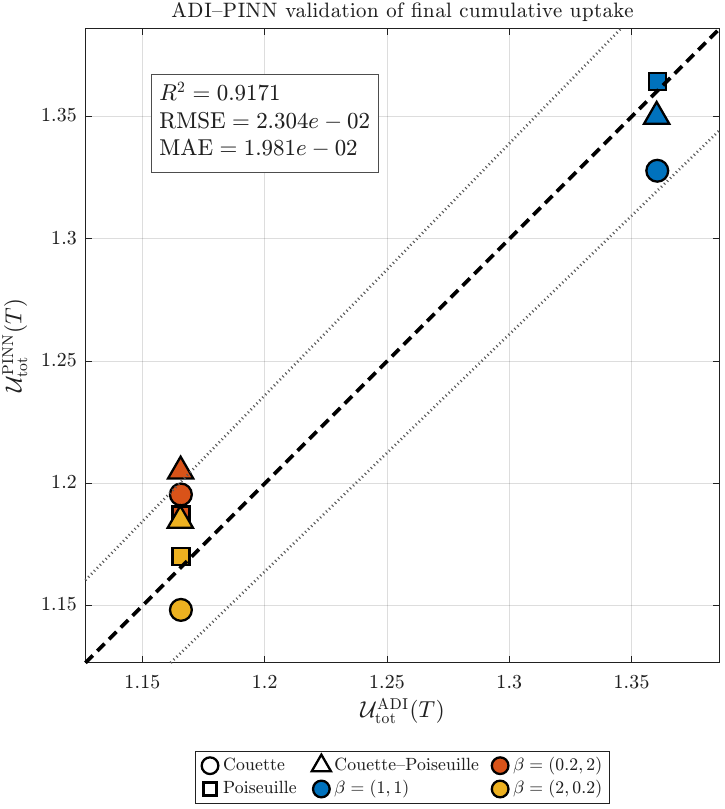}
        \caption{Scatter validation of \(\mathcal{U}_{\mathrm{tot}}(T)\).}
        \label{fig:utot-scatter-validation}
    \end{subfigure}

    \caption{
    ADI--PINN validation of cumulative wall uptake for the point-like Gaussian source \(C_0^{(\mathrm{P})}(x,y)=\exp(-x^2-y^2)\). 
    In panel (a), black solid curves denote the ADI finite-difference benchmark, while dashed curves denote the PINN predictions. 
    Panel (b) shows the final-time relative error \(E_{\mathcal{U}}\), with the dashed horizontal line denoting a \(5\%\) reference level. 
    Panel (c) compares \(\mathcal{U}_{\mathrm{tot}}^{\mathrm{PINN}}(T)\) with \(\mathcal{U}_{\mathrm{tot}}^{\mathrm{ADI}}(T)\); the dashed line denotes perfect agreement, and the dotted lines denote \(\pm 3\%\) agreement bands. Marker shapes denote the shear profiles, while colours denote the reactive-wall configurations. The reported \(R^2\), RMSE and MAE quantify the final-time agreement between the PINN and ADI predictions.
    }
    \label{fig:adi-pinn-validation}
\end{figure}

We next validate the reactive-wall diagnostics for the point-like Gaussian source \eqref{eq:ic-point-source}. This source is localised in both the streamwise and wall-normal directions. It therefore provides a stringent test of wall-removal dynamics, because the solute must first spread via transverse diffusion and shear-driven transport before being absorbed by the reactive walls. The validation is performed for the three reactive-wall configurations
\[
(\beta_1,\beta_2)=(1,1),\qquad
(\beta_1,\beta_2)=(0.2,2),\qquad
(\beta_1,\beta_2)=(2,0.2),
\]
and for the three mean-centred shear flows: Couette, Poiseuille and Couette--Poiseuille flow. Thus, the validation set contains nine flow--reaction cases.

This validation provides a more stringent test than concentration-field comparisons alone, as the total cumulative uptake relies heavily on the temporal accumulation of the boundary concentrations. Agreement in the total cumulative wall uptake, \(\mathcal{U}_{\mathrm{tot}}(t)\) (as defined in equation \eqref{eq:pinn-total-cumulative-uptake}), requires the PINN to accurately predict not only the advective-diffusive evolution of the interior plume but also the localised wall concentration history governing reactive removal at both boundaries.

In this validation subsection only, superscripts \(\mathrm{PINN}\) and \(\mathrm{ADI}\) are used to distinguish quantities computed from the trained neural-network solution and from the finite-difference benchmark. Figure~\ref{fig:adi-pinn-validation} demonstrates that the PINN accurately reproduces the ADI-predicted cumulative wall uptake. The time histories in figure~\ref{fig:adi-pinn-validation}(a) are in close agreement for all selected shear-flow and reactive-wall configurations, indicating that the trained network captures the temporal accumulation of wall removal. The relative-error plot in figure~\ref{fig:adi-pinn-validation}(b) shows that the final-time error in \(\mathcal{U}_{\mathrm{tot}}(T)\) remains below \(5\%\) for all cases, with a maximum error of approximately \(3.38\%\). The scatter comparison in figure~\ref{fig:adi-pinn-validation}(c) provides a compact final-time validation: each marker corresponds to one flow--reaction case, with the horizontal coordinate representing \(\mathcal{U}_{\mathrm{tot}}^{\mathrm{ADI}}(T)\) and the vertical coordinate representing \(\mathcal{U}_{\mathrm{tot}}^{\mathrm{PINN}}(T)\). The dashed diagonal line denotes perfect agreement, while the dotted lines indicate \(\pm 3\%\) deviations from this line. The reported \(R^2\) measures the linear agreement between the ADI and PINN final-time uptake values, whereas the root-mean-square error (RMSE) and mean absolute error (MAE) quantify the absolute discrepancy in \(\mathcal{U}_{\mathrm{tot}}(T)\). Together, these results confirm that the trained PINN reproduces both the time-dependent and final-time cumulative wall-removal dynamics with sufficient accuracy for the subsequent reactive-dispersion analysis.

\subsection{Reactive dispersion under wall-resolved absorption}
\label{subsec:reactive-dispersion-wall-resolved-absorption}

\begin{figure}
    \centering

    \begin{subfigure}[t]{0.98\linewidth}
        \centering
        \includegraphics[width=\linewidth]{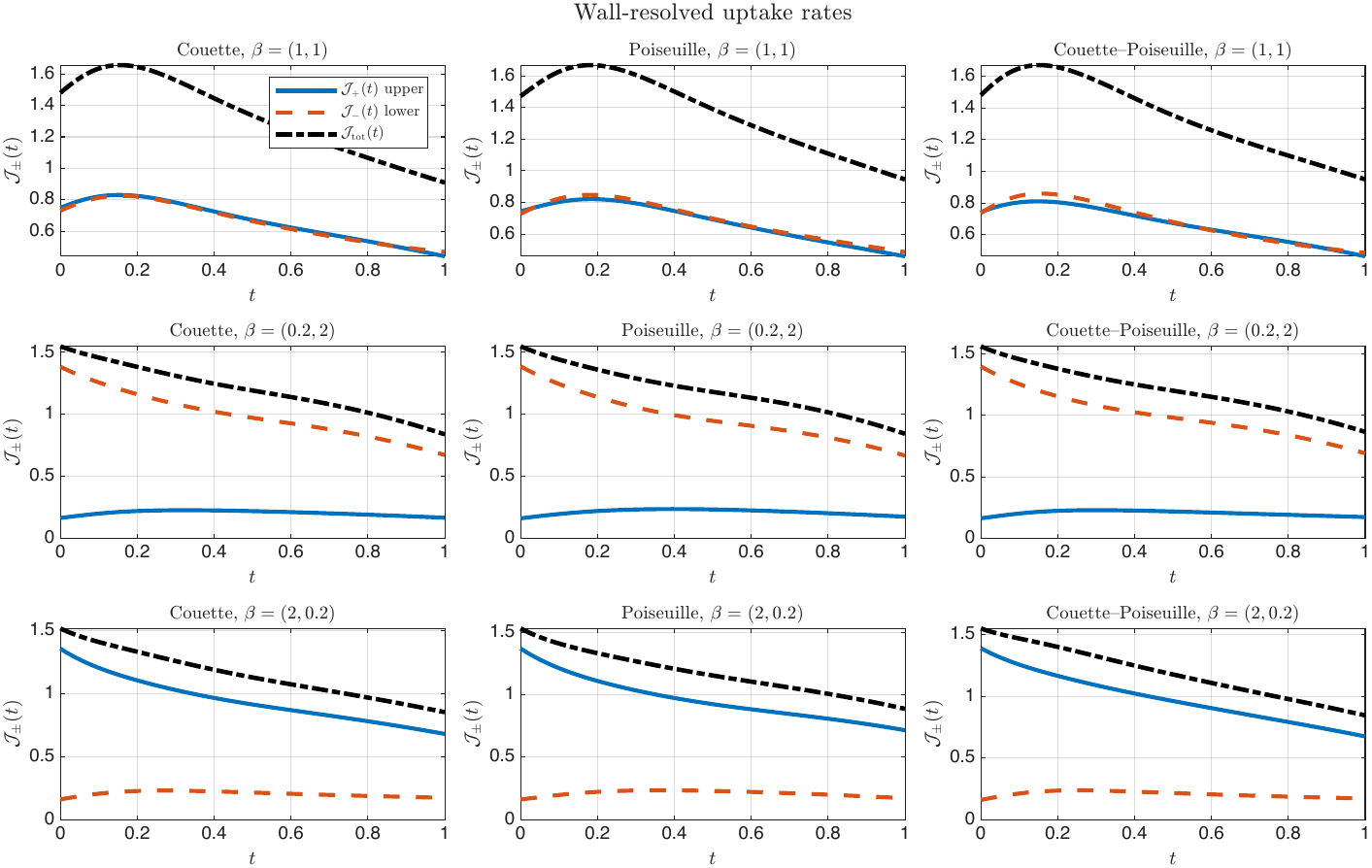}
        \caption{Wall-resolved uptake rates.}
        \label{fig:wall-uptake-rates}
    \end{subfigure}

    \vspace{0.25cm}

    \begin{subfigure}[t]{0.98\linewidth}
        \centering
        \includegraphics[width=\linewidth]{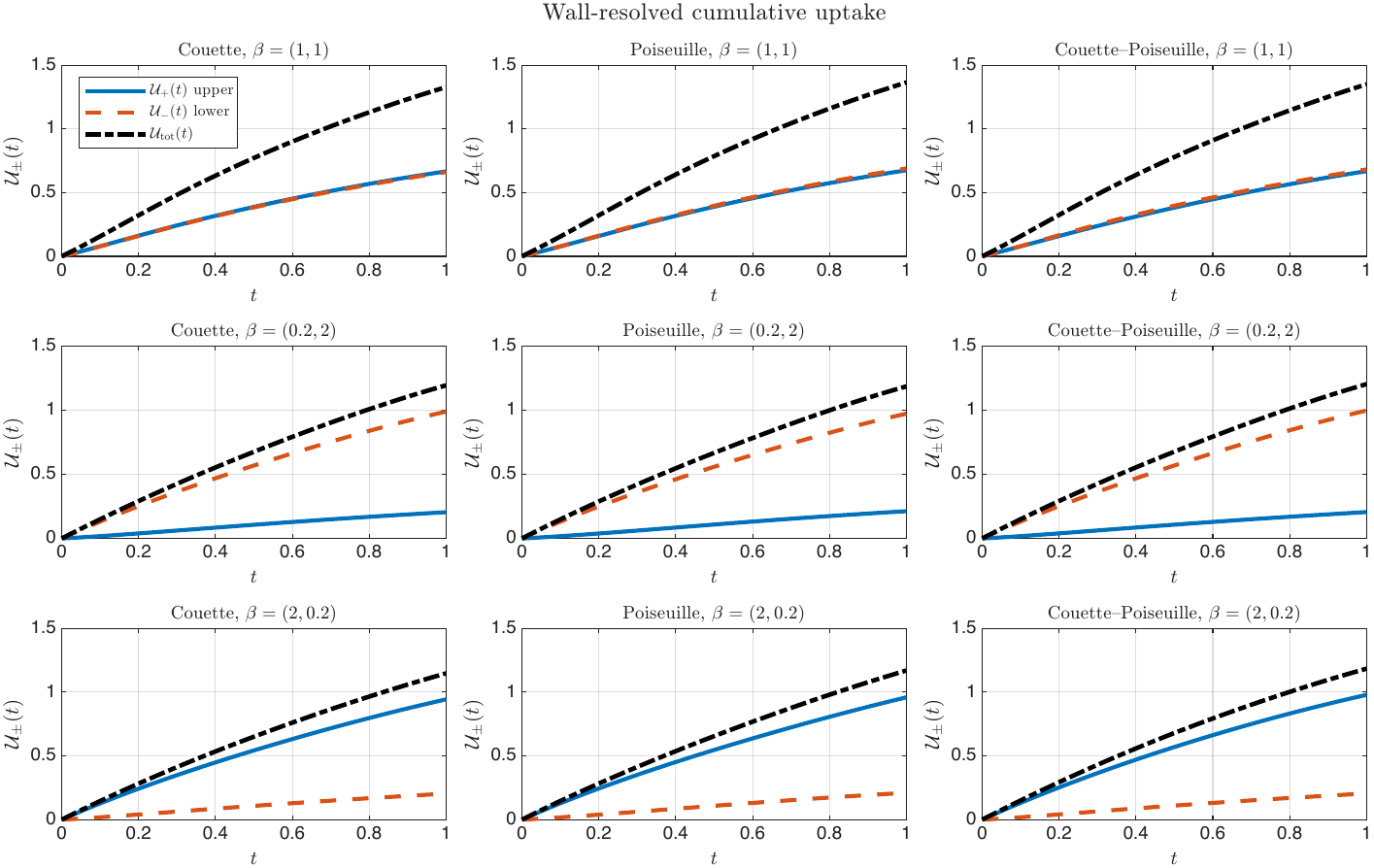}
        \caption{Wall-resolved cumulative uptake.}
        \label{fig:wall-cumulative-uptake}
    \end{subfigure}

    \caption{
    Wall-resolved reactive removal extracted from the trained PINN solution for the point-like Gaussian source. 
    Panel (a) shows the instantaneous integrated uptake rates \(\mathcal{J}_{+}(t)\), \(\mathcal{J}_{-}(t)\), and \(\mathcal{J}_{\mathrm{tot}}(t)\). 
    Panel (b) shows the corresponding cumulative uptakes \(\mathcal{U}_{+}(t)\), \(\mathcal{U}_{-}(t)\), and \(\mathcal{U}_{\mathrm{tot}}(t)\). 
    These quantities show how shear-driven dispersion and wall-dependent reactivity combine to determine total and wall-resolved solute removal.
    }
    \label{fig:wall-resolved-dynamics}
\end{figure}

\begin{figure}
    \centering
    \includegraphics[width=0.9\linewidth]{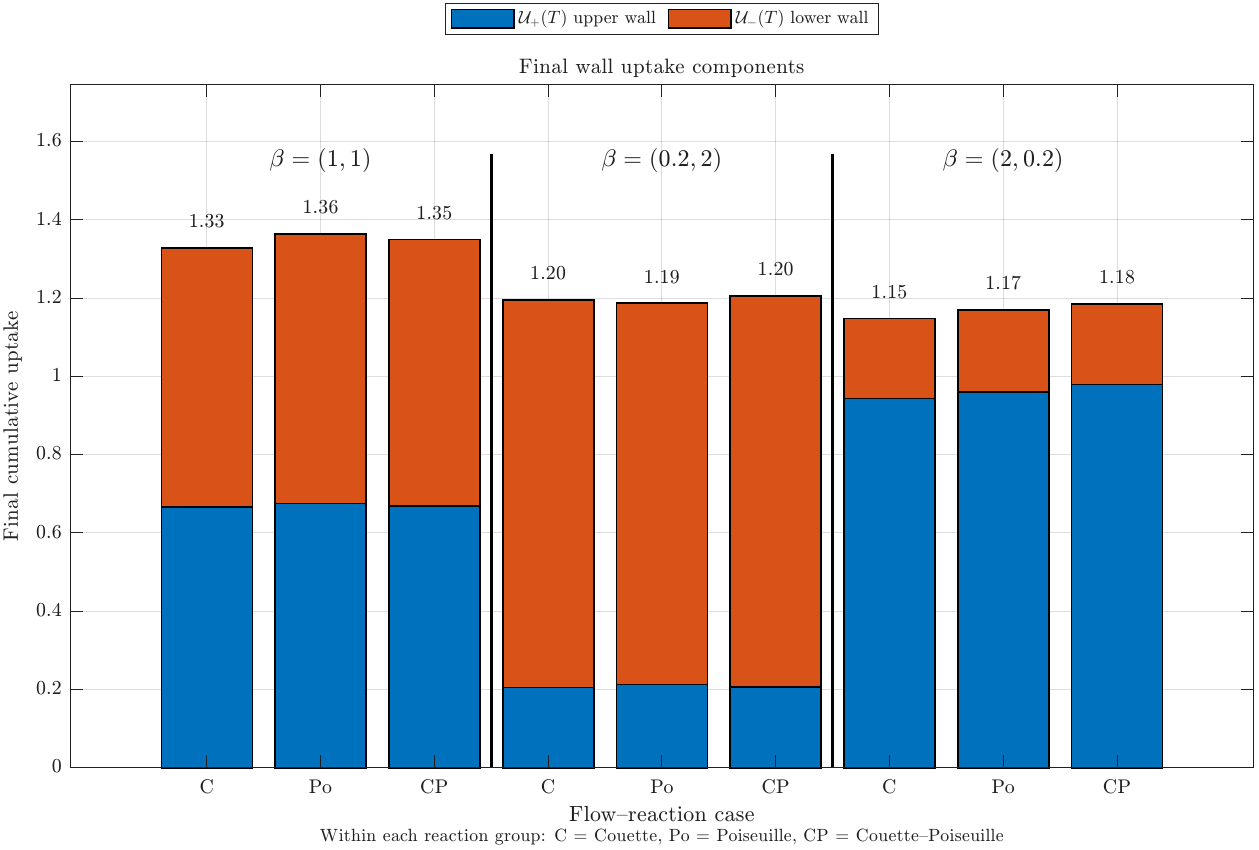}
    \caption{
    Final wall uptake components extracted from the trained PINN solution for the point-like Gaussian source. 
    The stacked bars show \(\mathcal{U}_{+}(T)\) and \(\mathcal{U}_{-}(T)\), so that the total bar height gives \(\mathcal{U}_{\mathrm{tot}}(T)\). 
    The plot provides a compact final-time summary of the wall-resolved removal contributions across the selected shear-flow and reactive-wall configurations.
    }
    \label{fig:final-wall-uptake-bar}
\end{figure}

Having validated the cumulative wall-removal dynamics, we now analyse how the removed solute is partitioned between the two reactive walls. Unless otherwise stated, the wall-uptake diagnostics in the remainder of this section are evaluated from the PINN-predicted concentration field for the point-like Gaussian source.

The total cumulative uptake \(\mathcal{U}_{\mathrm{tot}}(t)\) measures the combined removal by both walls. However, this total quantity alone does not indicate whether the removal is balanced between the upper and lower walls or dominated by one of them. We therefore examine the wall-resolved uptake rates \(\mathcal{J}_{+}(t)\), \(\mathcal{J}_{-}(t)\) and the corresponding cumulative uptakes \(\mathcal{U}_{+}(t)\), \(\mathcal{U}_{-}(t)\). This separation is important because symmetric and asymmetric wall reactions can produce different upper--lower uptake balances even when their total removal is comparable.

Figure~\ref{fig:wall-resolved-dynamics} shows three physically distinct reactive-dispersion regimes. For the symmetric case \((\beta_1,\beta_2)=(1,1)\), the upper- and lower-wall contributions remain nearly balanced. This is expected because both the wall reactivity and the initial point-like source are symmetric about the channel centreline. This case, therefore, acts as a physical consistency check: the learned solution should not introduce artificial wall preference when the imposed problem is symmetric.

For \((\beta_1,\beta_2)=(0.2,2)\), the lower wall is ten times more reactive than the upper wall. The lower-wall uptake rate \(\mathcal{J}_{-}(t)\) and cumulative uptake \(\mathcal{U}_{-}(t)\) therefore dominate the removal. When the wall reactivities are reversed to \((\beta_1,\beta_2)=(2,0.2)\), the dominant wall-removal pathway is also reversed. These two cases demonstrate reaction-induced breaking of transverse symmetry: even though the source is initially centred, the unequal wall reactions bias the net removal toward the more reactive boundary.

The uptake curves also show that reactive dispersion is not governed by the wall coefficients alone. The wall flux is the product of a reaction coefficient and a wall concentration. Hence, the wall with the larger \(\beta_i\) tends to dominate, but the instantaneous magnitude and timing of uptake depend on how advection and diffusion deliver solute to that wall. The three shear profiles, therefore, create different hydrodynamic pathways for wall contact. Couette flow introduces antisymmetric linear shear, Poiseuille flow introduces a symmetric parabolic shear, and Couette--Poiseuille flow combines both effects. These differences affect the streamwise stretching of the plume and the distribution of near-wall concentration, which then influence \(\mathcal{J}_{\pm}(t)\) and \(\mathcal{U}_{\pm}(t)\).

The final-time decomposition in figure~\ref{fig:final-wall-uptake-bar} condenses the same information into a compact wall-partition summary. The symmetric case gives comparable upper- and lower-wall contributions, while the asymmetric cases show clear dominance of the wall with the larger uptake coefficient. The total bar height also varies with the imposed shear flow, showing that the total removal is controlled jointly by wall reactivity and shear-driven solute redistribution.

\subsection{Wall selectivity and reaction-induced transverse asymmetry}
\label{subsec:wall-selectivity-asymmetry}

\begin{figure}
    \centering

    \begin{subfigure}[t]{0.44\linewidth}
        \centering
        \includegraphics[width=\linewidth]{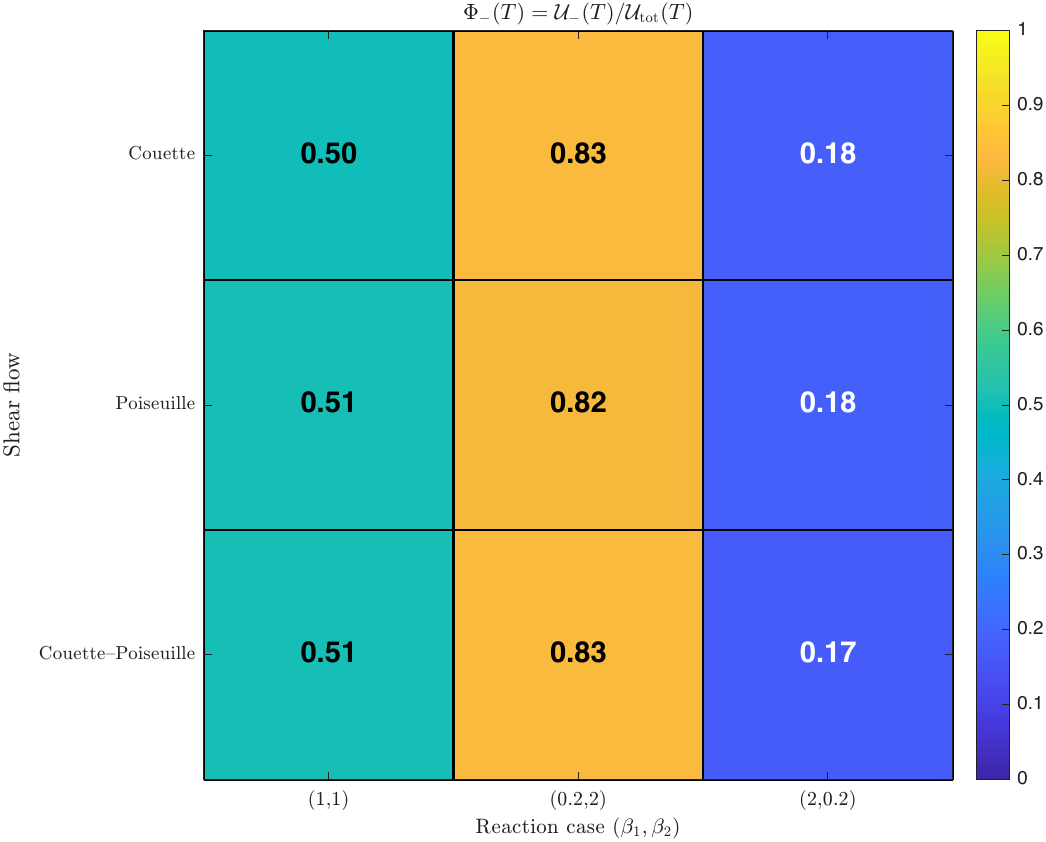}
        \caption{Final lower-wall uptake fraction.}
        \label{fig:lower-wall-fraction}
    \end{subfigure}
    \hfill
    \begin{subfigure}[t]{0.55\linewidth}
        \centering
        \includegraphics[width=\linewidth]{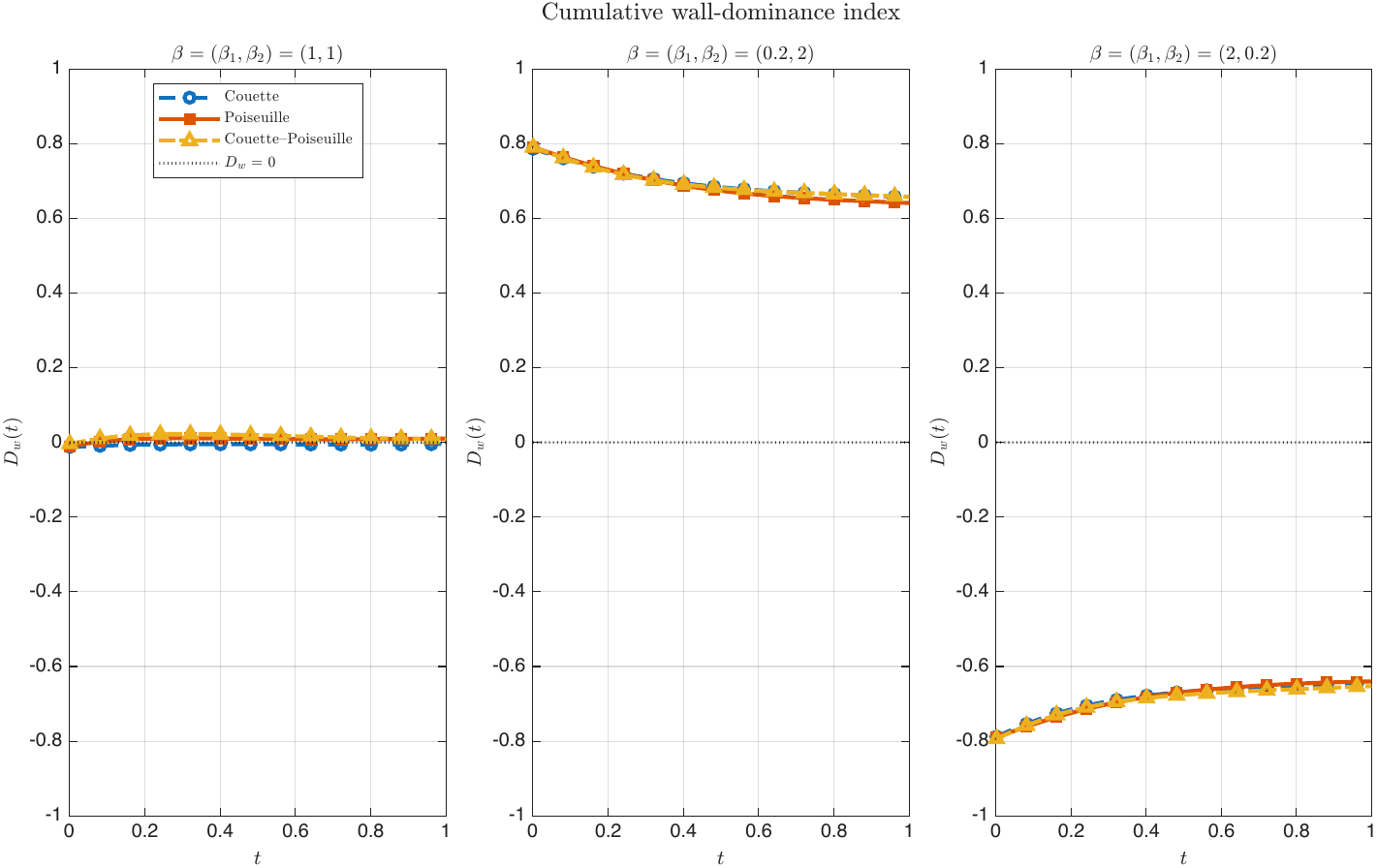}
        \caption{Cumulative wall-dominance index.}
        \label{fig:wall-dominance-index}
    \end{subfigure}

    \caption{
    Wall-selective reactive-dispersion diagnostics extracted from the trained PINN solution for the point-like Gaussian source. 
    Panel (a) shows the final lower-wall uptake fraction 
    \(\Phi_-(T)=\mathcal{U}_-(T)/\mathcal{U}_{\mathrm{tot}}(T)\). 
    Panel (b) shows the cumulative wall-dominance index 
    \(D_w(t)=[\mathcal{U}_-(t)-\mathcal{U}_+(t)]/\mathcal{U}_{\mathrm{tot}}(t)\). 
    Positive values indicate lower-wall-dominated uptake, negative values indicate upper-wall-dominated uptake, and values near zero indicate balanced wall removal.
    }
    \label{fig:wall-selective-diagnostics}
\end{figure}

The wall-resolved uptake curves quantify absolute removal by each boundary. To isolate the relative partition of removal, we use the normalised diagnostics
\[
\Phi_-(T)
=
\frac{\mathcal{U}_-(T)}
{\mathcal{U}_{\mathrm{tot}}(T)}
\]
and
\[
D_w(t)
=
\frac{\mathcal{U}_-(t)-\mathcal{U}_+(t)}
{\mathcal{U}_{\mathrm{tot}}(t)}.
\]
These quantities separate wall preference from the overall magnitude of uptake. Thus, they are especially useful for identifying reaction-induced transverse asymmetry.

Figure~\ref{fig:wall-selective-diagnostics}(a) gives a compact final-time representation of wall selectivity. For \((\beta_1,\beta_2)=(1,1)\), the values remain close to \(\Phi_-(T)=1/2\), indicating balanced removal. For \((\beta_1,\beta_2)=(0.2,2)\), \(\Phi_-(T)>1/2\), confirming lower-wall-dominated uptake. For \((\beta_1,\beta_2)=(2,0.2)\), \(\Phi_-(T)<1/2\), indicating upper-wall-dominated uptake. The dynamic index in figure~\ref{fig:wall-selective-diagnostics}(b) gives the corresponding time-dependent picture: \(D_w(t)\) remains near zero for symmetric reactivity, becomes positive when the lower wall is more reactive, and becomes negative when the upper wall is more reactive.

This result highlights an important feature of reactive dispersion: the total cumulative uptake measures how much solute is removed, but it fails to capture the spatial distortion of the plume. The diagnostics \(\Phi_-(T)\) and \(D_w(t)\) reveal that unequal wall reactivities (e.g., $\beta_1 \neq \beta_2$) break the transverse symmetry of the concentration field. This wall-selective uptake effectively creates a depleted concentration boundary layer near the highly reactive wall, steepening the local transverse concentration gradient and driving a continuous, asymmetric diffusive flux from the bulk. Capturing this transverse non-uniformity is essential, as classical one-dimensional models inherently average out the cross-sectional variations that dictate the slow-decaying transient effects in highly asymmetric reactive environments \cite{Dhar2021, Jiang2022, Poddar2024a}.

\subsection{Shear-dependent streamwise organisation of reactive uptake}
\label{subsec:shear-dependent-uptake-organization}

\begin{figure}
    \centering
    \includegraphics[width=\linewidth]{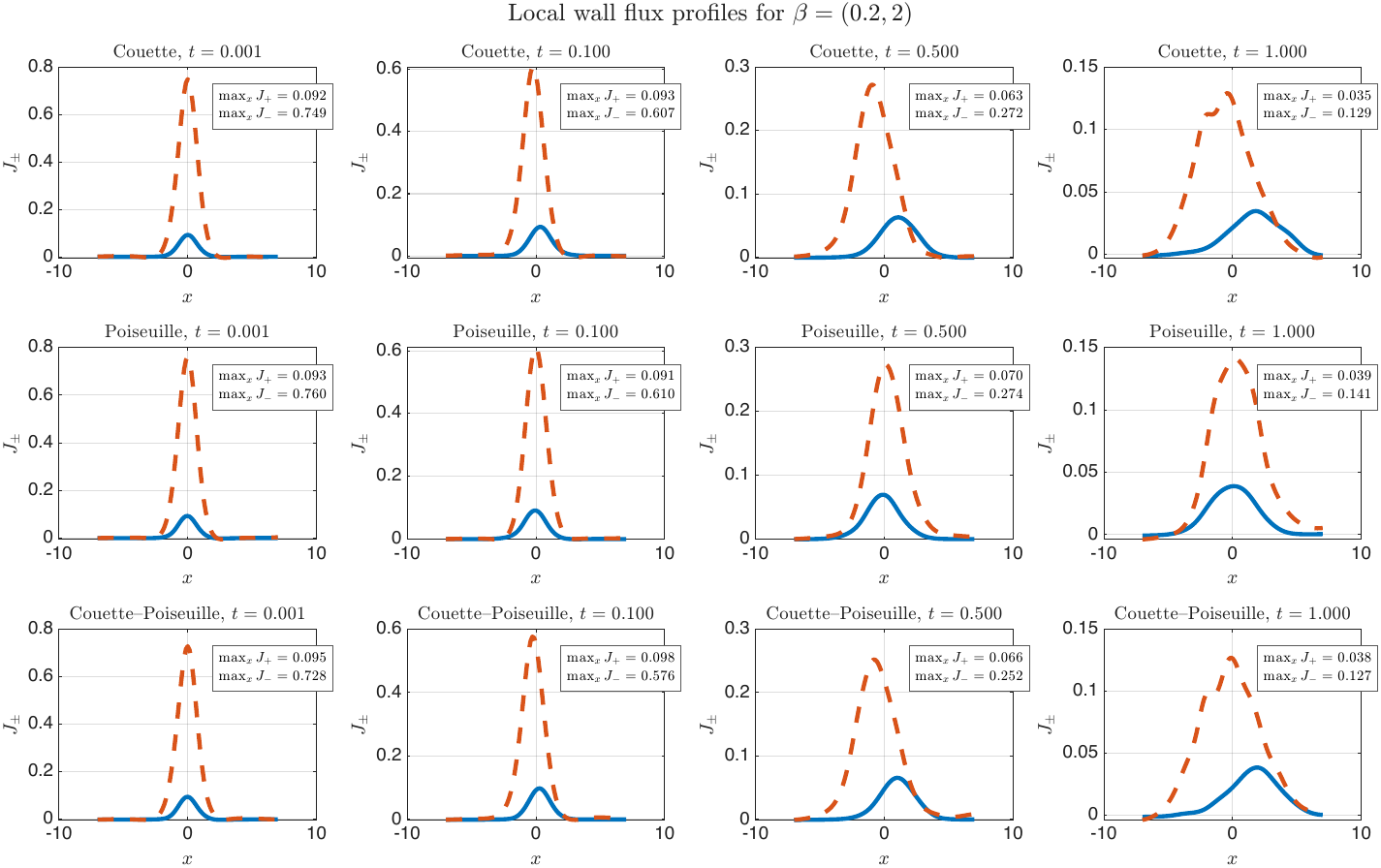}
    \caption{
    Local wall uptake flux profiles for the point-like Gaussian source in the asymmetric reactive case \((\beta_1,\beta_2)=(0.2,2)\). 
    Blue solid curves denote \(J_+(x,t)\), while red dashed curves denote \(J_-(x,t)\). 
    The annotation boxes report \(\max_x J_+(x,t)\) and \(\max_x J_-(x,t)\), highlighting the stronger lower-wall removal associated with the larger lower-wall reactivity.
    }
    \label{fig:local-wall-flux-profiles}
\end{figure}

\begin{figure}
    \centering
    \includegraphics[width=0.92\linewidth]{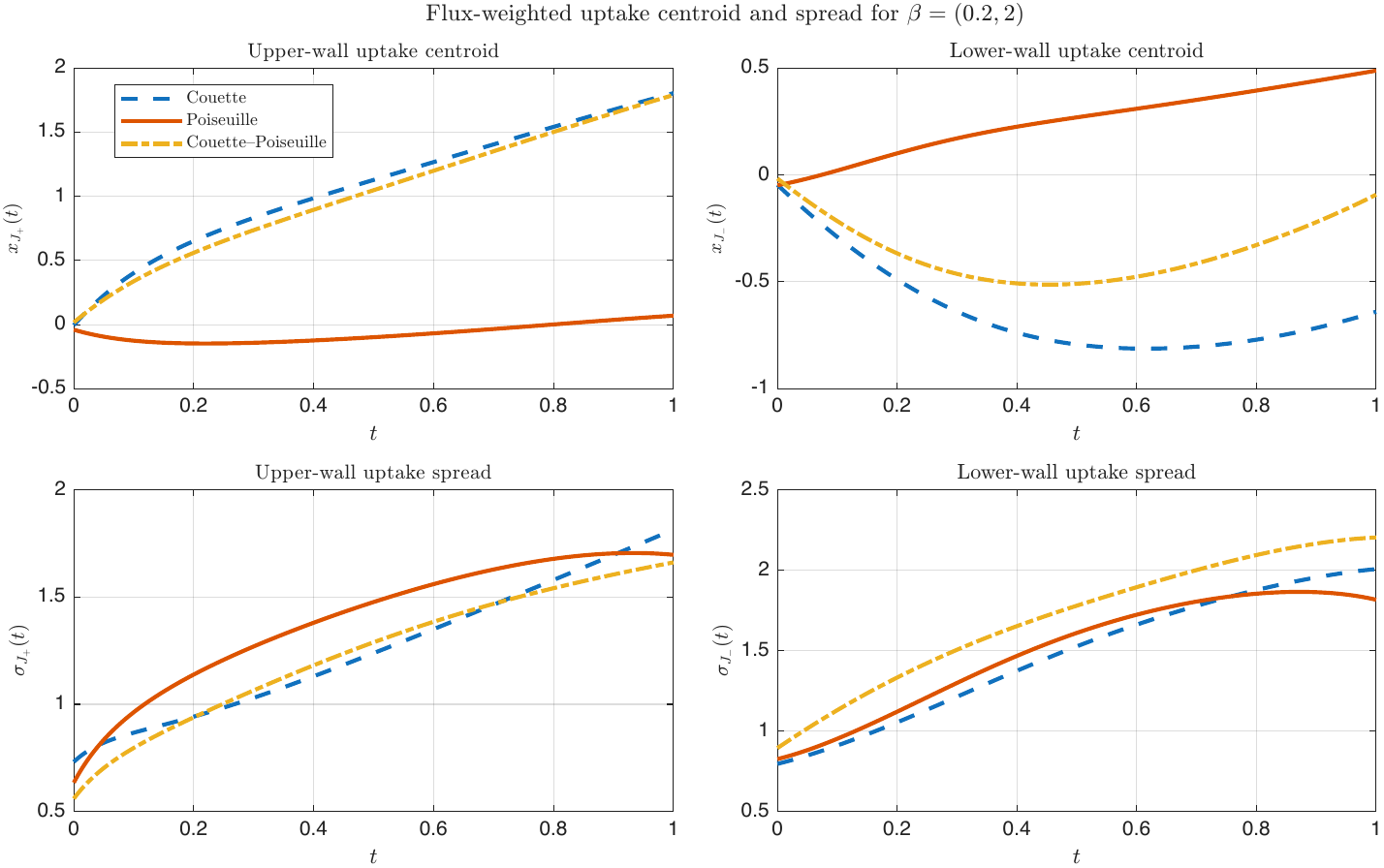}
    \caption{
    Flux-weighted uptake centroid and spread for the point-like Gaussian source in the asymmetric reactive case \((\beta_1,\beta_2)=(0.2,2)\). 
    The centroids \(x_{J_\pm}(t)\) identify the streamwise locations where the upper- and lower-wall uptake profiles are concentrated, while \(\sigma_{J_\pm}(t)\) measures their spatial spread. 
    These diagnostics reveal flow-dependent spatial organisation of wall removal even when the total cumulative uptake is similar across the mean-centred shear profiles.
    }
    \label{fig:uptake-centroid-spread}
\end{figure}

The preceding diagnostics quantify the amount and wall-partitioned removal. We now examine where along the channel this uptake occurs. This is a distinct physical question, because two flows may remove a similar total amount of solute but concentrate the removal in different streamwise regions.

Figure~\ref{fig:local-wall-flux-profiles} shows the local wall-flux profiles for the asymmetric reactive case \((\beta_1,\beta_2)=(0.2,2)\). Since the lower wall is ten times more reactive than the upper wall, the lower-wall flux \(J_-(x,t)\) is larger than the upper-wall flux \(J_+(x,t)\). However, the streamwise position and width of these flux profiles depend on the imposed shear flow, because the velocity field controls how the solute cloud is stretched and delivered to the wall.

The centroid and spread diagnostics in figure~\ref{fig:uptake-centroid-spread} quantify this spatial organisation. The centroid \(x_{J_\pm}(t)\) gives the effective streamwise location where wall uptake is concentrated, while \(\sigma_{J_\pm}(t)\) measures the streamwise width of the uptake region. In Couette flow, the mean-centred velocity is antisymmetric, so solute near the upper and lower walls is transported in opposite streamwise directions. This produces a separation between the effective uptake locations at the two walls. In Poiseuille flow, the mean-centred velocity is symmetric about the centre line and has the same value at both walls, so the wall-flux distributions remain more symmetrically organised. In Couette--Poiseuille flow, the linear and parabolic components combine to produce a stronger imbalance between the two wall-adjacent velocities, leading to a more pronounced displacement of the uptake centroid, especially at the more reactive lower wall.

Thus, the reaction coefficient determines how strongly a wall absorbs solute once the solute reaches it, while the shear profile determines where along the channel the solute is delivered to the wall. Figure~\ref{fig:uptake-centroid-spread} therefore demonstrates that wall-reactive dispersion must be characterised not only by the total cumulative uptake \(\mathcal{U}_{\mathrm{tot}}(t)\), but also by the streamwise location and spread of the wall-removal region.

\subsection{Coupled concentration--shear--reaction structure}
\label{subsec:coupled-concentration-shear-reaction}

\begin{figure}
    \centering
    \includegraphics[width=\linewidth]{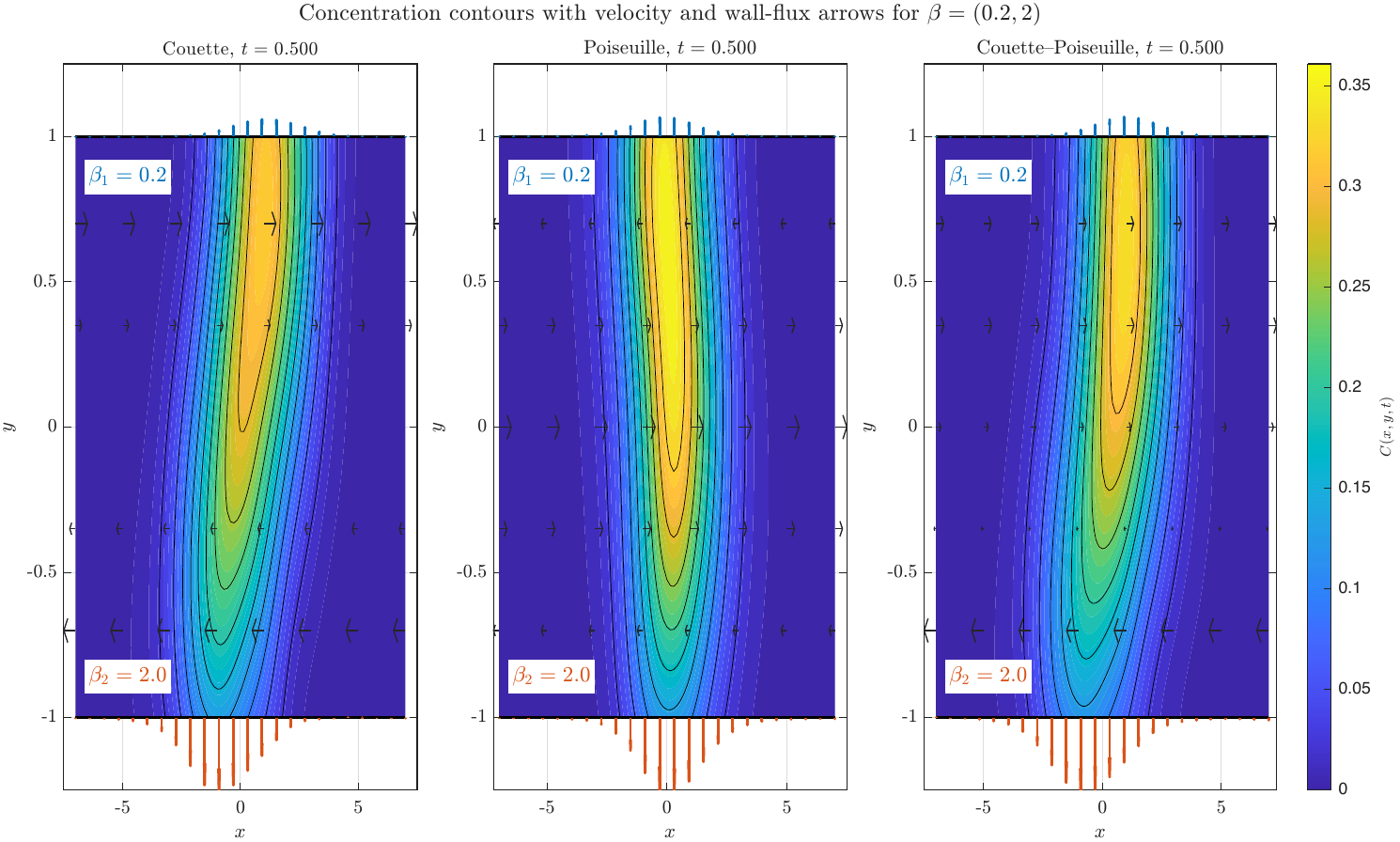}
    \caption{
    Concentration contours with superposed mean-centred velocity arrows and wall-flux arrows for the point-like Gaussian source in the asymmetric reactive case \((\beta_1,\beta_2)=(0.2,2)\). 
    Black arrows inside the channel indicate the imposed shear velocity \(u^{(s)}(y)\). 
    Blue and red/orange wall-normal arrows indicate the local upper- and lower-wall uptake fluxes \(J_+(x,t)\) and \(J_-(x,t)\), respectively. 
    The stronger lower-wall arrows visualise the enhanced lower-wall uptake caused by \(\beta_2>\beta_1\).
    }
    \label{fig:contour-velocity-wall-flux-arrows}
\end{figure}

We connect the wall-flux diagnostics to the underlying concentration field and imposed shear profile. Figure~\ref{fig:contour-velocity-wall-flux-arrows} shows the PINN-predicted concentration contours together with mean-centred velocity arrows and wall-normal uptake arrows for the point-like Gaussian source with \((\beta_1,\beta_2)=(0.2,2)\). This representation gives a direct physical picture of the coupled concentration--shear--reaction mechanism.

The mean-centred shear profile redistributes the solute cloud in the streamwise direction. Transverse diffusion transports solute from the channel interior toward the reactive boundaries. Once the solute reaches the wall, the local uptake is determined by the product of wall concentration and wall reactivity. Hence, the wall-flux arrows are stronger at the lower wall in the asymmetric case because \(\beta_2>\beta_1\). The figure visually confirms the same mechanism quantified by the uptake curves, wall-dominance index, local flux profiles and centroid/spread diagnostics.

This combined view emphasises the central physical contribution of the present study. The PINN is not used only as a mesh-free approximation of \(C(x,y,t)\). Its differentiable representation enables systematic extraction of wall-resolved quantities from the learned solution. These diagnostics show that reactive dispersion in shear flows is governed by three coupled mechanisms: hydrodynamic redistribution of the solute cloud, transverse diffusive delivery to the walls, and wall-dependent absorption. The resulting wall removal is therefore characterised not only by the total amount removed, but also by the dominant wall, the time at which dominance develops, and the streamwise location and spread of the uptake region.

\section{Conclusions}
\label{sec:conclusions}

In this study, a physics-informed neural network (PINN) framework is developed to investigate the two-dimensional dispersion of reactive solutes in canonical shear flows bounded by first-order absorbing walls. By embedding the dimensionless convection--diffusion equation, localised initial source distributions, and reactive Robin boundary conditions into a unified composite loss function, the PINN provides a mesh-free framework for imposing the governing equation and boundary constraints within a single optimisation problem.

The physical fidelity of the mesh-free framework was rigorously validated against an alternating-direction implicit (ADI) finite-difference benchmark. The network accurately captured the spatiotemporal evolution of the two-dimensional concentration fields, the cross-sectionally averaged profiles, and the total surviving solute mass across Couette, Poiseuille, and combined Couette--Poiseuille flows. More critically, the continuous and fully differentiable nature of the trained PINN surrogate enabled the precise extraction of derived integral transport characteristics, such as the time-dependent effective dispersion coefficient and the axial variance, without introducing numerical truncation errors typical of discrete methods.

Beyond validating the concentration fields, this study utilised the differentiable network to uncover the detailed mechanics of wall-resolved reactive dispersion. The analysis demonstrated that unequal wall reactivities (e.g., $\beta_1 \neq \beta_2$) fundamentally break the transverse symmetry of the concentration field, leading to highly selective, asymmetric wall-removal processes. The introduction of flux-weighted uptake centroids and spread diagnostics revealed that the imposed hydrodynamic shear profile controls the streamwise spatial organisation of this removal. Specifically, the antisymmetric shear of Couette flow creates a distinct spatial divergence in upper- and lower-wall uptake locations, whereas the symmetric Poiseuille flow confines reactive depletion to overlapping streamwise coordinates. These findings confirm that macroscopic reactive dispersion cannot be fully characterised by total solute removal alone; the specific advective-diffusive pathways dictating local wall contact are equally critical.

Ultimately, this work establishes that PINNs provide an interpretable mesh-free framework for analysing complex mass transport problems. Nevertheless, the precise evaluation of the longitudinal dispersion coefficient remains challenging because its computation relies on derivatives of the second moment of the expected concentration field, which are prone to local approximation errors. Future research will consequently focus on improving PINN formulations that enhance derivative accuracy, thereby permitting more reliable prediction of dispersion properties. By combining PINNs with the classical technique of moments and mean concentration expansion, further work may eliminate the need to explicitly resolve the singular initial condition. Diffusiophoretic transport, nonlinear phase-exchange kinetics, and non-Newtonian or oscillatory flow regimes can be studied using this framework. The PINN framework can also be extended to study tracer dispersion in advection-dominated flows.

\backsection[Acknowledgements]{
The authors thank their respective institutions for providing a supportive research environment.
}

\backsection[Declaration of interests]{
The authors report no conflict of interest.
}

\backsection[Data availability statement]{
The codes which developed to generate the results in this study are available from the corresponding author upon reasonable request.
}

\bibliographystyle{jfm}
\bibliography{Reference}

\end{document}


\maketitle

\section{Supplementary concentration-field validation}
\label{sec:supp-concentration-validation}

The main manuscript reports representative reactive validation results for the point-like Gaussian source, including blockwise absolute errors and cross-sectionally averaged concentration comparisons. The present supplementary material provides the corresponding reactive validation for the line-like Gaussian source. Together, the main manuscript and the supplementary material validate the learned concentration field across complementary source configurations, the three canonical mean-centred shear profiles and the representative reactive-wall configurations considered in the study.

The line-like Gaussian source is
\begin{equation}
C_0^{(\mathrm{L})}(x,y)=\exp(-x^2).
\label{eq:supp-line-source}
\end{equation}
This source is localised in the streamwise direction and uniform in the wall-normal direction. Therefore, it mainly tests the ability of the PINN to reproduce shear-induced axial spreading under reactive wall absorption. This is complementary to the point-like Gaussian source used in the main reactive-dispersion diagnostics, which additionally tests wall-normal localisation and transverse diffusion.

The blockwise absolute error $B_{\mathrm{err}}(x,y,t)$ and the cross-sectionally averaged mean-concentration error $E_{\langle C \rangle_y}(x,t)$ are evaluated exactly as defined in Section 4.1 of the main manuscript. These diagnostics provide a dual validation: $B_{\mathrm{err}}$ acts as a local space--time validation metric, whereas $E_{\langle C \rangle_y}$ confirms the accuracy of the streamwise transport after averaging over the channel height.

The supplementary validation is presented across the three canonical mean-centred shear profiles $s\in\{\mathrm{C},\mathrm{Po},\mathrm{CP}\}$ and the three primary reactive-wall configurations: symmetric absorption $(\beta_1,\beta_2)=(1,1)$, lower-wall-dominated absorption $(\beta_1,\beta_2)=(0.2,2)$, and upper-wall-dominated absorption $(\beta_1,\beta_2)=(2,0.2)$.

\section{Reactive blockwise absolute error for the line-like source}
\label{sec:supp-line-like-reactive-error}
\begin{figure}
\centering
\includegraphics[width=0.98\linewidth]{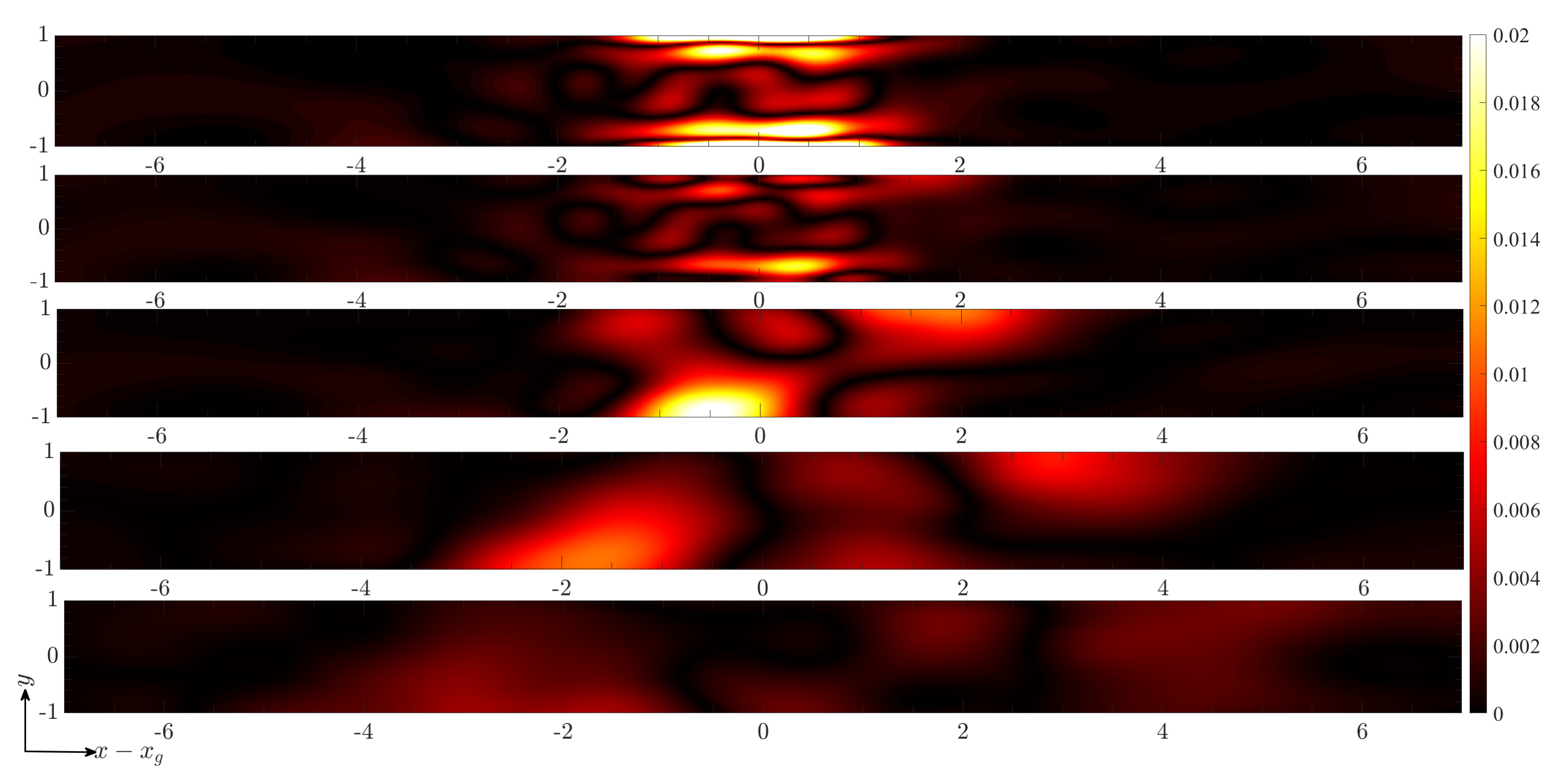}
\caption{
Blockwise absolute error for Couette flow with the line-like source $C_0^{(\mathrm{L})}(x,y)=\exp(-x^2)$ and symmetric wall absorption $(\beta_1,\beta_2)=(1,1)$. The temporal blocks show the local PINN--ADI discrepancy during the evolution of the concentration field.
}
\label{fig:supp-berr-c-s1-beta11}
\end{figure}

\begin{figure}
\centering
\includegraphics[width=0.98\linewidth]{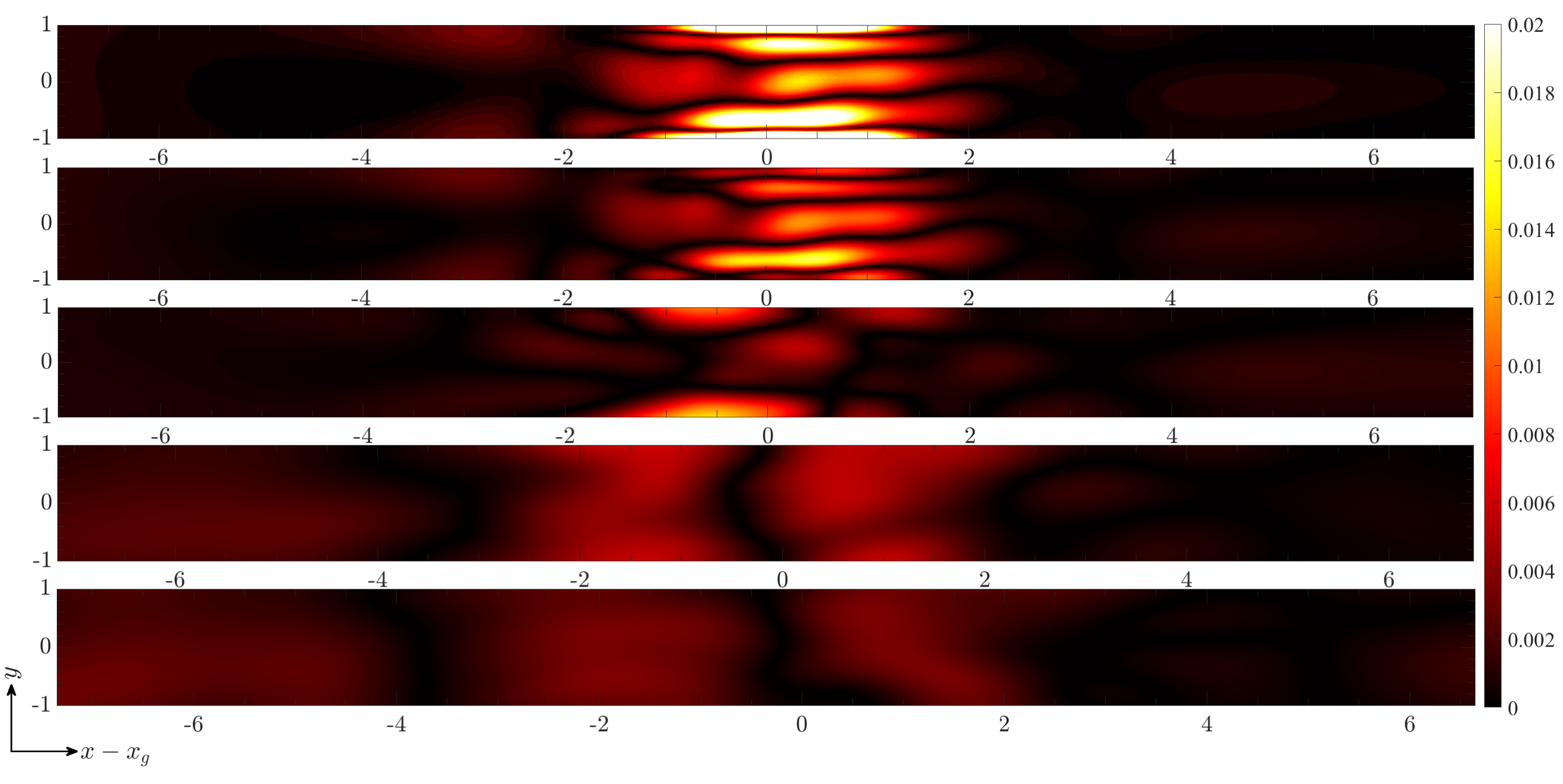}
\caption{
Blockwise absolute error for Poiseuille flow with the line-like source $C_0^{(\mathrm{L})}(x,y)=\exp(-x^2)$ and symmetric wall absorption $(\beta_1,\beta_2)=(1,1)$. The localised error structure indicates that the PINN remains close to the ADI benchmark under parabolic shear and equal wall reactivity.
}
\label{fig:supp-berr-p-s1-beta11}
\end{figure}

\begin{figure}
\centering
\includegraphics[width=0.98\linewidth]{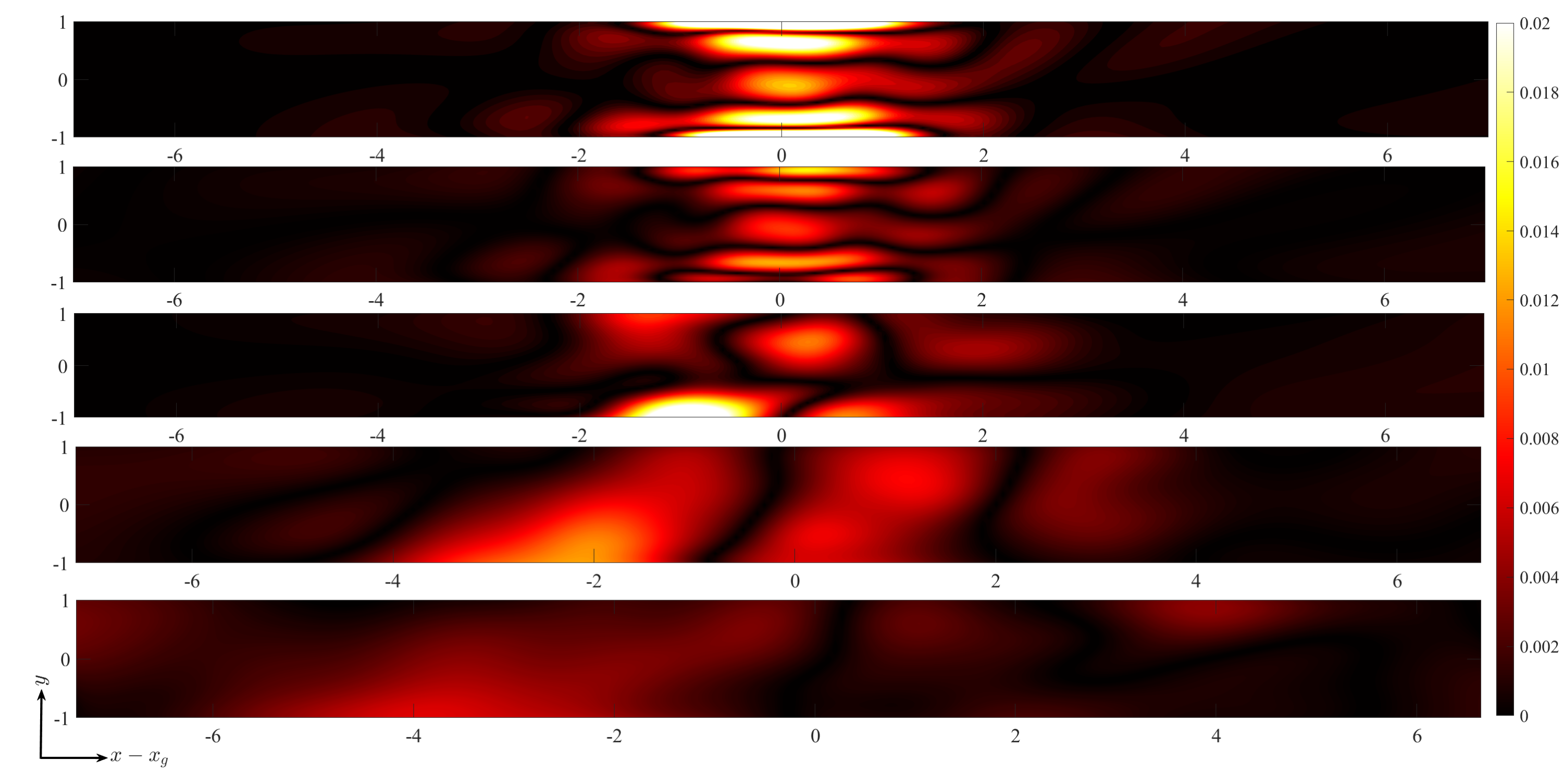}
\caption{
Blockwise absolute error for Couette--Poiseuille flow with the line-like source $C_0^{(\mathrm{L})}(x,y)=\exp(-x^2)$ and symmetric wall absorption $(\beta_1,\beta_2)=(1,1)$. The result tests the learned concentration field under combined linear and parabolic shear with balanced wall removal.
}
\label{fig:supp-berr-cp-s1-beta11}
\end{figure}

\begin{figure}
\centering
\includegraphics[width=0.98\linewidth]{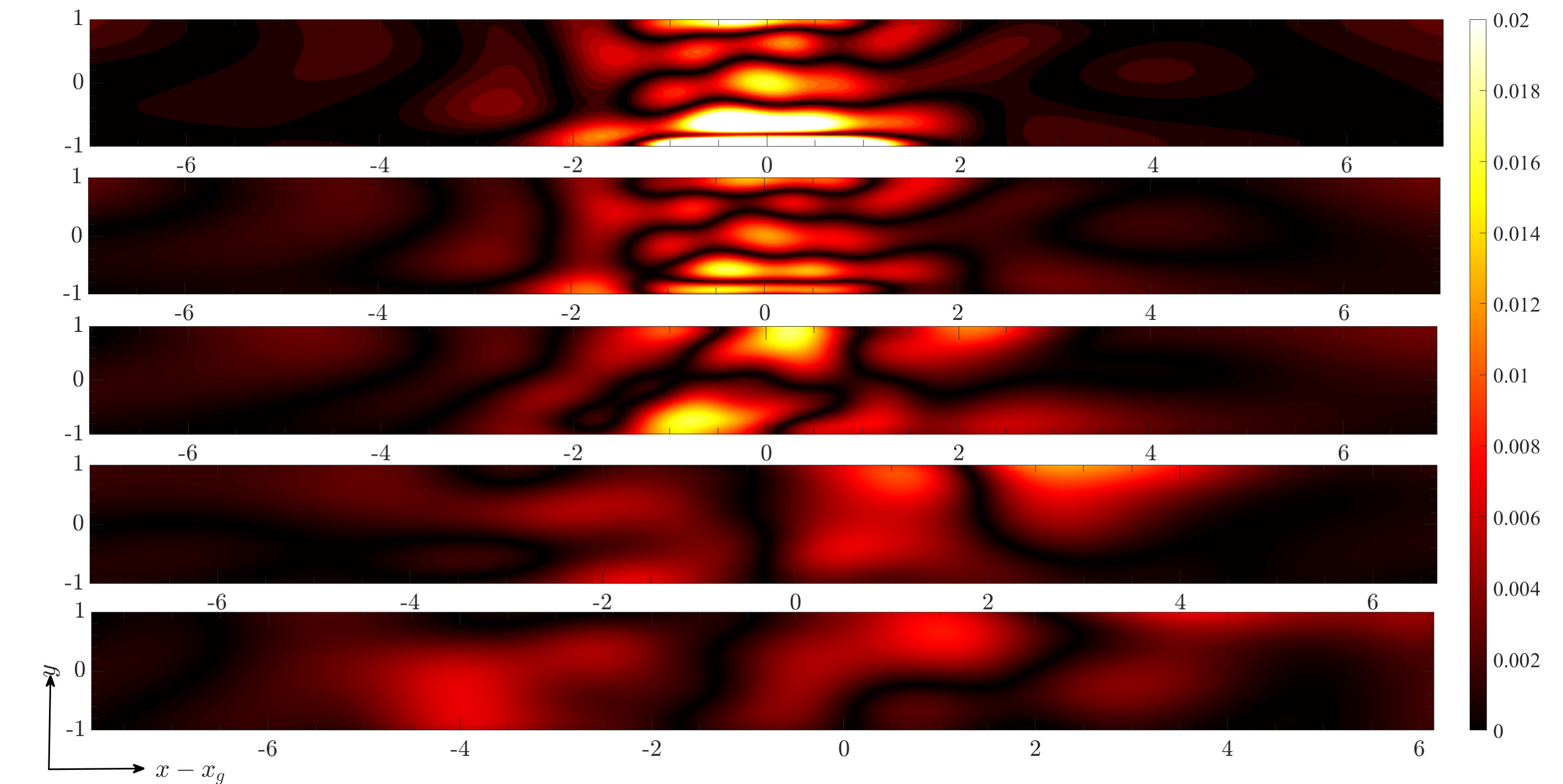}
\caption{
Blockwise absolute error for Couette flow with the line-like source $C_0^{(\mathrm{L})}(x,y)=\exp(-x^2)$ and asymmetric wall absorption $(\beta_1,\beta_2)=(0.2,2)$. In this case, the lower wall is more reactive than the upper wall.
}
\label{fig:supp-berr-c-s1-beta022}
\end{figure}

\begin{figure}
\centering
\includegraphics[width=0.98\linewidth]{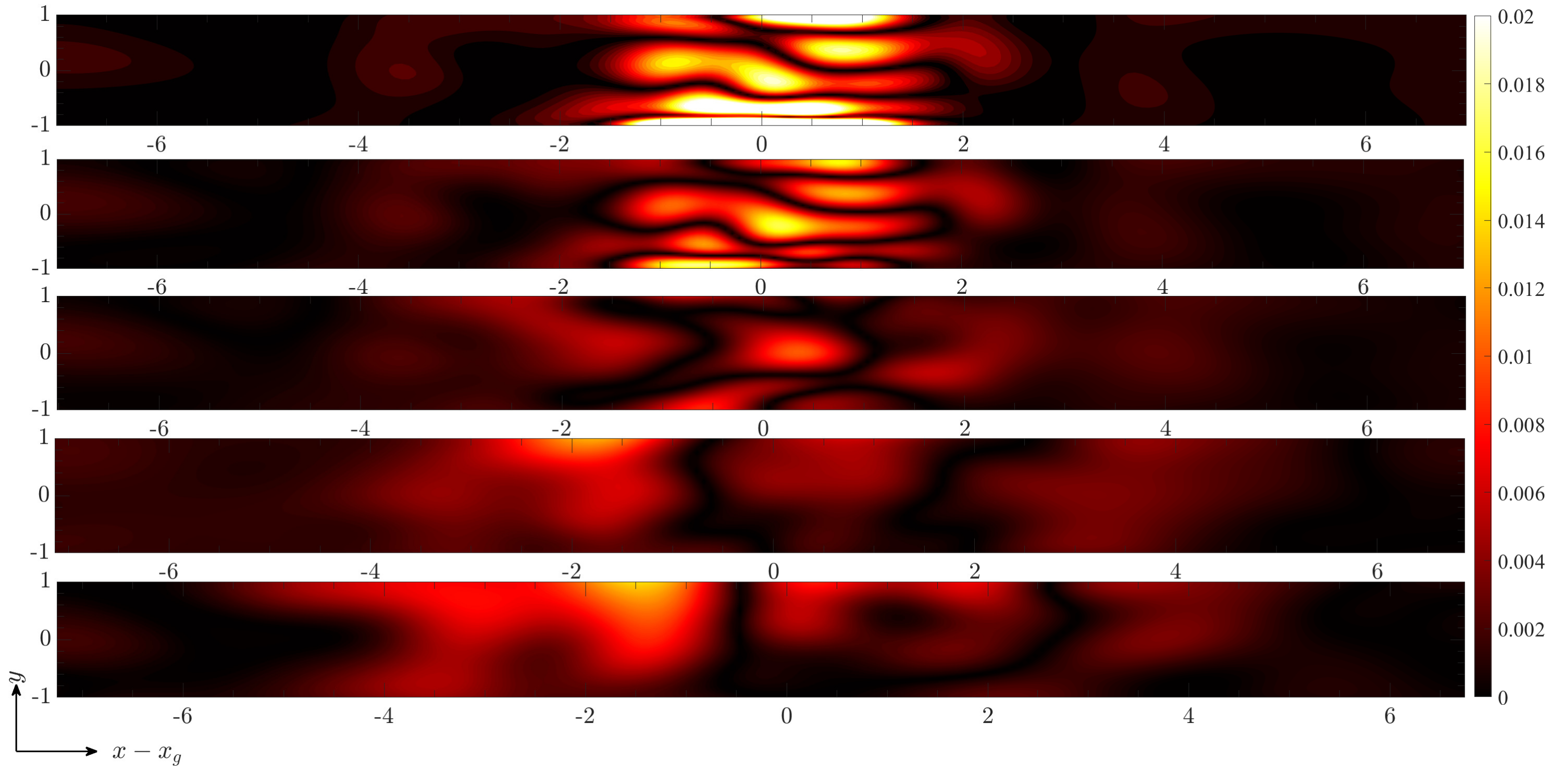}
\caption{
Blockwise absolute error for Poiseuille flow with the line-like source $C_0^{(\mathrm{L})}(x,y)=\exp(-x^2)$ and asymmetric wall absorption $(\beta_1,\beta_2)=(0.2,2)$. The figure validates the learned concentration field when lower-wall uptake dominates under symmetric parabolic shear.
}
\label{fig:supp-berr-p-s1-beta022}
\end{figure}

\begin{figure}
\centering
\includegraphics[width=0.98\linewidth]{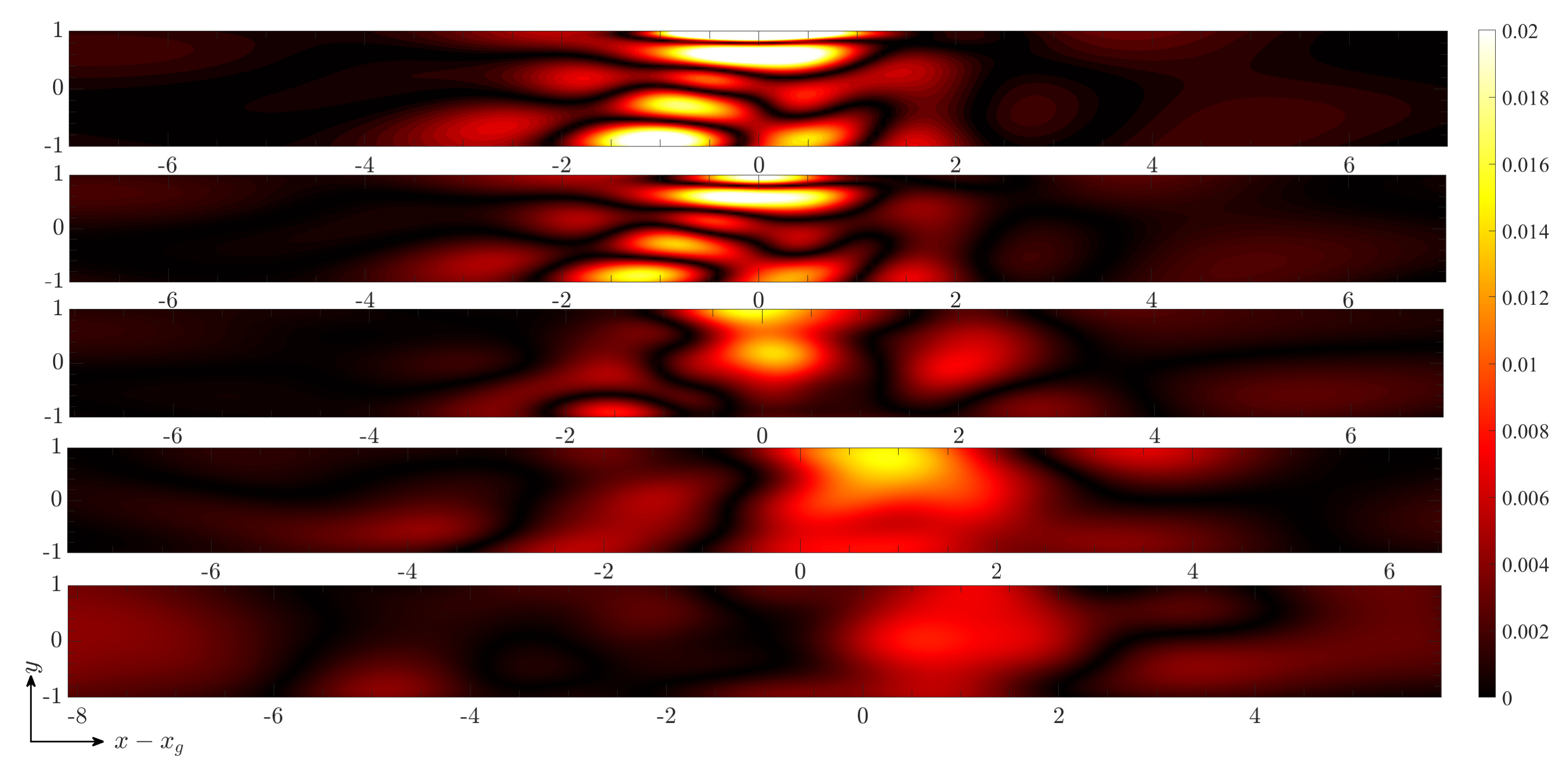}
\caption{
Blockwise absolute error for Couette--Poiseuille flow with the line-like source $C_0^{(\mathrm{L})}(x,y)=\exp(-x^2)$ and asymmetric wall absorption $(\beta_1,\beta_2)=(0.2,2)$. The case combines lower-wall-dominated reaction with the mixed shear structure of Couette--Poiseuille flow.
}
\label{fig:supp-berr-cp-s1-beta022}
\end{figure}

\begin{figure}
\centering
\includegraphics[width=0.98\linewidth]{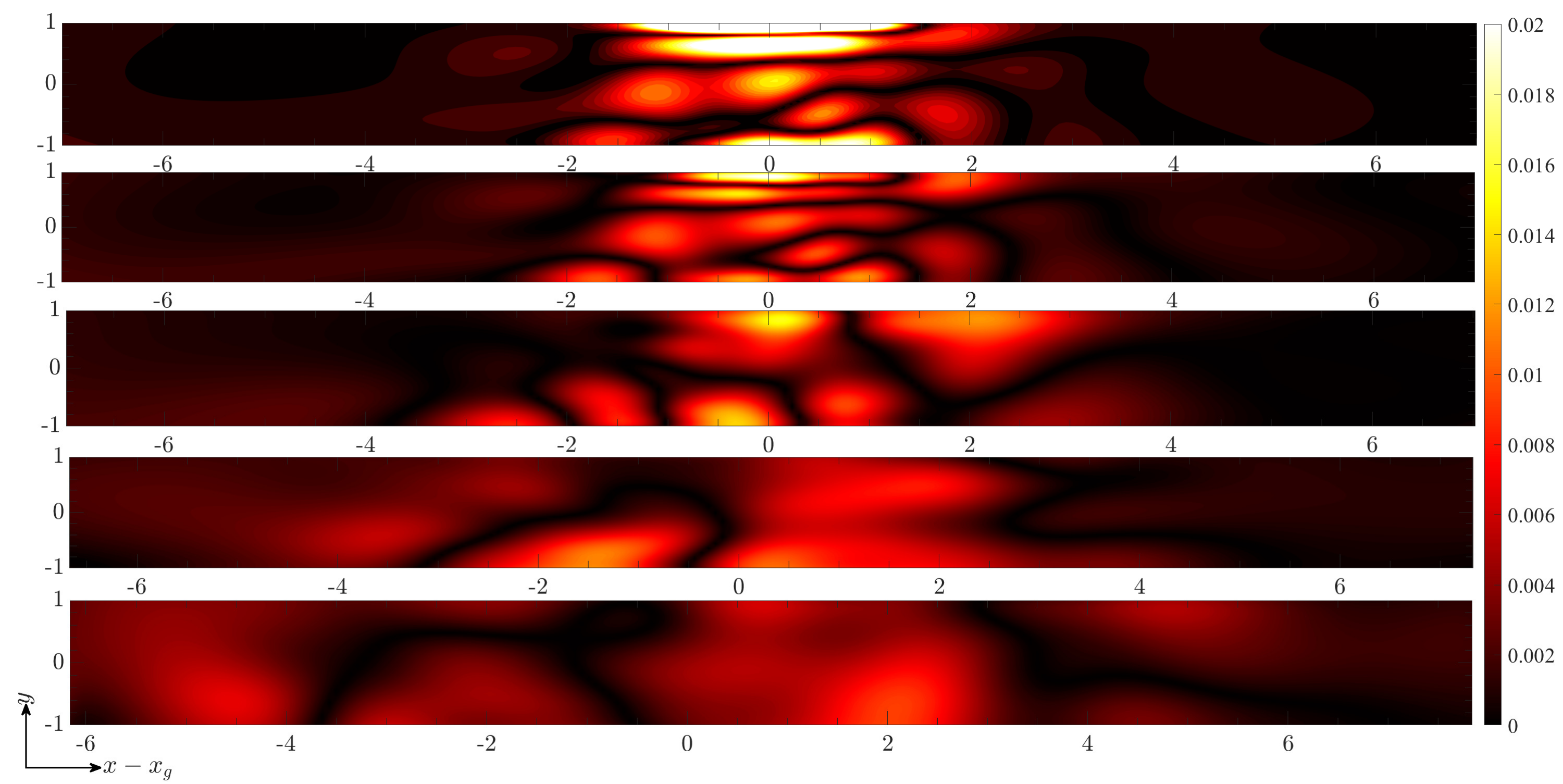}
\caption{
Blockwise absolute error for Couette flow with the line-like source $C_0^{(\mathrm{L})}(x,y)=\exp(-x^2)$ and asymmetric wall absorption $(\beta_1,\beta_2)=(2,0.2)$. In this case, the upper wall is more reactive than the lower wall.
}
\label{fig:supp-berr-c-s1-beta202}
\end{figure}

\begin{figure}
\centering
\includegraphics[width=0.98\linewidth]{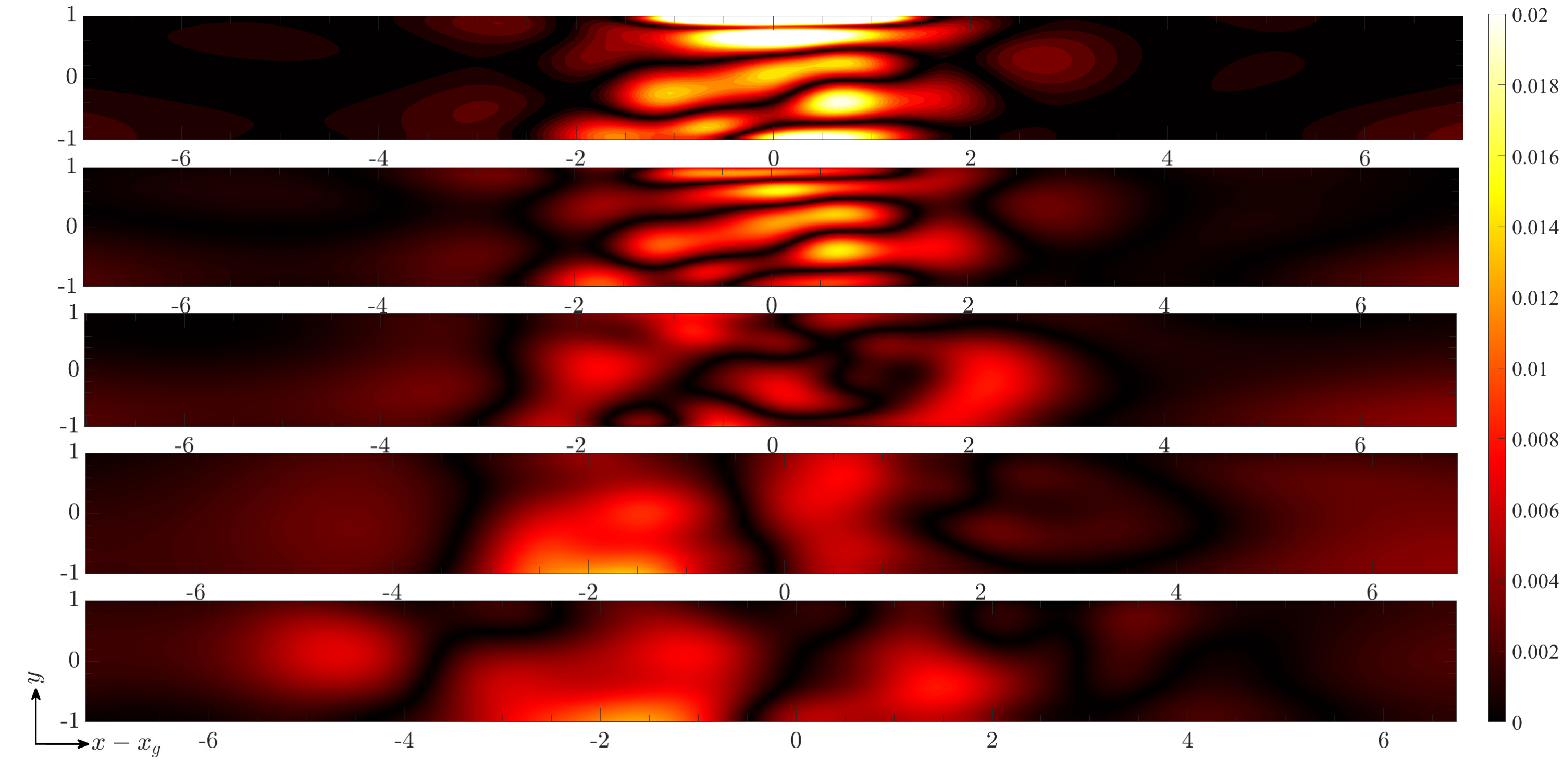}
\caption{
Blockwise absolute error for Poiseuille flow with the line-like source $C_0^{(\mathrm{L})}(x,y)=\exp(-x^2)$ and asymmetric wall absorption $(\beta_1,\beta_2)=(2,0.2)$. The result validates the concentration field when the wall-selective absorption bias is reversed.
}
\label{fig:supp-berr-p-s1-beta202}
\end{figure}

\begin{figure}
\centering
\includegraphics[width=0.98\linewidth]{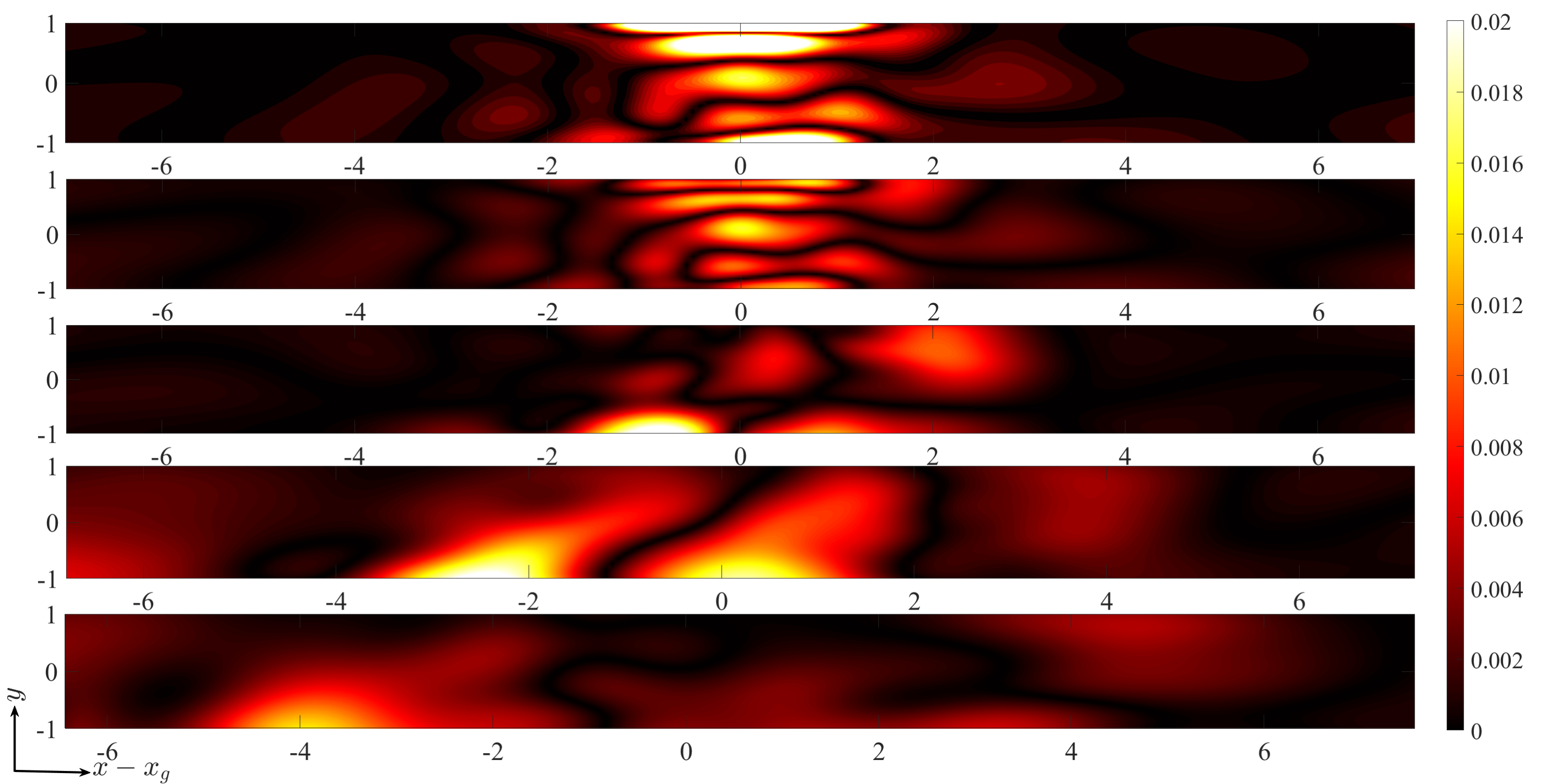}
\caption{
Blockwise absolute error for Couette--Poiseuille flow with the line-like source $C_0^{(\mathrm{L})}(x,y)=\exp(-x^2)$ and asymmetric wall absorption $(\beta_1,\beta_2)=(2,0.2)$. The figure shows that the PINN remains close to the ADI benchmark under combined shear and upper-wall-dominated absorption.
}
\label{fig:supp-berr-cp-s1-beta202}
\end{figure}

Figures~\ref{fig:supp-berr-c-s1-beta11}--\ref{fig:supp-berr-cp-s1-beta202} show the reactive blockwise absolute error for the line-like source across the three canonical shear profiles and the three reactive-wall configurations. Instead of placing all nine panels in a single compressed matrix, the results are separated into individual figures. This arrangement keeps each temporal block clearly visible while preserving the systematic comparison across wall reactivity and shear profile. Figures~\ref{fig:supp-berr-c-s1-beta11}--\ref{fig:supp-berr-cp-s1-beta11} correspond to symmetric wall absorption, \((\beta_1,\beta_2)=(1,1)\). Figures~\ref{fig:supp-berr-c-s1-beta022}--\ref{fig:supp-berr-cp-s1-beta022} correspond to lower-wall-dominated absorption, \((\beta_1,\beta_2)=(0.2,2)\). Figures~\ref{fig:supp-berr-c-s1-beta202}--\ref{fig:supp-berr-cp-s1-beta202} correspond to upper-wall-dominated absorption, \((\beta_1,\beta_2)=(2,0.2)\).



\section{Cross-sectionally averaged concentration validation for the line-like source}
\label{sec:supp-line-like-mean-concentration-error}

\begin{figure}
\centering
\includegraphics[width=0.98\linewidth]{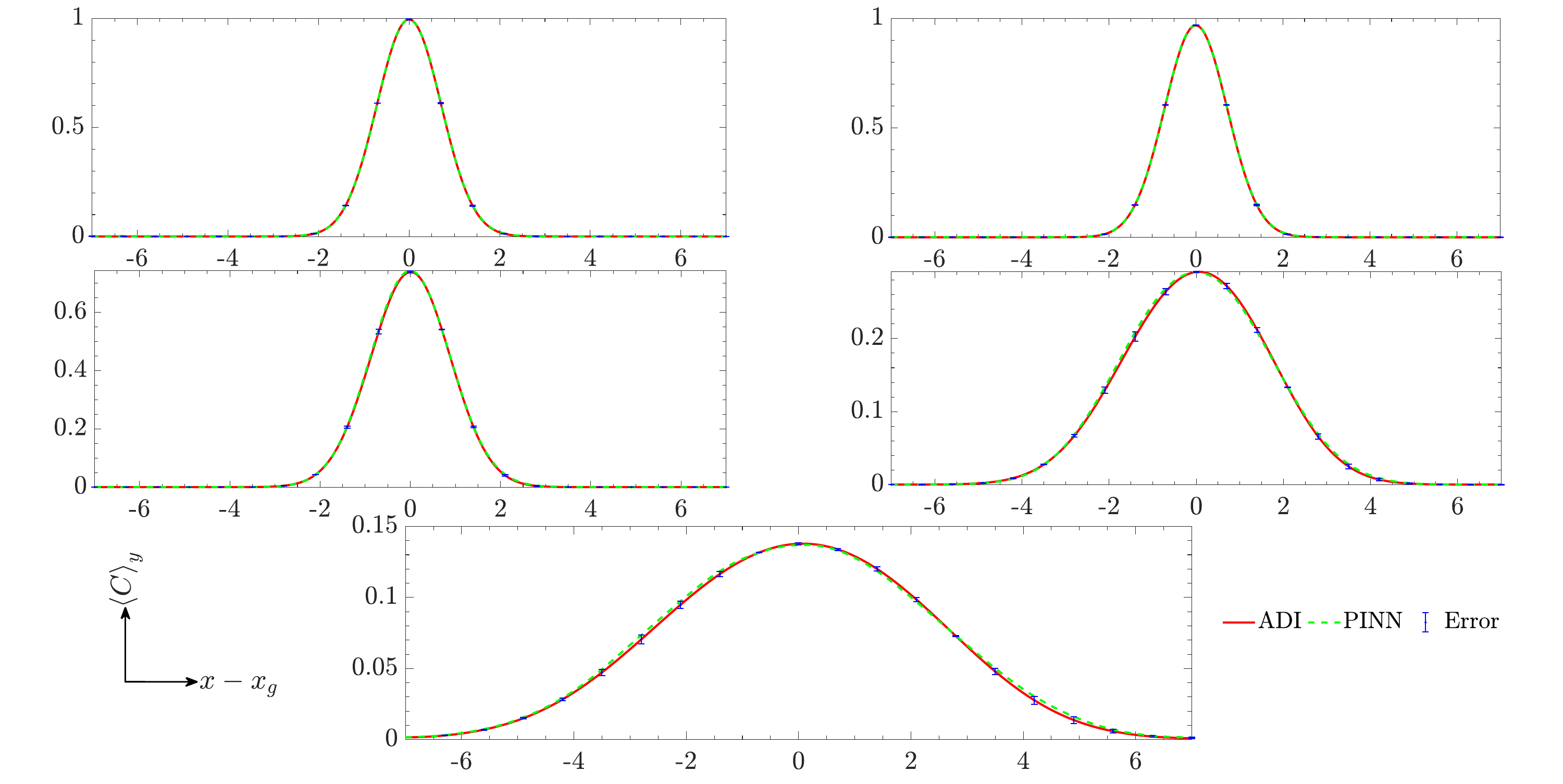}
\caption{
Cross-sectionally averaged concentration comparison and mean-concentration error for Couette flow with the line-like source $C_0^{(\mathrm{L})}(x,y)=\exp(-x^2)$ and symmetric wall absorption $(\beta_1,\beta_2)=(1,1)$. The figure compares $\langle C \rangle_{y,\mathrm{PINN}}(x,t)$ with $\langle C \rangle_{y,\mathrm{ADI}}(x,t)$.
}
\label{fig:supp-cmwe-c-s1-beta11}
\end{figure}

\begin{figure}
\centering
\includegraphics[width=0.98\linewidth]{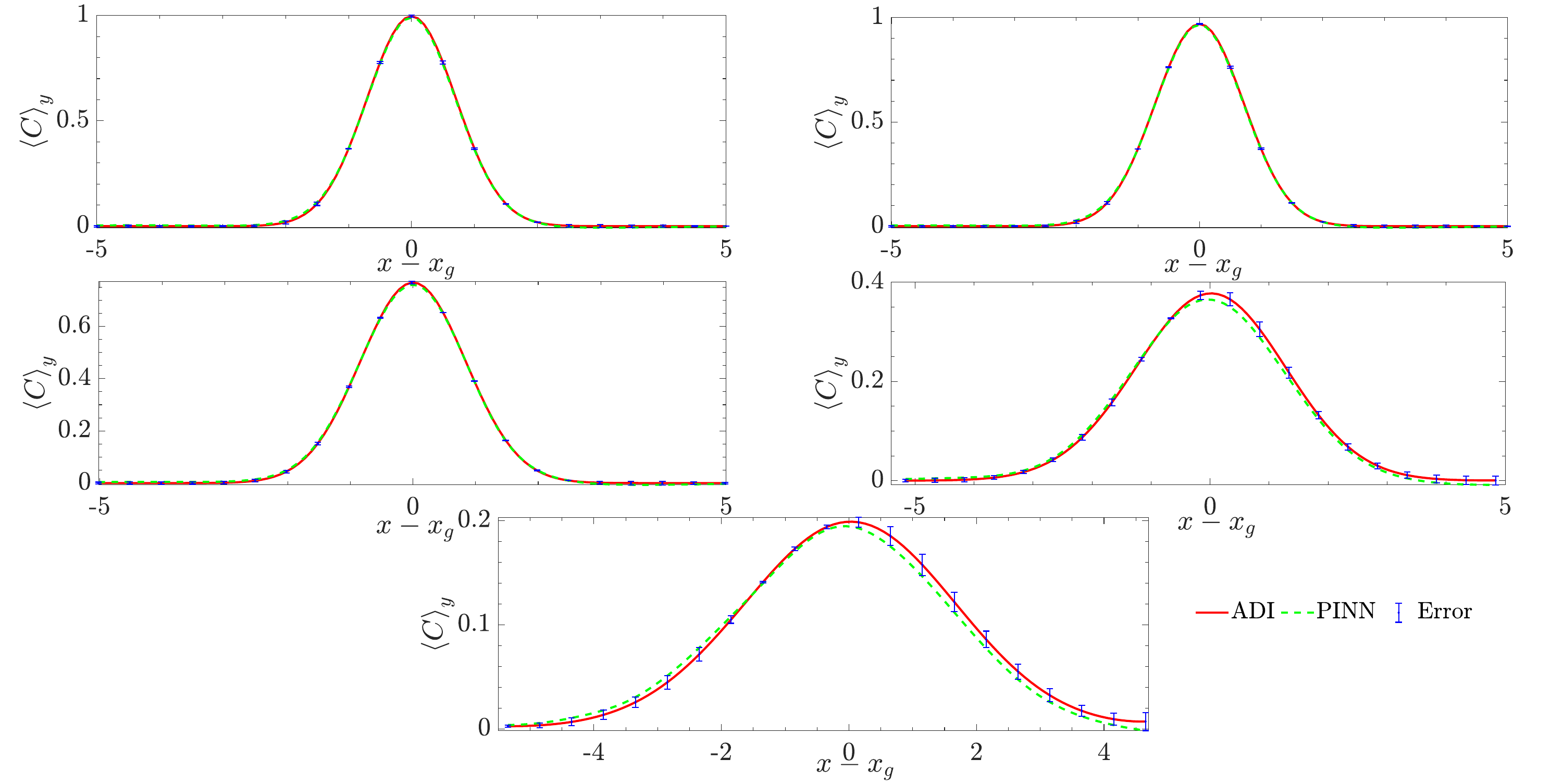}
\caption{
Cross-sectionally averaged concentration comparison and mean-concentration error for Poiseuille flow with the line-like source $C_0^{(\mathrm{L})}(x,y)=\exp(-x^2)$ and symmetric wall absorption $(\beta_1,\beta_2)=(1,1)$. The figure compares $\langle C \rangle_{y,\mathrm{PINN}}(x,t)$ with $\langle C \rangle_{y,\mathrm{ADI}}(x,t)$.
}
\label{fig:supp-cmwe-p-s1-beta11}
\end{figure}

\begin{figure}
\centering
\includegraphics[width=0.98\linewidth]{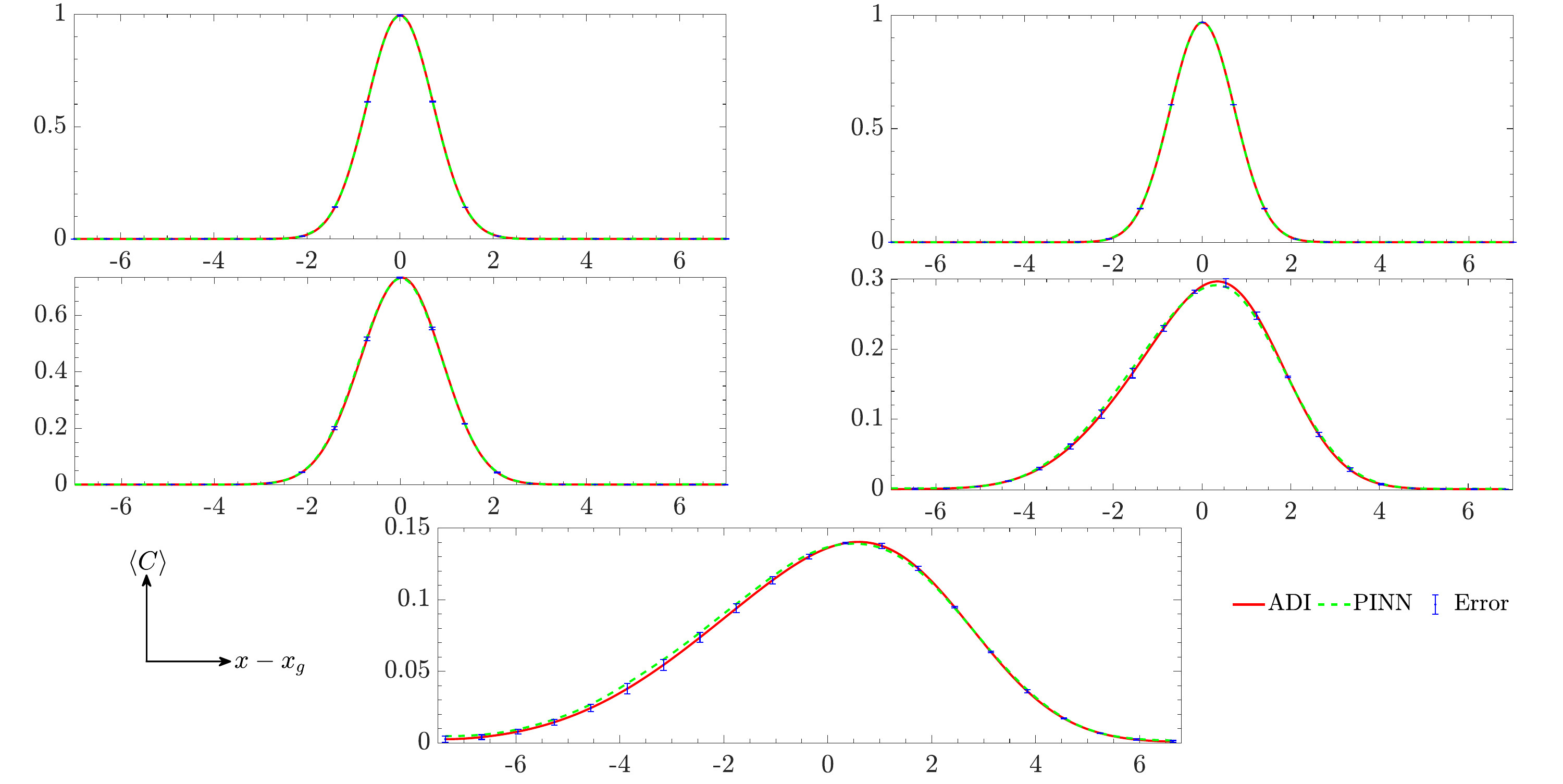}
\caption{
Cross-sectionally averaged concentration comparison and mean-concentration error for Couette--Poiseuille flow with the line-like source $C_0^{(\mathrm{L})}(x,y)=\exp(-x^2)$ and symmetric wall absorption $(\beta_1,\beta_2)=(1,1)$. The figure compares $\langle C \rangle_{y,\mathrm{PINN}}(x,t)$ with $\langle C \rangle_{y,\mathrm{ADI}}(x,t)$.
}
\label{fig:supp-cmwe-cp-s1-beta11}
\end{figure}

\begin{figure}
\centering
\includegraphics[width=0.98\linewidth]{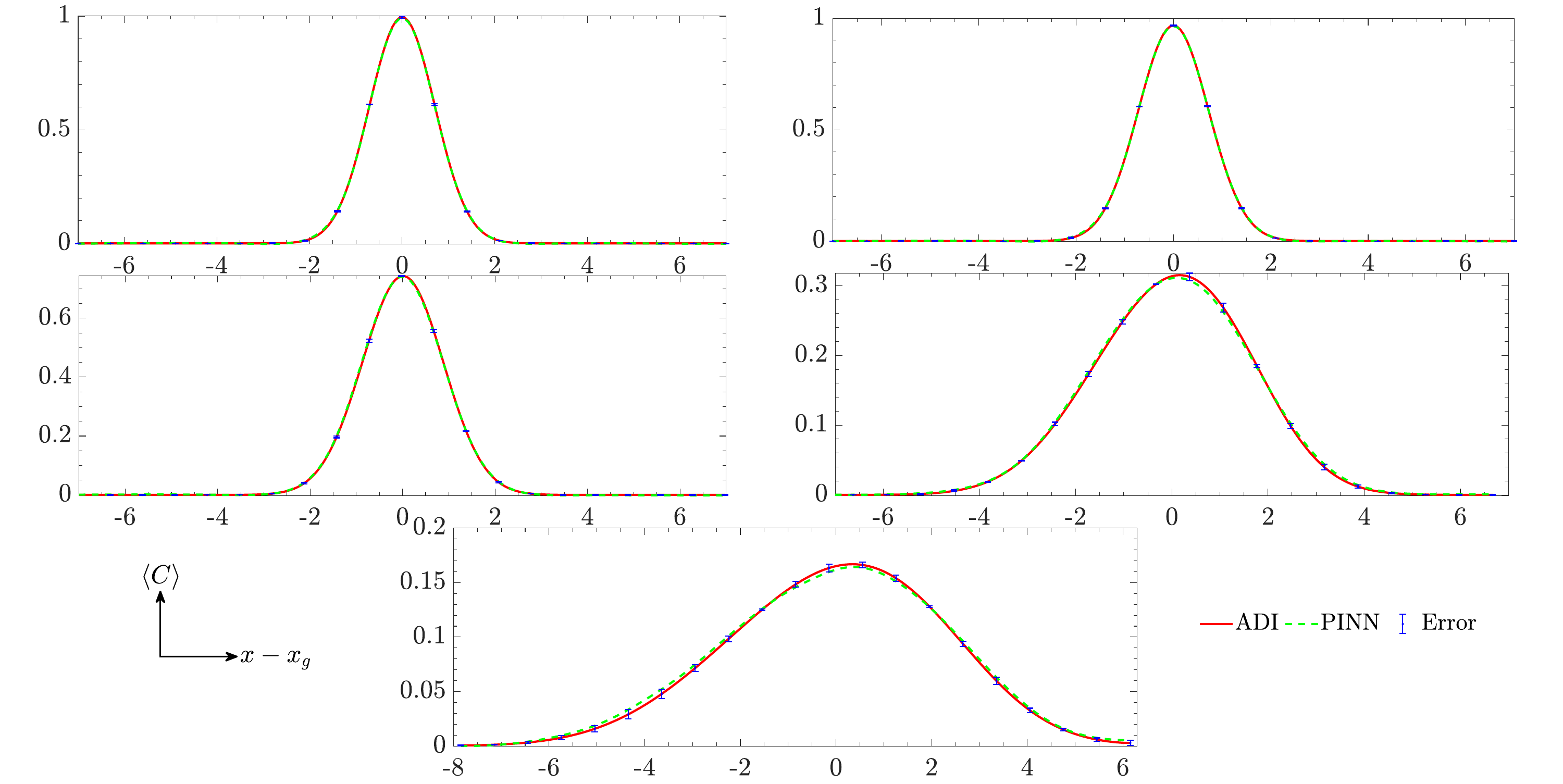}
\caption{
Cross-sectionally averaged concentration comparison and mean-concentration error for Couette flow with the line-like source $C_0^{(\mathrm{L})}(x,y)=\exp(-x^2)$ and lower-wall-dominated absorption $(\beta_1,\beta_2)=(0.2,2)$. The figure compares $\langle C \rangle_{y,\mathrm{PINN}}(x,t)$ with $\langle C \rangle_{y,\mathrm{ADI}}(x,t)$.
}
\label{fig:supp-cmwe-c-s1-beta022}
\end{figure}

\begin{figure}
\centering
\includegraphics[width=0.98\linewidth]{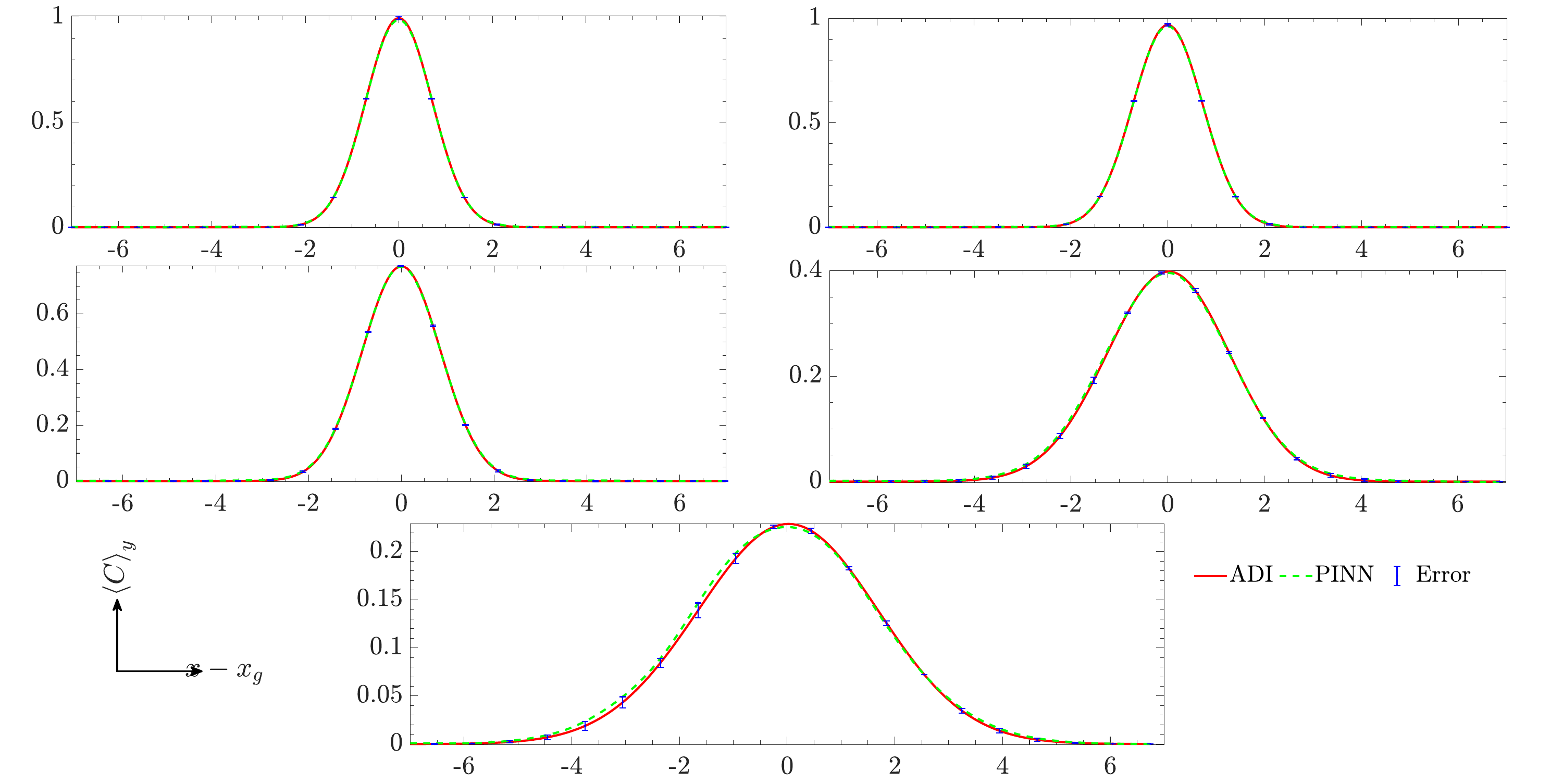}
\caption{
Cross-sectionally averaged concentration comparison and mean-concentration error for Poiseuille flow with the line-like source $C_0^{(\mathrm{L})}(x,y)=\exp(-x^2)$ and lower-wall-dominated absorption $(\beta_1,\beta_2)=(0.2,2)$. The figure compares $\langle C \rangle_{y,\mathrm{PINN}}(x,t)$ with $\langle C \rangle_{y,\mathrm{ADI}}(x,t)$.
}
\label{fig:supp-cmwe-p-s1-beta022}
\end{figure}

\begin{figure}
\centering
\includegraphics[width=0.98\linewidth]{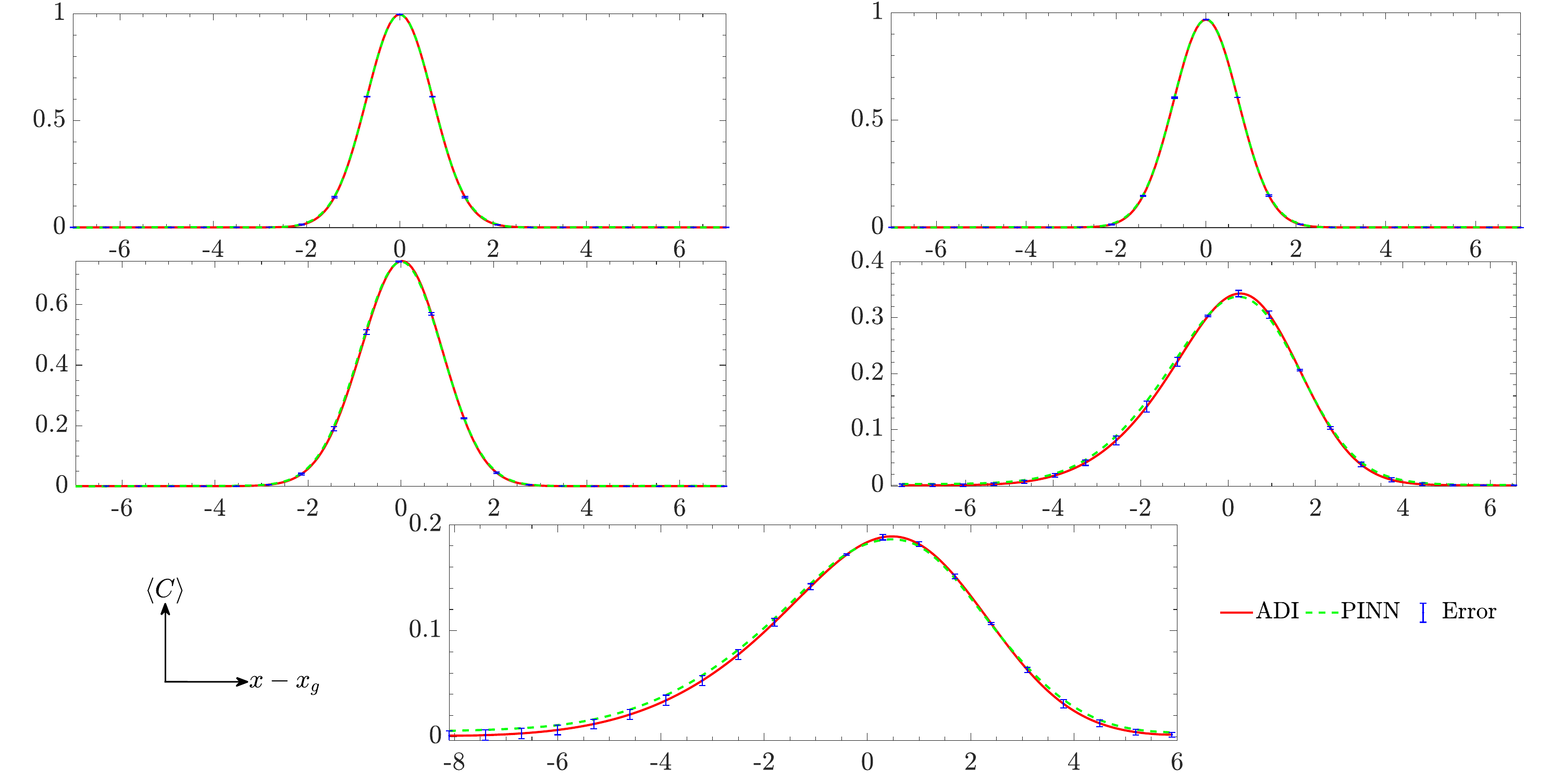}
\caption{
Cross-sectionally averaged concentration comparison and mean-concentration error for Couette--Poiseuille flow with the line-like source $C_0^{(\mathrm{L})}(x,y)=\exp(-x^2)$ and lower-wall-dominated absorption $(\beta_1,\beta_2)=(0.2,2)$. The figure compares $\langle C \rangle_{y,\mathrm{PINN}}(x,t)$ with $\langle C \rangle_{y,\mathrm{ADI}}(x,t)$.
}
\label{fig:supp-cmwe-cp-s1-beta022}
\end{figure}

\begin{figure}
\centering
\includegraphics[width=0.98\linewidth]{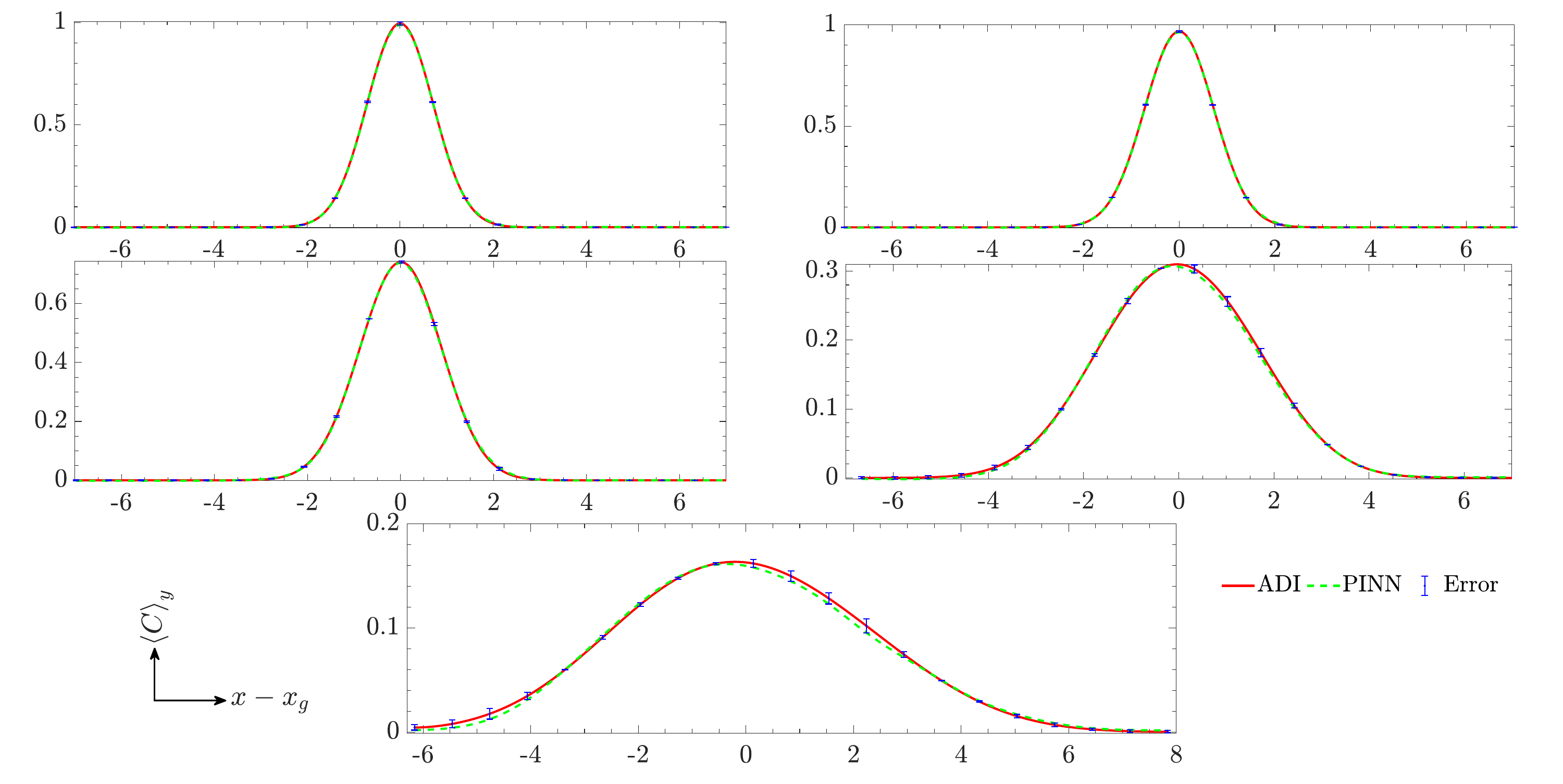}
\caption{
Cross-sectionally averaged concentration comparison and mean-concentration error for Couette flow with the line-like source $C_0^{(\mathrm{L})}(x,y)=\exp(-x^2)$ and upper-wall-dominated absorption $(\beta_1,\beta_2)=(2,0.2)$. The figure compares $\langle C \rangle_{y,\mathrm{PINN}}(x,t)$ with $\langle C \rangle_{y,\mathrm{ADI}}(x,t)$.
}
\label{fig:supp-cmwe-c-s1-beta202}
\end{figure}

\begin{figure}
\centering
\includegraphics[width=0.98\linewidth]{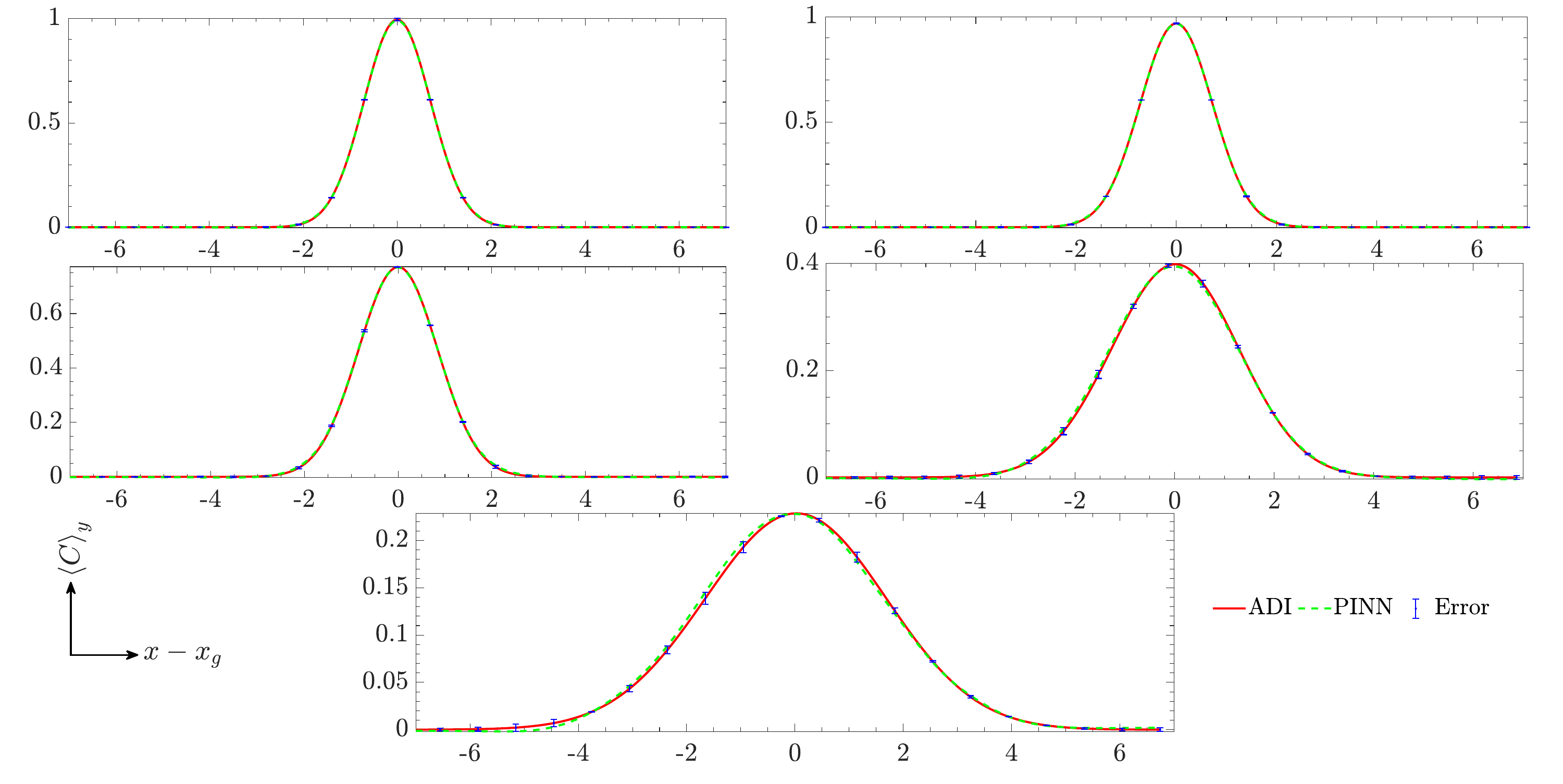}
\caption{
Cross-sectionally averaged concentration comparison and mean-concentration error for Poiseuille flow with the line-like source $C_0^{(\mathrm{L})}(x,y)=\exp(-x^2)$ and upper-wall-dominated absorption $(\beta_1,\beta_2)=(2,0.2)$. The figure compares $\langle C \rangle_{y,\mathrm{PINN}}(x,t)$ with $\langle C \rangle_{y,\mathrm{ADI}}(x,t)$.
}
\label{fig:supp-cmwe-p-s1-beta202}
\end{figure}

\begin{figure}
\centering
\includegraphics[width=0.98\linewidth]{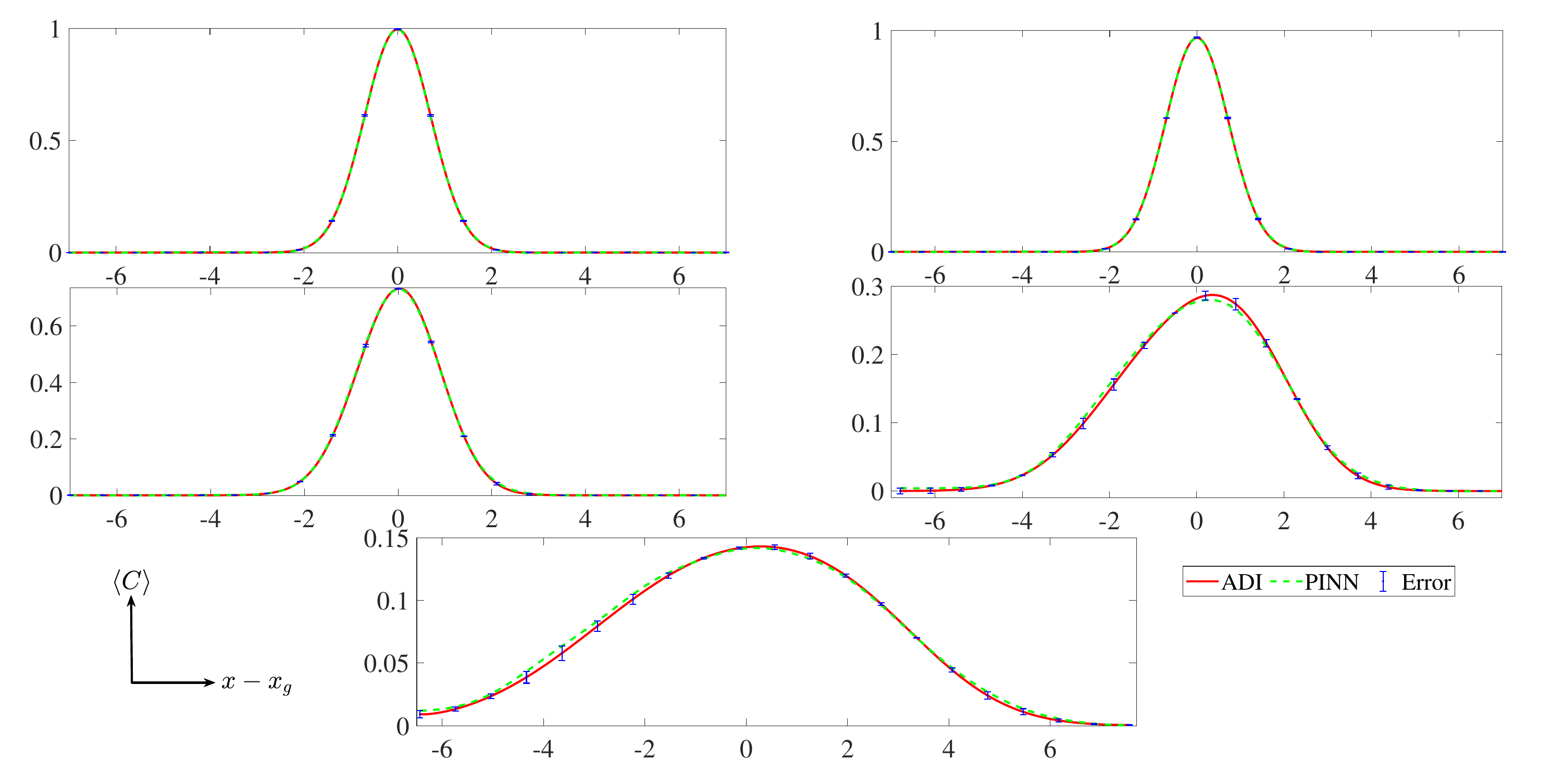}
\caption{
Cross-sectionally averaged concentration comparison and mean-concentration error for Couette--Poiseuille flow with the line-like source $C_0^{(\mathrm{L})}(x,y)=\exp(-x^2)$ and upper-wall-dominated absorption $(\beta_1,\beta_2)=(2,0.2)$. The figure compares $\langle C \rangle_{y,\mathrm{PINN}}(x,t)$ with $\langle C \rangle_{y,\mathrm{ADI}}(x,t)$.
}
\label{fig:supp-cmwe-cp-s1-beta202}
\end{figure}

Figures~\ref{fig:supp-cmwe-c-s1-beta11}--\ref{fig:supp-cmwe-cp-s1-beta202} show the cross-sectionally averaged concentration profiles and the corresponding mean-concentration errors for the line-like source. These results complement the blockwise absolute-error validation by demonstrating the accuracy of the streamwise-averaged transport field. Figures~\ref{fig:supp-cmwe-c-s1-beta11}--\ref{fig:supp-cmwe-cp-s1-beta11} correspond to symmetric wall absorption, \((\beta_1,\beta_2)=(1,1)\). Figures~\ref{fig:supp-cmwe-c-s1-beta022}--\ref{fig:supp-cmwe-cp-s1-beta022} correspond to lower-wall-dominated absorption, \((\beta_1,\beta_2)=(0.2,2)\). Figures~\ref{fig:supp-cmwe-c-s1-beta202}--\ref{fig:supp-cmwe-cp-s1-beta202} correspond to upper-wall-dominated absorption, \((\beta_1,\beta_2)=(2,0.2)\).

\section{Interpretation of supplementary validation}
\label{sec:supp-interpretation}

Figures~\ref{fig:supp-berr-c-s1-beta11}--\ref{fig:supp-berr-cp-s1-beta202} complement the main manuscript by showing that the blockwise concentration-field agreement is not restricted to the point-like Gaussian source. For the line-like source, the initial concentration field is uniform across the channel height and localised only in the streamwise direction. This source emphasises shear-induced axial spreading and therefore provides a different validation challenge from the point-like source.

Figures~\ref{fig:supp-cmwe-c-s1-beta11}--\ref{fig:supp-cmwe-cp-s1-beta202} provide the corresponding cross-sectionally averaged concentration validation. These plots show whether the PINN accurately reproduces the streamwise mean solute distribution after averaging over the channel height. Thus, the supplementary validation checks both local two-dimensional accuracy through $B_{\mathrm{err}}$ and averaged streamwise transport accuracy through $E_{\langle C \rangle_y}$.

For the symmetric reactive cases $(\beta_1,\beta_2)=(1,1)$, both walls remove solute with equal strength. The blockwise errors and mean-concentration errors remain localised across Couette, Poiseuille and Couette--Poiseuille flows, indicating that the PINN captures the concentration evolution arising from the combined effects of shear, diffusion and symmetric wall absorption.

For $(\beta_1,\beta_2)=(0.2,2)$, the lower wall is more reactive than the upper wall. This configuration introduces a transverse asymmetry in the wall-removal process. The comparison with the ADI solution shows that the PINN remains accurate even when wall absorption is strongly biased toward one boundary. For $(\beta_1,\beta_2)=(2,0.2)$, the transverse absorption bias is reversed, and the supplementary figures show that the PINN is also able to reproduce this upper-wall-dominated concentration dynamics.

The supplementary validation confirms that the concentration-field agreement between the PINN and ADI benchmark holds for the line-like source under both symmetric and asymmetric reactive-wall configurations. Together with the point-like source validation reported in the main manuscript, these results support the use of the trained PINN solution for extracting wall-resolved reactive-dispersion diagnostics, including $\mathcal{J}_{\pm}(t)$, $\mathcal{U}_{\pm}(t)$, $D_w(t)$, $x_{J_\pm}(t)$ and $\sigma_{J_\pm}(t)$.